\documentclass[]{aa}  

\usepackage{graphicx}
\usepackage{txfonts}
\usepackage{natbib}
\usepackage{xcolor}
\usepackage{float}
\usepackage{lscape}
\usepackage{upgreek}
\usepackage{longtable}
\bibpunct{(}{)}{;}{a}{}{,}
\usepackage{mathrsfs}
\usepackage{fontenc} 
\usepackage{txfonts} 
\usepackage{placeins}
\newcounter{subfigure}
\newcounter{subtable}

\begin{document} 

\renewcommand{\thefigure}{\arabic{figure}\alph{subfigure}}
\renewcommand{\thetable}{\arabic{table}\alph{subtable}}

\title{The {\em Gaia}-ESO survey: A spectroscopic study of the young open cluster NGC 3293
  \thanks{Based on observations made with the ESO/VLT, at Paranal Observatory, under program 188.B-3002 (the {\em Gaia}-ESO public spectroscopic survey, PIs G. Gilmore and S. Randich). Also based on observations under programs 171.0237 and 073.0234.}
  \thanks{The full Tables \ref{tab_observations}, \ref{tab_results}, and \ref{tab_rotation} are only available in electronic form at the CDS via anonymous ftp to cdsarc.u-strasbg.fr (130.79.128.5) or via
  }
}
   \titlerunning{The {\em Gaia}-ESO survey: A spectroscopic study of NGC 3293}
  \authorrunning{T. Morel et al.}
   
     \author{T. Morel
         \inst{\ref{inst:01}}\thanks{\email{tmorel@uliege.be}}
         \and
         A. Blaz\`ere
         \inst{\ref{inst:01},\ref{inst:02}}
	 \and
         T. Semaan
         \inst{\ref{inst:01},\ref{inst:03}}
         \and
         E. Gosset
         \inst{\ref{inst:01}}
	 \and         
         J. Zorec
         \inst{\ref{inst:04}}
	 \and
         Y. Fr\'emat
         \inst{\ref{inst:05}}
         \and
         R. Blomme
         \inst{\ref{inst:05}}        
	 \and
         S. Daflon
         \inst{\ref{inst:06}}
	 \and
         A. Lobel
         \inst{\ref{inst:05}}
	 \and
         M. F. Nieva
         \inst{\ref{inst:07}}
	 \and
         N. Przybilla
         \inst{\ref{inst:07}}
	 \and
         M.~Gebran
         \inst{\ref{inst:71},\ref{inst:75}} 
         \and
         A.~Herrero
         \inst{\ref{inst:24},\ref{inst:25}} 
         \and
         L.~Mahy
         \inst{\ref{inst:05},\ref{inst:72},\ref{inst:01}} 
         \and         
         W. Santos
         \inst{\ref{inst:06}}
         \and         
         G. Tautvai$\check{\rm s}$ien\.e 
         \inst{\ref{inst:08}}
	 \and
         G. Gilmore
         \inst{\ref{inst:09}}
	 \and
         S. Randich
         \inst{\ref{inst:10}}
	 \and
         E.~J. Alfaro
         \inst{\ref{inst:23}}
	 \and
         M. Bergemann
         \inst{\ref{inst:21},\ref{inst:211}}
	 \and
         G. Carraro
         \inst{\ref{inst:38}}
	 \and
         F. Damiani
         \inst{\ref{inst:18}}
	 \and
         E. Franciosini
         \inst{\ref{inst:10}}
	 \and
         L. Morbidelli
         \inst{\ref{inst:10}}
	 \and
         E. Pancino
         \inst{\ref{inst:10},\ref{inst:31}}
	 \and
         C.~C. Worley
         \inst{\ref{inst:09}}          
	 \and
         S. Zaggia
         \inst{\ref{inst:22}}
     }

  \institute{
Space sciences, Technologies and Astrophysics Research (STAR) Institute, Universit\'e de Li\`ege, Quartier Agora, All\'ee du 6 Ao\^ut 19c, B\^at. B5c, B4000-Li\`ege, Belgium\label{inst:01}
\and
Centre National d’Etudes Spatiales, Toulouse, France\label{inst:02}
\and
Observatoire de Gen\`eve, Universit\'e de Gen\`eve, Chemin Pegasi 51, 1290 Versoix, Switzerland\label{inst:03}
\and
Sorbonne Université CNRS, UMR 7095, Institut d’Astrophysique de Paris, 75014 Paris, France\label{inst:04}
\and
Royal Observatory of Belgium, Ringlaan 3, B-1180 Brussels, Belgium\label{inst:05}
\and
Observat\'orio Nacional/MCTIC, R. Gal. Jos\'e Cristino 77, S\~ao Cristov\~ao, 20921-400 Rio de Janeiro/RJ, Brazil\label{inst:06}
\and
Institut f\"ur Astro- und Teilchenphysik, Universit{\"a}t Innsbruck, Technikerstr. 25/8, 6020 Innsbruck, Austria\label{inst:07}
\and
Department of Physics and Astronomy, Notre Dame University-Louaize, PO Box 72, Zouk Mika\"el, Lebanon \label{inst:71}
\and
Department of Chemistry and Physics, Saint Mary’s College, Notre Dame, IN 46556, USA \label{inst:75}
\and
Instituto de Astrof\'{\i}sica de Canarias, E-38205 La Laguna, Tenerife, Spain\label{inst:24}
\and
Universidad de La Laguna, Dept. Astrof\'{\i}sica, E-38206 La Laguna, Tenerife, Spain\label{inst:25}
\and
Institute of Astronomy, KU Leuven, Celestijnenlaan 200D, B-3001 Leuven, Belgium\label{inst:72}
\and
Institute of Theoretical Physics and Astronomy, Vilnius University, Sauletekio av. 3, 10257, Vilnius, Lithuania\label{inst:08}
\and
Institute of Astronomy, University of Cambridge, Madingley Road, Cambridge CB3 0HA, United Kingdom\label{inst:09}
\and
INAF - Osservatorio Astrofisico di Arcetri, Largo E. Fermi 5, 50125, Florence, Italy\label{inst:10}
\and
INAF - Osservatorio Astronomico di Palermo, Piazza del Parlamento 1, 90134, Palermo, Italy\label{inst:18}
\and
Max-Planck Institut f\"{u}r Astronomie, K\"{o}nigstuhl 17, 69117 Heidelberg, Germany\label{inst:21}
\and
Niels Bohr International Academy, Niels Bohr Institute, University of Copenhagen Blegdamsvej 17, DK-2100 Copenhagen, Denmark\label{inst:211}
\and
INAF - Padova Observatory, Vicolo dell'Osservatorio 5, 35122 Padova, Italy\label{inst:22}
\and
Instituto de Astrof\'{i}sica de Andaluc\'{i}a-CSIC, Apdo. 3004, 18080, Granada, Spain\label{inst:23}
\and
Space Science Data Center - Agenzia Spaziale Italiana, via del Politecnico, s.n.c., I-00133, Roma, Italy\label{inst:31}
\and
Dipartimento di Fisica e Astronomia, Universit\`a di Padova, Vicolo dell'Osservatorio 3, 35122 Padova, Italy\label{inst:38}
}

   \date{Received 25 May 2022 ; accepted 19 July 2022}

   \abstract{We present a spectroscopic analysis of the GIRAFFE and UVES data collected by the {\em Gaia}-ESO survey for the young open cluster NGC 3293. Archive spectra from the same instruments obtained in the framework of the `VLT-FLAMES survey of massive stars' are also analysed. Atmospheric parameters, non-local thermodynamic equilibrium (LTE) chemical abundances for six elements (He, C, N, Ne, Mg, and Si), or variability information are reported for a total of about 160 B stars spanning a wide range in terms of spectral types (B1 to B9.5) and rotation rate (up to 350 km s$^{-1}$). Our analysis leads to about a five-fold increase in the number of cluster members with an abundance determination and it characterises the late B-star population in detail for the first time. We take advantage of the multi-epoch observations on various timescales and a temporal baseline, sometimes spanning $\sim$15 years, to detect several binary systems or intrinsically line-profile variables. A deconvolution algorithm is used to infer the current, true (deprojected) rotational velocity distribution. We find a broad, Gaussian-like distribution peaking around 200--250 km s$^{-1}$. Although some stars populate the high-velocity tail, most stars in the cluster appear to rotate far from critical. We discuss the chemical properties of the cluster, including the low occurrence of abundance peculiarities in the late B stars and the paucity of objects showing CN-cycle burning products at their surface. We argue that the former result can largely be explained by the inhibition of diffusion effects because of fast rotation, while the latter is generally in accord with the predictions of single-star evolutionary models under the assumption of a wide range of initial spin rates at the onset of main-sequence evolution. However, we find some evidence for a less efficient mixing in two quite rapidly rotating stars that are among the most massive objects in our sample. Finally, we obtain a cluster age of $\sim$20 Myrs through a detailed, star-to-star correction of our results for the effect of stellar rotation (e.g. gravity darkening). This is significantly older than previous estimates from turn-off fitting that fully relied on classical, non-rotating isochrones.}
   \keywords{Open clusters and associations: individual: NGC 3293 -- Stars: fundamental parameters -- Stars: abundances}

\maketitle
%

\section{Introduction}\label{sect_introduction}

Young open clusters have long been recognised as key testbeds for our understanding of the physics and evolution of massive stars because they provide a snapshot of a stellar population sharing the same distance and initial chemical composition, but with members that span a very wide mass range. The cluster \object{NGC 3293} (also known as the Gem Nebula) belongs to the small cohort of not too distant, well-populated ensembles of massive stars that are particularly well suited for that purpose. The cluster lies in the north-western outskirts of the Carina Nebula (\object{NGC 3372}), which is one of the most interesting and intense sites of star formation relatively nearby \citep{smith08}. Its distance, which was recently estimated to be 2.3--2.4 kpc based on {\em Gaia} EDR3 astrometric data, is fully compatible with that of the Carina association to which it is thus likely physically associated \citep{goppl22}. Although it is not expected to host any O stars owing to its moderate age \citep[$\sim$10--15 Myrs; e.g.][]{baume03,preibisch17,bisht21}, it is actually one of the most populous stellar aggregates in the Carina Nebula region \citep[e.g.][]{preibisch17}. It contains tens of relatively unevolved early B stars \citep{evans05}, along with a few blue and red supergiants, including \object{HD 91969} (B0 Ib) and \object{V361 Car} (M1.5 Iab), for instance. It also hosts some objects of particular interest, such as a chemically peculiar, strongly magnetic B2 star \citep[\object{CPD --57$^{\degr}$3509};][]{przybilla16} or several multiperiodic $\beta$ Cep pulsating variables, among which one in an eclipsing binary \citep[\object{HD 92024};][]{engelbrecht86}.

The {\em Gaia}-ESO public survey (hereafter GES) is a recently completed, ambitious spectroscopic survey of $\sim$10$^5$ stars in the Milky Way. The two main components of the project consist in observations of the field population and open clusters. As discussed by \citet{bragaglia22}, NGC 3293 was selected as one of the young southern clusters to be intensively observed as part of the survey. The observations are made with the multi-object instrument Fibre Large Array Multi-Element Spectrograph \citep[FLAMES;][]{pasquini02} installed on the Very Large Telescope (VLT), enabling the simultaneous observation of the fields with the GIRAFFE and UVES spectrographs. The main aim of the GES is to supplement the {\em Gaia} space mission \citep{gaia_a} in order to address several issues related to the formation and evolution of the Milky Way \citep{gilmore22,randich22}. Complementary ground-based observations are particularly advantageous when studying massive stars because {\em Gaia} offers less diagnostic power for stellar characterisation, especially in terms of abundances. A discussion of the stellar parameters recently released as part of {\em Gaia} DR3 can be found in \citet{fouesneau22} and \citet{blomme_gaia22}. The former study includes a comparison with the GES results for OB stars where a large dispersion is usually observed\footnote{See also \url{https://gea.esac.esa.int/archive/documentation/GDR3/Data_analysis/chap_cu8par/sec_cu8par_apsis/ssec_cu8par_apsis_esphs.html}}.

NGC 3293 has a long history of photometric measurements \citep[e.g.][]{feinstein80,baume03,bisht21}, but spectroscopic investigations are much less common. Pioneering studies of this kind include \citet{feast58}, who obtained radial velocities (RVs) and spectral classification for the brightest stars, and \citet{balona75} who determined their projected rotational velocities. Most abundance studies in the literature are restricted to the brightest cluster members \citep{mathys02,niemczura09_ngc3293}. However, this cluster was also chosen by the GES because it was studied in some detail by the large ESO programme `VLT-FLAMES survey of massive stars' \cite[hereafter FS;][]{evans05}\footnote{\url{https://star.pst.qub.ac.uk/~sjs/flames/}}. As such, it can be used for benchmarking.

We take advantage of the largest spectroscopic dataset gathered to date for NGC 3293 to study the properties of its stellar B-type population in terms of spectral variability, chemical abundances, and rotational velocities. The age of the cluster is also revisited thanks to a thorough correction of our results for the effects of fast rotation. The FS led to the determination of fundamental stellar parameters and abundances for a sizeable number of early B-type stars. Our study can be regarded as being complementary to that of the FS and a leap forward towards a comprehensive characterisation of this cluster. In particular, we extend the determination of homogeneous parameters and chemical abundances to much lower masses (down to B9.5). Our study also brings about a number of improvements. For instance, to increase the sample size, the conclusions drawn by the FS about the rotational and chemical properties of this cluster \citep{dufton06,hunter09} were based on the combination of the results with those of two other Galactic clusters, \object{NGC 4755} and \object{NGC 6611}, even though the latter is much younger. In contrast, our results are entirely based on a statistically sound sample of stars whose membership to NGC 3293 can be established on a firmer footing thanks to the recently available {\it Gaia} data. For these various reasons, we revisit the results obtained by the FS. In addition, we reanalyse their GIRAFFE and UVES spectra for completeness and validation purposes.

 This paper is structured as follows. Section ~\ref{sect_context} clarifies how our work fits into the context of the GES. Section ~\ref{sect_membership} discusses the selection of targets and their membership, while Sect.~\ref{sect_data} presents the observations. Our results concerning the atmospheric parameters, variability and binarity, and chemical abundances are provided in Sect.~\ref{sect_analysis}. Section \ref{sect_discussion_pnrc} is devoted to a discussion of the rotational velocity distribution and age of the cluster after accounting for the effects of rapid rotation. Our main findings about the chemical properties of our targets are presented in Sect.~\ref{sect_discussion_abundances}. Finally, our main conclusions are given in Sect.~\ref{sect_conclusions}.

   \section{This work in the context of the GES}\label{sect_context}

 The GES consortium is divided into several working groups (WGs). WG13 is in charge of the analysis of the OBA stars \citep{blomme11,blomme22}. As other WGs in the GES consortium, a number of research groups (called `nodes') within WG13 independently analysed the spectra using their own techniques and codes. Following a critical evaluation of the quality of the data products from each node, the individual results are weighted and eventually combined to produce recommended, homogenised parameters and abundances. An important point is that, unlike the case of the cool stars treated by the other WGs, the abundances are not computed adopting the recommended parameters. Instead, to ensure self-consistency, the abundance determination performed by each node is based on its own set of atmospheric parameters. Full details about the scientific objectives, data collected for young clusters, organisation, and data processing procedures implemented in WG13 can be found in \citet{blomme22}.

 The present paper presents the results obtained for NGC 3293 by the `Li\`{e}ge node' based on the final data release of the survey (iDR6). Preliminary results based on iDR3 have been discussed by \citet{semaan15}. As a cautionary note, our results slightly differ from those to be delivered by the GES to ESO for subsequent archiving and public release to the community because of the homogenisation phase described above. The comparison is discussed in Appendix \ref{appendix_comparison_homogeneised_results} where it is shown that the differences are minor, except for the projected rotational velocity because of a quite poor agreement between the various nodes (\citealt{blomme22}; see also Sect.~\ref{sect_external_validation_other_nodes}). The main purpose of the complex GES homogenisation procedures is to ensure optimal consistency across the various WGs, minimise systematics with respect to similar ongoing or forthcoming spectroscopic surveys, and facilitate the global interpretation of the catalogue. The ultimate objective being to fulfil the top-level goals of the survey, which are deciphering the formation history and evolution of the various populations (thin and thick discs, bulge, and halo) making up our Galaxy. In contrast, all the results discussed in this paper are not recalibrated in any way and -- more importantly -- are obtained in a much more homogeneous and self-consistent way. As such, they are more suitable for a dedicated study of NGC 3293. For this particular cluster, it can be noted that the Li\`{e}ge node provided a significant fraction of all the parameters delivered by WG13, was assigned the highest weight during the homogenisation phase (1.50 compared to 0.58--0.81 for the other nodes that analysed this cluster), and is the only one providing abundance data \cite[see][]{blomme22}. This paper is the first of a series of WG13 publications presenting the analysis of the GES data collected for young open clusters.

\section{Target selection and cluster membership}\label{sect_membership}

Starting with the list of stars observed by the GES in the field of NGC 3293 (Sect.~\ref{sect_membership_preselection}), we first selected those suitable for a spectral analysis (Sect.~\ref{sect_selection_spectral_analysis}) and finally identified in this sample clear non cluster-members (Sect.~\ref{sect_membership_gaia}).

\subsection{Initial selection of GES targets in the cluster}\label{sect_membership_preselection}

The pre-selection of the cluster members was performed by the GES prior to the first release of the {\em Gaia} data and solely relied on photometric criteria \citep[see][]{bragaglia22}. The 2MASS catalogue was the main starting point. High-quality observations\footnote{As defined in \url{www.ipac.caltech.edu/2mass/releases/allsky/doc/sec1_6b.html}.} were first selected from the full dataset in a large area of 12.5$\arcmin$ radius (corresponding to the FLAMES field of view) around the cluster centre quoted in the WEBDA database\footnote{\url{http://webda.physics.muni.cz}}. High-quality, near-infrared (IR) data in all three bands ($JHK_\mathrm{s}$) were required for a star to be included in this large pool of candidates. This list was next cross correlated with various optical catalogues available in the literature \citep{delgado11,baume03,dias06,netopil07,evans05}. They provide detailed photometry, as well as membership information. The cross-match with the catalogue of \citet{delgado11} is straightforward, as it already lists 2MASS cross IDs. For the others, a positional match within 1$\arcsec$ was required. A star was considered further if {\em at least one} of the optical catalogues classifies it as a member, whereas it was excluded if it is identified as an interloper in {\em all} catalogues in which it is listed.

The apparent cluster radius is $\sim$6--7$\arcmin$ corresponding to a physical radius of $\sim$5 pc \citep{bisht21,preibisch17}. To mitigate contamination, only stars within a radius of 4.1$\arcmin$ \citep{baume03} were initially kept. A number of stars studied by the FS \citep{evans05} without any membership information from the optical catalogues are located much farther away (up to $\sim$10$\arcmin$) than the generally accepted radius. Whether the cluster is more spatially extended than commonly believed deserves further investigation, but it was decided to add them back in. In addition, less secure members were also observed to avoid having some FLAMES fibres not being allocated, which increases the proportion of contaminants (see Sect.~\ref{sect_membership_gaia}). 

To refine the selection, two dereddened colour-magnitude diagrams (CMDs) were used: $J_0$ vs ($J-H$)$_0$ and $V_0$ vs ($B-V$)$_0$. A pre-{\em Gaia} distance modulus of 12.2 mag \citep{baume03}, $E$($B-V$) = 0.263 mag as quoted in WEBDA, and a canonical extinction law with $R_\mathrm{V}$ = 3.1 \citep{cardelli89} were assumed. Stars were kept if they fulfilled the following criteria:

 \begin{equation}
  \begin{cases} (J-H)_0 < 0.4, & \mbox{if } J_0\mbox{ < 13} \\ (J-H)_0 < 0.143 J_0 - 1.457, & \mbox{if } J_0\mbox{ \gid \, 13} \end{cases}
 \end{equation}
 and
 \begin{equation}
  \begin{cases} (B-V)_0 < 0.85, & \mbox{if } V_0\mbox{ < 14.5} \\ (B-V)_0 < 0.136 V_0 - 1.127, & \mbox{if } V_0\mbox{ \gid \, 14.5.} \end{cases}
 \end{equation}

 \subsection{Stars selected for spectral analysis}\label{sect_selection_spectral_analysis}

 We determine the parameters and chemical abundances of stars covering the full $T_\mathrm{eff}$ range of B stars, that is from 10 to 32 kK. The stars to be processed at the lower $T_\mathrm{eff}$ boundary were selected by a visual inspection of the blend formed by \ion{Ti}{ii} $\lambda$4468.5 and \ion{He}{i} $\lambda$4471.5: the \ion{Ti}{ii} feature dominates for A stars. After selection of the B-type spectra, we have at this stage data for 186 stars.

As a next step, we screened out spectra that are too noisy to be analysed or suffer from severe instrumental problems (i.e. a picket-fence pattern in the case of UVES). Finally, after visual inspection, we discarded objects displaying {\em obvious} spectral peculiarities, either a composite morphology pointing to a spectroscopic binary (SB$n$, with $n$ $\geq$ 2) or a strong, double-peaked emission profile in Balmer lines that could safely be attributed to a massive circumstellar disc. Continuum emission from the disc in Be stars requires a specific treatment \citep[e.g.][]{ahmed17}. However, the incidence of this type of objects is discussed in Sect.~\ref{sect_discussion_rotational_velocity}.

Single-lined (SB1) and tentative double-lined (SB2) binaries were treated further under the assumption that the secondary does not significantly bias our results through, for instance, continuum dilution. We find that the abundance distributions for the stars identified as single and SB1's are indistinguishable.

As to whether parameters are provided in the presence of line-profile variations (LPVs) depends on the strength of the LPVs and sampling of the observations. The variations arise from pulsations or, in the late B stars, from rotational modulation of a spotted photosphere that is presumably a common phenomenon in this $T_\mathrm{eff}$ regime \citep{balona19}. Stars with strongly distorted line profiles were dropped unless the changes are reasonably well sampled (see example in Appendix \ref{appendix_impact_pulsations}). Because our results rely on the co-addition of all (RV corrected) exposures, they may be regarded in this case as representative of the mean values averaged over the variability cycle. In contrast, stars with nearly symmetric profiles were kept irrespective of the number of observations.

\subsection{Check of cluster membership based on Gaia EDR3 data}\label{sect_membership_gaia}

There are efforts within the consortium to assign cluster membership probabilities from {\em Gaia} astrometric data supplemented by GES RVs \citep[][]{jackson20,jackson22}. However, \object{NGC 3293} was not considered because the analysis relies on GIRAFFE HR15N spectra that are not available. Furthermore, there are too few objects with a measured RV for the methods to be suitable.

Carrying out a full statistical modelling of the 3D kinematics is beyond the scope of this paper. However, we decided to examine the astrometric properties of our sample because the fraction of contaminants is anticipated to be quite high \citep{bragaglia22}. In particular, the cluster lies in a crowded region very close to the Galactic plane ($b$ $\sim$ 0.07$\degr$). We first cross-matched the GES and {\em Gaia} EDR3 \citep{gaia21} catalogues using a search radius of 2$\arcsec$. Duplicate entries in {\em Gaia} EDR3 were found in a few cases, but a spurious association can safely be rejected thanks to a mismatch in coordinate, parallax, or $G$ magnitude. The parallax, $\varpi$, is well determined, with on average $\varpi$/$\sigma_{\varpi}$ $\sim$ 20--25. Because the quality of the {\em Gaia} data does not afford in that case to confidently assess membership, we conservatively kept stars with possible issues with the processing of the astrometric measurements or an ill-behaved solution: either a {\tt duplicated\_source} flag raised or a renormalised unit weight error, {\tt RUWE}, above 1.4 \citep[e.g.][]{lindegren20}. The number of visibility periods, {\tt visibility\_periods\_used}, is always sufficient \citep[i.e. above 8; see][]{arenou18}.

\begin{figure}[h!]
\centering
\includegraphics[trim=90 200 60 170,clip,width=0.8\hsize]{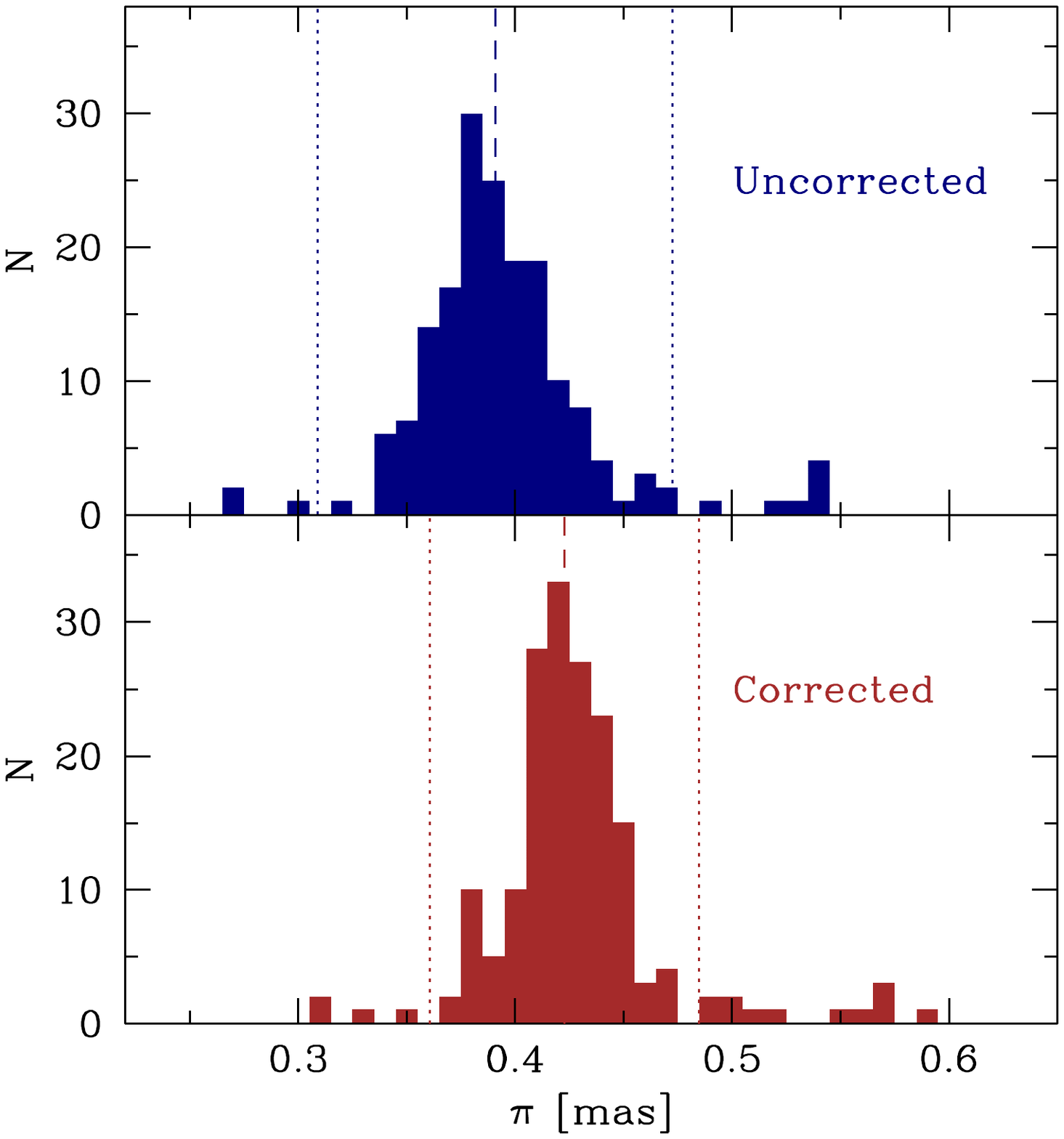}
\caption{Breakdown of the parallaxes for the stars with well-behaved astrometric solutions before ({\em top panel}) and after ({\em bottom panel}) applying the zero-point offsets of \citet{lindegren21}. A foreground star (\object{GES\,10344563--5813091}) with a much larger parallax is off scale. The dashed lines show the mean values, while the dotted lines indicate the 3-$\sigma$ thresholds.}
\label{fig_parallaxes}
\end{figure}

 The {\em Gaia} EDR3 parallaxes are affected by small zero-point biases, which are a complex function of the stellar brightness and colour, for instance. We applied corrections on a star-to-star basis following \citet{lindegren21}. Although the offsets are small ($\sim$ --33 $\upmu$as on average for the sub-sample with reliable astrometric solutions), they are significant at the distance of the cluster and lead to a more peaked parallax distribution (Fig.~\ref{fig_parallaxes}). It supports the reliability and usefulness of these corrections for bright, blue sources.

\begin{figure*}[h!]
\centering
\includegraphics[trim=60 325 10 260,clip,width=\hsize]{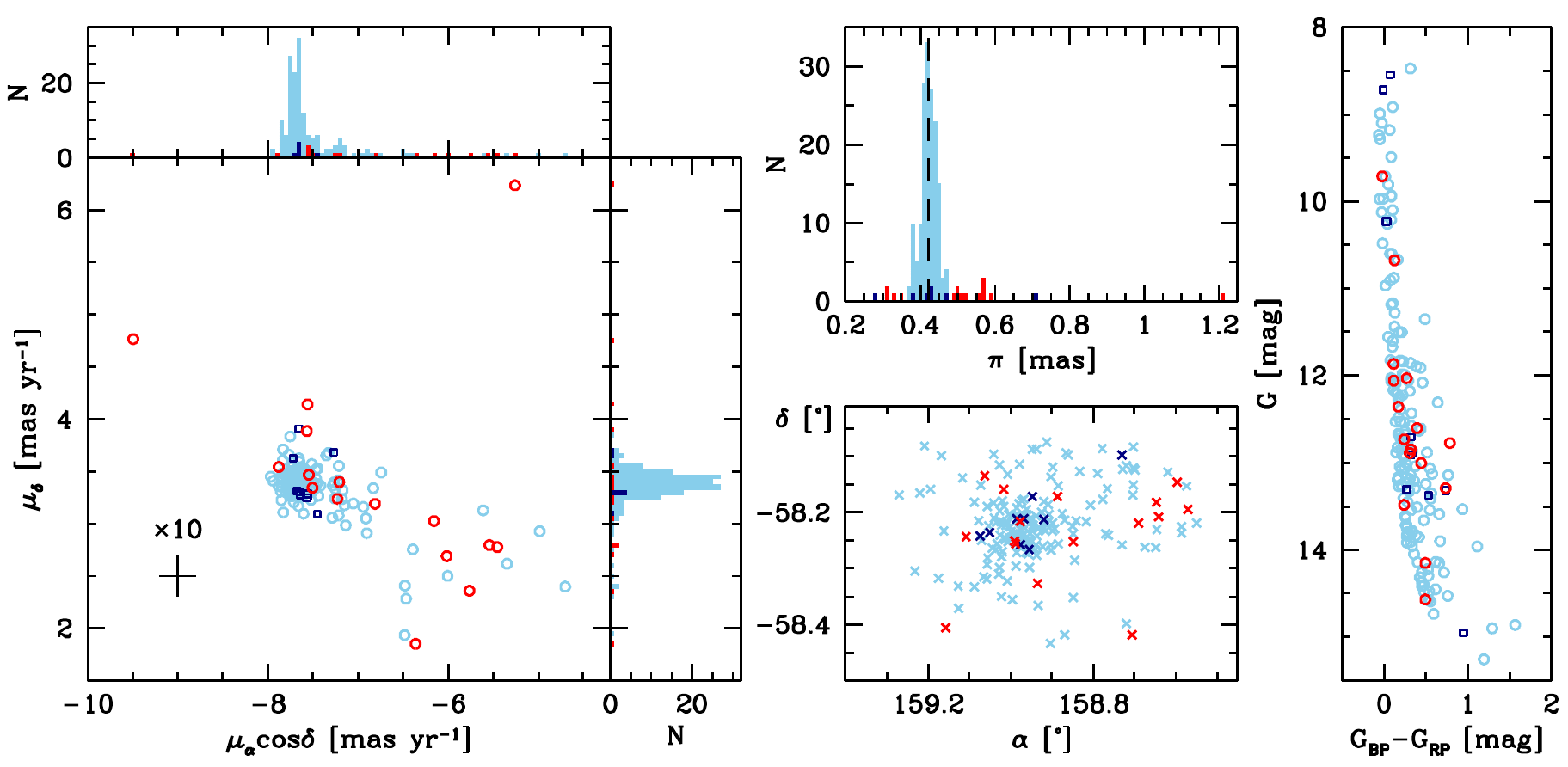}
\caption{Astrometric and photometric properties of our sample. Red symbols highlight in all panels stars assumed not to be cluster members, while dark blue symbols indicate stars with possible issues with the {\em Gaia} EDR3 data. {\em Left panels}: proper motions. A typical error bar with a size multiplied by ten is shown. {\em Top middle panel}: parallax distribution. The dashed line shows our mean value. {\em Bottom middle panel}: coordinates (epoch J2000). {\em Right panel}: CMD.}
\label{fig_membership}
\end{figure*}

The astrometric data are shown in Fig.~\ref{fig_membership}. A total of 16 likely foreground or background late B stars were identified by their parallax deviating by more than 3$\sigma$ from the mean of the distribution that is found to be $\langle \varpi \rangle$ = 0.423$\pm$0.021 mas after iterative 3-$\sigma$ clipping for the sub-sample with well-behaved astrometric solutions. Although results are provided, they are not considered further when discussing the properties of the cluster (Sects.~\ref{sect_discussion_pnrc} and \ref{sect_discussion_abundances}). \object{GES\,10343562--5815459} was retained because its parallax is only slightly above the threshold, while its proper motion is fully compatible with that of the cluster.

One can clearly notice in Fig.~\ref{fig_membership} (left panels) a group of nine presumed members with a right parallax, but large $\mu_{\alpha}$ and low $\mu_{\delta}$ values\footnote{Because the space velocities are of less importance here, we ignored the bias of the order of 40 $\upmu$as yr$^{-1}$ affecting the {\em Gaia} EDR3 proper motions for stars with $G$ $<$ 13 mag \citep{cantat_gaudin21}.}. These stars are preferentially located to the north of NGC 3293 and therefore at the very northern edge of the Carina complex. They also have RVs that often differ from the cluster systemic velocity (Sect.~\ref{sect_binarity}). Although most of them probably belong to the field despite the fact that they lie at the right distance, they are all kept because a few could be runaways. We note that these nine stars (among which two eventually do not have parameters determined) would contribute to a level of contamination not exceeding $\sim$10\%. Considering them in the following does not modify our conclusions in any appreciable way.

We end up with 149 stars for which we provide parameters (among which 141 have abundances). To this total, 16 stars with a variability or binarity flag can be added, which leads to a total of 165 objects with information of some sort. If only the members are counted, 137 and 130 stars have parameters and abundances, respectively. We estimate that about 120--130 B stars were identified by \citet{baume03} in the inner region (4.1$\arcmin$ circle radius) of the cluster through optical photometry. However, it only gives a very rough idea of our completeness level because the area we cover is much wider (Sect.~\ref{sect_membership_preselection}).

\section{Observational data}\label{sect_data}

The GIRAFFE ($R$ $\sim$ 20\,000) and UVES ($R$ $\sim$ 47\,000) GES settings used are described in \citet{blomme22}. Only the UVES U520 blue arm was considered here because much more information is encoded compared to the red arm. Data were also obtained with the GIRAFFE HR09B grating, but are not used either because of the lack of useful diagnostic lines in the B-star regime. The bulk of the data were obtained during the period February--April 2012 (i.e. prior to GIRAFFE upgrade), while a few UVES spectra were acquired in January 2013. The HR04 grating was considered at a much later stage of the project in order to use H$\gamma$ as an additional surface gravity indicator \citep{berlanas17}. As a result, fewer stars have this setting available. The data were secured in December 2017. A small fraction of the GIRAFFE spectra (not HR04) appeared to be contaminated by that of a calibration lamp in the adjacent MEDUSA fibre. Depending on the severity of the problem, these spectra were either ignored or the associated results were given a lower weight.

  For the FS data, HR05A and HR14A are replaced by HR05B and HR14B, respectively. The last two gratings have a better spectral resolution at the expense of a slightly narrower wavelength range \citep[see][]{blomme22}. The GIRAFFE data were supplemented by a few UVES spectra not discussed in \citet{evans05}. All the raw FS data were retrieved from the ESO archives and pre-processed using exactly the same reduction procedures as for the GES data \citep[][]{sacco14,gilmore22}. The FS HR02 data were not treated because they cover a wavelength region bluewards of any GES settings. We also draw attention to the fact that the GES ignored the FS observations of the brightest stars in the cluster (down to $V$ $\sim$ 6.5) acquired with FEROS. Therefore, those data are not included in the present analysis.

\begin{figure}[h!]
\centering
\includegraphics[trim=120 175 80 140,clip,width=0.8\hsize]{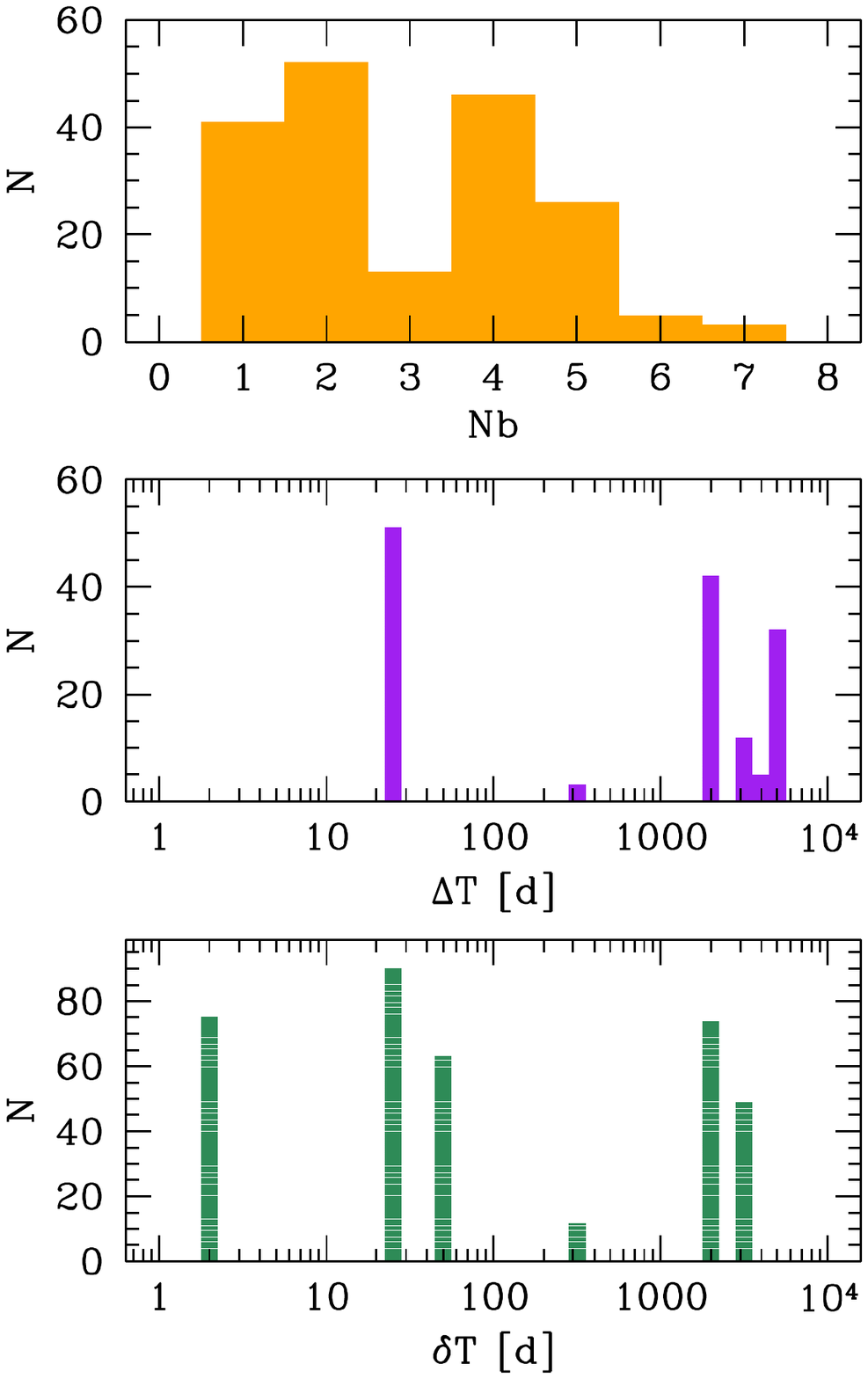}
\caption{Time sampling. Breakdown of number of independent epochs, $Nb$ ({\em top panel}), total time span of observations, $\Delta T$ ({\em middle panel}), and time interval between consecutive epochs, $\delta T$ ({\em bottom panel}).}
\label{fig_sampling}
\end{figure}

 Multi-epoch observations (secured $\sim$3 weeks apart) are often available for HR05A, while data for other settings may have been obtained during different nights. The availability of repeated observations (quite often up to four or five) allows us to carry out, to our knowledge, the first modern binary detection programme through spectroscopy since \citet{feast58}. In particular, the FS data were obtained over a single night. We take advantage of these data acquired much earlier \citep[14 April 2003;][]{evans05} to extend the time span of the observations to a baseline (9--14 years) appropriate for the detection of binaries with relatively long periods. The observations are described in Table \ref{tab_observations}. A timescale relevant to binarity was assumed to define the epochs: they are separated by more than one day. Observations secured on an hourly timescale that is commensurate with, for instance, pulsations were only obtained with UVES (see example in Appendix \ref{appendix_impact_pulsations}). The breakdown of the number of independent epochs, total time span of the observations, and time interval between consecutive epochs is shown in Fig.~\ref{fig_sampling}. The histogram of the time span is dominated by three peaks: one corresponding to the two GES HR05A observations gathered $\sim$25 days apart, as well as two at large values arising from the late acquisition of the GES HR04 data and objects with both FS and GES spectra. Despite a time sampling that appears in principle suitable for the detection of binaries with a quite wide range of orbital periods (see bottom panel of Fig.~\ref{fig_sampling}), it is important to bear in mind that the cadence strongly varies across the sample. In addition, the RV time series were obtained with a variety of instrumental configurations.

\begin{figure}[h!]
\centering
\includegraphics[trim=90 400 65 170,clip,width=0.8\hsize]{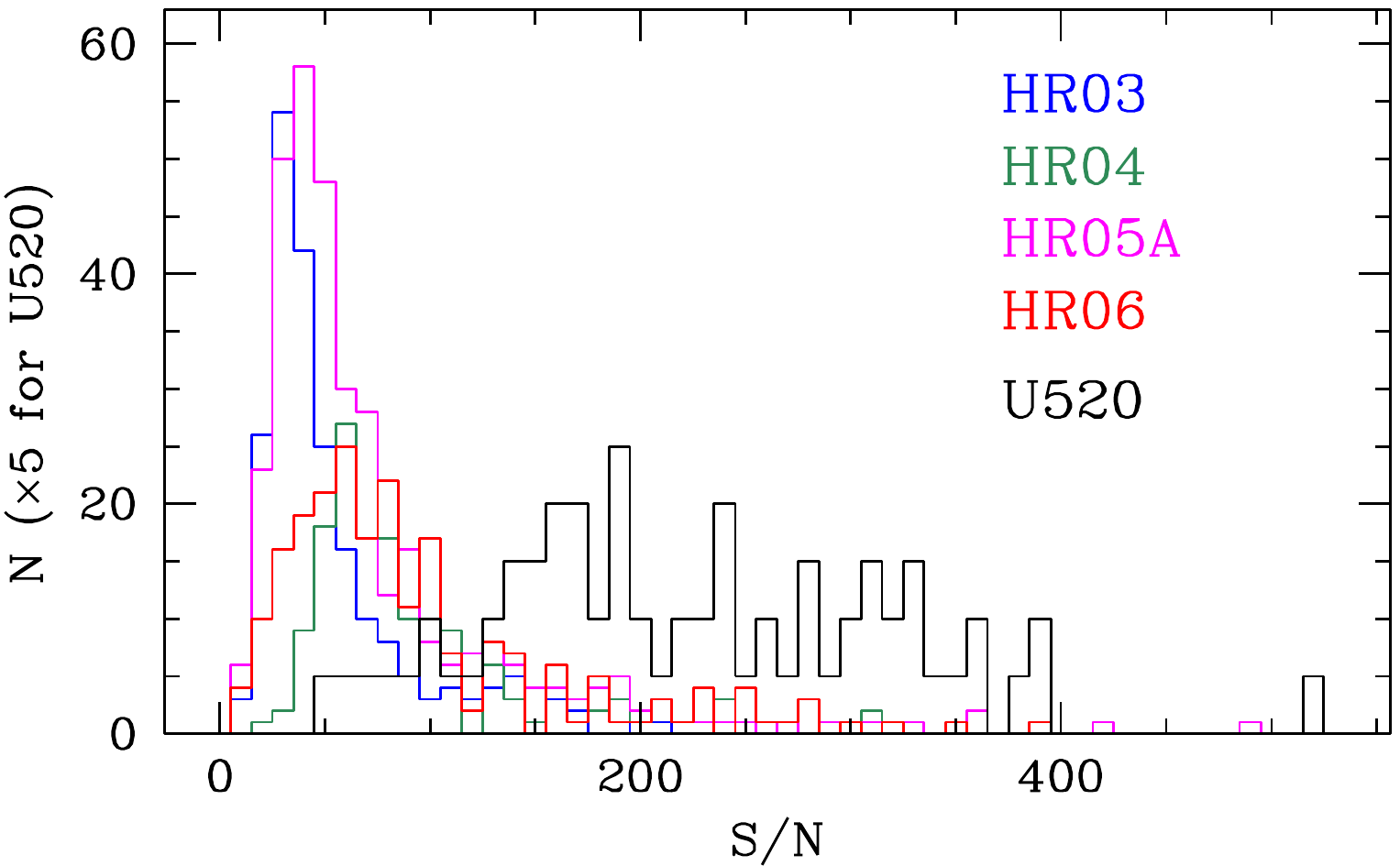}
\caption{Breakdown of the mean S/N of the epoch spectra. The S/N is computed by the GES and refers to the value averaged across the entire wavelength range. The histogram for U520 is multiplied by five for better visibility.}
\label{fig_snr}
\end{figure}

The mean signal-to-noise ratio (S/N) of the epoch spectra is shown for each setting in Fig.~\ref{fig_snr}. The wide range of values reflects the fact that all stars in a single FLAMES pointing were observed with the same exposure time. As a result, the data quality for the faintest targets (i.e. late B-type dwarfs) is much lower. The S/N of the HR05A spectra eventually used for the parameter determination is generally better by a factor $\sim$1.4 because it is often the combination of two epoch spectra (Sect.~\ref{sect_parameters}). HR14A is not used for that purpose, but only for inferring the \ion{Si}{ii} and \ion{Ne}{i} abundances. For the limited number of stars for which either of the two can be measured, the S/N lies in the range 55--490 with a mean of $\sim$180. For the stars with FS data reprocessed, we find that the quality of the corresponding GES HR06 spectra is similar. However, the S/N of the GES spectra is larger by a factor ranging from $\sim$1.1 (HR03) to $\sim$1.5 (HR04) for the other GIRAFFE gratings.

Finally, a number of benchmark OBA stars were observed during the survey \citep{pancino17,blomme22}. To evaluate the reliability of our results, we analysed the high-quality data of \object{134 Tau} (B9 IV), \object{HD 56613} (B8 V), \object{HD 35912} (B2 V), \object{$\gamma$ Peg} (B2 IV), \object{$\tau$ Sco} (B0.2 V), and \object{$\theta$ Car} (B0 Vp). Regrettably, no HR04 spectra were obtained for this sample.

\begin{table*}[h!]
\small
\caption{Summary of observations and sampling properties.}
\centering
\hspace*{-0.4cm}
\begin{tabular}{lll|cc|cc|cc|cc|cc|c|c}
\hline\hline
       &                &       & \multicolumn{10}{c|}{Number of time-resolved spectra} & & \\
       &                &       & \multicolumn{2}{c|}{HR03} & \multicolumn{2}{c|}{HR04} & \multicolumn{2}{c|}{HR05} & \multicolumn{2}{c|}{HR06} & \multicolumn{2}{c|}{U520} & & \\
GES ID & Main SIMBAD ID & FS ID & FS & GES & FS & GES & FS & GES & FS & GES & FS & GES & \multicolumn{1}{|c}{$Nb$}  & \multicolumn{1}{|c}{$\Delta T$ [d]}\\
\hline
                           \multicolumn{15}{c}{NGC 3293} \\   
\hline
 \object{GES\,10341195--5813066} &  ...                    &  ...      & ... &  1   &  ... &  ... &  ... &  2  &  ... &  1   &  ... &  ... &  2 &     25.8\\
 \object{GES\,10341702--5811419} &  ...                    &  ...      & ... &  1   &  ... &  ... &  ... &  2  &  ... &  1   &  ... &  ... &  2 &     25.8\\
 \object{GES\,10341774--5809101} &  ...                    &  ...      & ... &  1   &  ... &  ... &  ... &  2  &  ... &  1   &  ... &  ... &  2 &     24.8\\
 \object{GES\,10342068--5814107} &  ...                    &  ...      & ... &  1   &  ... &  1   &  ... &  2  &  ... &  1   &  ... &  ... &  4 &   2140.0\\
 \object{GES\,10342078--5813305} &  CPD --57$^{\degr}$3450   & 3293-049 & 1   &  1   &  1   &  1   &  1   &  2  &  1   &  1   &  ... &  ... &  5 &   5368.2\\
 \object{GES\,10342325--5808448} &  ...                    &  ...      & ... &  1   &  ... &  ... &  ... &  2  &  ... &  1   &  ... &  ... &  2 &     24.0\\
 \object{GES\,10342859--5807396} &  ...                    &  ...      & ... &  1   &  ... &  ... &  ... &  2  &  ... &  1   &  ... &  ... &  2 &     24.0\\
...                              &  ...                    &  ...      & ... &  ... &  ... & ...  &  ... & ... &  ... &  ... &  ... &  ... & ... & ...\\
\hline                  
                          \multicolumn{15}{c}{Benchmarks} \\   
\hline                 
\object{GES\,00131415+1511008}   & $\gamma$ Peg  & ... & ... &  2 &  ... &  ... &  ... &  2 &  ... &  2 &  ... &  1   &  3 &    174.7\\
\object{GES\,05280146+0117537}   & HD 35912      & ... & ... &  1 &  ... &  ... &  ... &  1 &  ... &  1 &  ... &  1   &  2 &     30.0\\
\object{GES\,05493290+1239044}   & 134 Tau       & ... & ... &  1 &  ... &  ... &  ... &  1 &  ... &  1 &  ... &  1   &  2 &    125.6\\
\object{GES\,07173159--0549215}  & HD 56613      & ... & ... &  1 &  ... &  ... &  ... &  1 &  ... &  1 &  ... &  1   &  1 &      0.0\\
\object{GES\,10425736--6423398}  & $\theta$ Car  & ... & ... &  1 &  ... &  ... &  ... &  1 &  ... &  1 &  ... &  ... &  1 &      0.0\\
\object{GES\,16355294--2812579}  & $\tau$ Sco    & ... & ... &  2 &  ... &  ... &  ... &  1 &  ... &  1 &  ... &  2   &  3 &    272.2\\
\hline
\end{tabular}
\label{tab_observations} 
\tablefoot{The FS ID from \citet{evans05} is given in the third column. $Nb$ is the number of independent epochs, while $\Delta T$ is the total time span of the observations. The table is available in its entirety through the CDS. A portion is shown here for guidance regarding its form and content.}
\end{table*}

\section{Analysis and results}\label{sect_analysis}

Following a pre-processing of their spectra (Sect.~\ref{sect_pre_processing}), the objects eventually selected after filtering (Sect.~\ref{sect_membership}) had their stellar parameters (Sect.~\ref{sect_parameters}), variability status (Sect.~\ref{sect_binarity}), and chemical abundances (Sect.~\ref{sect_abundances}) determined.

\subsection{Data pre-processing}\label{sect_pre_processing}

All reduction steps (e.g. extraction of the spectra from the CCD chip, wavelength calibration) are performed by WG7 prior to delivery of the spectra to WG13. No nebular correction was applied to the NGC 3293 data. The GES internally produces stacks of all the spectra obtained for a given target and instrumental setting over the whole survey \citep[see][]{sacco14}. We extracted all the individual exposures and grouped them into epoch spectra: consecutive exposures were co-added and time-resolved spectra obtained over more than one day were treated separately.

By default, the spectra are normalised to the continuum by the GES reduction pipeline. However, this automatic procedure is optimised for cool stars and appears to lead to unsatisfactory results (line wings truncated) for the broad features (e.g. Balmer and helium lines) present in the spectra of OBA stars \citep{blomme13}. All spectra were therefore normalised manually using low-order polynomials with the {\tt IRAF}\footnote{{\tt IRAF} is distributed by the National Optical Astronomy Observatories, operated by the Association of Universities for Research in Astronomy, Inc., under cooperative agreement with the National Science Foundation.} software.

\subsection{Determination of atmospheric parameters}\label{sect_parameters}

We used global least-square minimisation to derive the stellar parameters. Using a {\tt Python} code we developed, we fit the observed normalised spectra with a grid of solar-metallicity, synthetic spectra computed with the SYNSPEC48 program, along with local thermodynamic equilibrium (LTE) ATLAS9\footnote{Taken from POLLUX database available at \url{http://npollux.lupm.univ-montp2.fr/}} \citep{kurucz93} and non-LTE TLUSTY\footnote{BSTAR2006 grid available at \url{http://nova.astro.umd.edu/}} \citep{Lanz+Hubeny07} model atmospheres. The ATLAS9 and TLUSTY grids were employed for the stars with $T_\mathrm{eff}$ below and above 15 kK, respectively. In both cases, a microturbulence, $\xi$ = 2 km s$^{-1}$, was used. Our analysis relies on codes assuming plane-parallel atmospheres in hydrostatic equilibrium. It is a suitable assumption given that none of our targets is expected to have a very strong wind. As a consequence, we do not provide any wind parameters, such as the mass-loss rate.

The first step consists in determining the RV and projected rotational velocity, $V\sin i$, for all epoch spectra. The synthetic spectra are thus convolved with a rotational profile \citep{gray05}, and then shifted in velocity. Instrumental broadening is taken into account. We did not consider broadening by macroturbulence, but it is expected to be largely dominated by rotation in our relatively unevolved targets \citep{simon_diaz17}. Furthermore, as shown below, most of them are (very) fast rotators. We calculate the $\chi^2$ for each synthetic spectra and interpolate the $\chi^2$ map to determine the best-fitting values. As a next step, we corrected each epoch spectrum for its individual RV and combined all GIRAFFE settings of a given target into a single spectrum put in the laboratory rest frame. For the same target, FS and GES spectra were treated separately. Finally, the determination of $T_\mathrm{eff}$ and $\log g$ is performed over the whole wavelength domain. The synthetic spectra are convolved with the rotational velocity averaged over the values obtained for each epoch settings and the combined spectra. Some examples of fits are shown in Appendix \ref{appendix_spectral_fits}. After determining $T_\mathrm{eff}$ and $\log g$, we used them to calculate anew the RV and $V\sin i$ of the individual epoch settings.

An uncertainty\footnote{All the uncertainties quoted throughout this paper are 1-$\sigma$ error bars.} is associated to each measurement on the basis of the $\chi^2$ surface (1-$\sigma$ contour). The typical random uncertainties are $\sim$800 K for $T_\mathrm{eff}$, $\sim$0.12\,dex for $\log g$, $\sim$11 km\,s$^{-1}$ for $V\sin i$ and $\sim$3 km\,s$^{-1}$ for RV. However, these figures considerably vary from star to star depending on the stellar parameters and data quality. As an illustration, $V\sin i$ and its uncertainty grow in parallel according to a typical ratio of about 5--6\%.

\subsection{Variability and binarity analysis}\label{sect_binarity}

We emphasise that our approach to investigate the variability and binarity fundamentally differs from that adopted in the recent literature \citep[e.g.][]{sana13} in that we do not attempt to infer a robust binary fraction. It is because of the limitations affecting the GES observations that were not designed for this particular purpose and, above all, of the subjectivity in our pre-selection (Sect.~\ref{sect_selection_spectral_analysis}) that can hardly be quantified. It is likely that the low binary occurrence rate we infer ($\sim$15\%) is grossly underestimated. Therefore, we refrain from discussing to what extent our (lower limit to the) binary fraction in NGC 3293 compares with that for B-type stars in other clusters \citep[e.g.][]{dunstall15,banyard22}. Similarly to the approach followed by other studies \citep[e.g.][]{holgado18}, our less ambitious goal here is instead to primarily flag stars whose determination of stellar parameters could potentially be affected by their spectral variability. As a byproduct, secure binary candidates are nonetheless identified.

The individual RVs produced at the end of the processing described in Sect.~\ref{sect_parameters} were analysed to identify variable stars. The HR06 data were not used because the RVs are more uncertain and it would introduce some heterogeneity given that fewer objects were observed with this setting. The steps described in the following can also be regarded as some kind of validation of the RV measurements.

\subsubsection{Confronting HR03 and HR04 settings}\label{sect_binarity_HR03HR04}

An important fraction of the targets were observed with both the HR03 and HR04 settings. For each object, we matched them by the pair of measured RVs. In case of three observations, we made two pairs
with the isolated exposure being repeated. To avoid strong redundancy, in the case of two HR03 and two HR04 exposures,
we formed only two pairs, taking care to match together the exposures most separated in time. The error
in the RV difference
 within a pair, $\sigma_\mathrm{d}$, is given by the quadratic sum of the RV uncertainties.
If the individual standard deviations are correctly estimated, the normalised difference
should indeed be a normal variate.
 We listed 136 pairs of RVs including 67 with exposures acquired on the same day. The last cases are interesting because they are not supposed to be markedly different from zero and are thus a good check that allows us to validate the differences in RVs.

\begin{figure}[h!]
\centering
\includegraphics[trim=40 205 40 365,clip,width=\hsize]{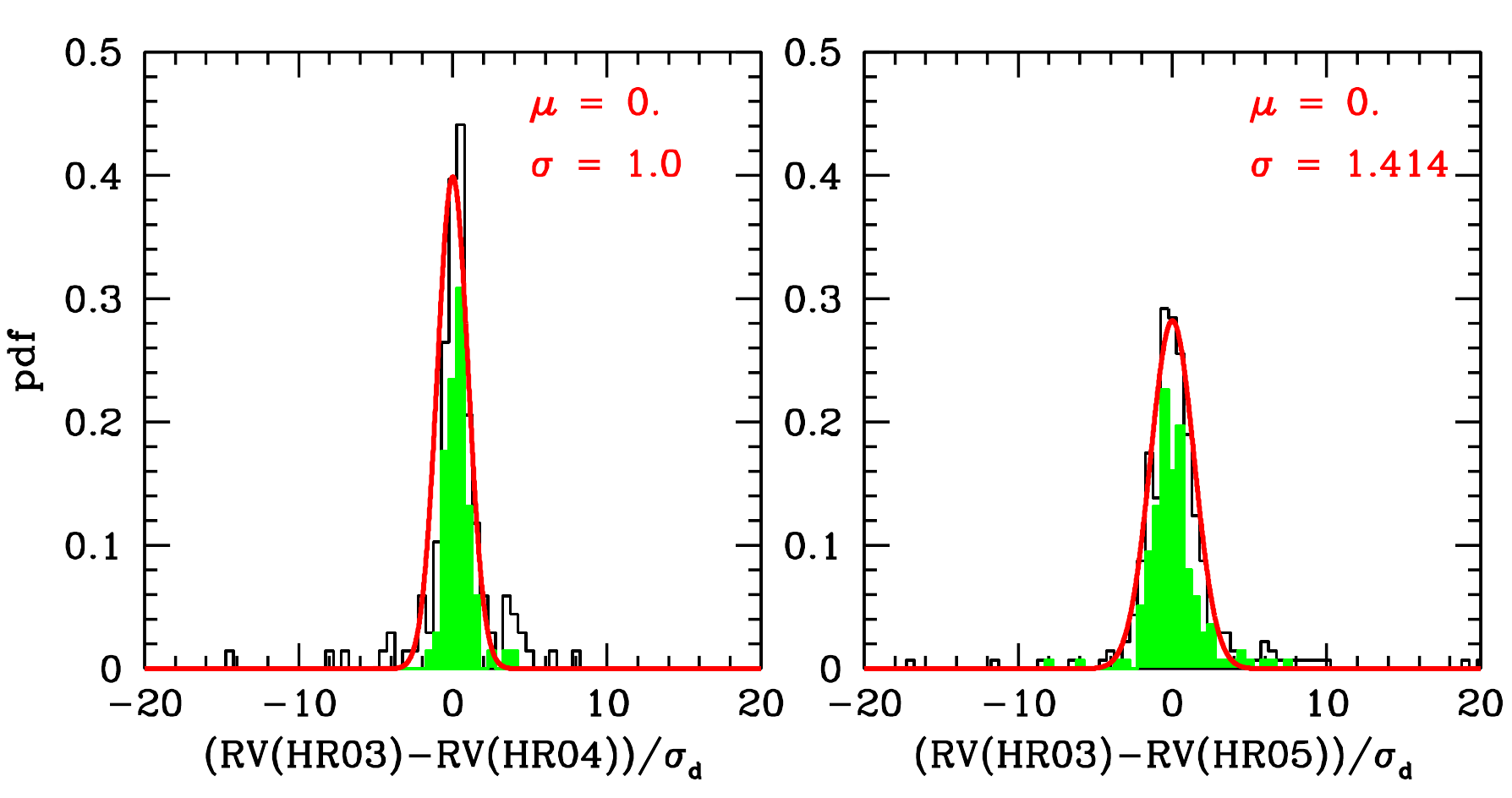}
\caption{Histograms of the normalised difference RV(HR03)--RV(HR04) ({\em left panel}) and RV(HR03)--RV(HR05) ({\em right panel}) for all the pairs. All the 
objects are included in the black histograms. The green histograms only include the 
contemporaneous pairs. The red continuous lines represent a Gaussian PDF of zero mean, $\mu$, and with either unit variance ({\em left panel}) or $\sigma \, = \, 1.414$ ({\em right panel}). \object{GES\,10352851--5812496} is off scale in the right panel.}
\label{egRV}
\end{figure}

The left panel of Fig.~\ref{egRV} illustrates the distribution of the normalised (by the expected error)
differences corresponding to each pairs. It is seen that their probability density function (PDF)
 is rather
clearly Gaussian with $\mu \, = \, 0$ and $\sigma \, = \, 1$. The mean is actually slightly
biased to $\mu$ = +0.4, but this is not to be considered as significant. It is interesting
to notice that the 67 contemporaneous pairs are pretty well in agreement with the Gaussian curve.
The errors in RV(HR03) and RV(HR04) are statistically similar. Thus $\sigma \, = \, 1$
is a good validation of the
 typical values for the errors in the individual RVs.

 Some rarer objects present discrepant pairs and
 can be suspected of variability.
Discrepant pairs go up to --14.6$\sigma$ on one side and up to +8.0$\sigma$ on the other side.
Not knowing if the +0.4 offset affecting $\mu$ is real or not, we cautiously considered as candidate variables
the pairs that are located outside of the $\pm \, 2.5$-$\sigma$ domain; we spotted out 20 pairs
that are discrepant 
 and retained for further investigation. They correspond
 to 15 different objects.

\subsubsection{Confronting HR03 and HR05 settings}\label{sect_binarity_HR03HR05}

The same kind of approach can be applied to the RV difference HR03 vs HR05, where the HR05 dataset includes
both HR05A and HR05B.
 We listed 274 pairs
 among which 159 are acquired on the same day.
 The distribution of the central part of the PDF is well Gaussian
 and centred on $\mu \, = \, 0$ (right panel of
 Fig.~\ref{egRV}). However, the dispersion is much larger with respect to the left panel leading to
 $\sigma \, = \, 1.414$. This is somewhat surprising since the standard deviation for
 HR05 is larger than for
 HR03 and HR04: this is taken into account, but the value could hardly be further increased. Thus the distribution should be narrower.
The problem is probably due to a rather bad estimation of the RV uncertainties determined from HR05 for an unknown reason.
Pairs are here also present in the tails of the distribution between --16.8$\sigma$ and +19.0$\sigma$,
except for \object{GES\,10352851--5812496}, which is at 68.8$\sigma$. We thought reasonable to extend a little the
threshold separating constant stars from variable candidates. A 4.0-$\sigma$ criterion
is producing 25 pairs related to potentially variable candidates that
are retained for further inquiry. They correspond to 19 different objects. We did not investigate the pair of setting HR04 vs HR05 to avoid redundancy.

\subsubsection{Confronting identical settings}\label{sect_binarity_ident}

Finally, we built pairs of observations acquired with the same setting, at different epochs.
We drew a list of seven
discrepant pairs for HR03, four for HR04 and, finally, 16
 for HR05A/B 
for which we have some suspicion of variability at the 3.0-$\sigma$ level. 
As usual, all these selected pairs were retained
for further analysis.

\subsubsection{Confronting UVES spectra}\label{sect_binarity_UVES}

In addition to the GIRAFFE spectra, we inspected a total of 69 FS and GES U520 spectra.
 They cover a wider wavelength domain encompassing many more
 lines. From a statistical point of view, treating them 
 in a similar way as
 the GIRAFFE spectra is much more difficult. This UVES set contains
  24 objects, among which eight only have UVES spectra.
 The 16 remaining ones have both UVES and GIRAFFE spectra in various proportion.

 Out of the eight objects, two only have one FS spectrum available. One of these two clearly exhibits numerous examples of line doubling pointing to a probable SB2 character; while the other one has a peculiar spectral morphology (double-peaked emission lines). Four objects exhibit variability well beyond the 3-$\sigma$ threshold; one is located at a more marginal level, but is confirmed variable by a detailed eye inspection. One object must be considered as constant. Finally, one has all its exposures severely affected by instrumental or reduction problems.
 
The situation is much more complex for the 16 objects
 that have both types of data
(UVES + GIRAFFE) where the classification is a mix of the work described in the preceding sub-sections, 
 that performed in case only UVES spectra are available,
 and detailed eye inspection. 
 Including the star discussed above, it resulted in the total detection of seven constant objects observed with UVES. The others
 are suspected variables.

\subsubsection{Flagging the variability and the binarity}\label{sect_binarity_flagging}

Stars with changes in RVs that could be assigned to binary motion and/or a variable line shape were identified. 
The main criterion is an outlying value with respect to the distribution
of the RV differences between pairs of GIRAFFE settings, namely, HR03 vs HR04 and HR03 vs HR05A/B. 
As a second step, objects presenting variations in the epoch spectra of a given setting were sought. 
In all cases, the relevant
 spectra were visually examined. 
Stars presenting significant variability on the basis of at least two criteria 
(i.e. between pairs of settings or for the same wavelength domain) are classified as true variables with a good significance level. An additional visual inspection helps to discriminate between SB1 
(or previously unrecognised SB2) and intrinsic line-profile variables (due to pulsations or any other cause).
 The decision for objects with both UVES and GIRAFFE data is first based on the GIRAFFE spectra and is then aided by the information extracted from the UVES spectra. Among the stars clearly identified as binaries (confidence level `A' or `B'), only \object{GES\,10361791--5814296} (secure SB1 and tentative SB2) has an anomalously high {\em Gaia} EDR3 {\tt RUWE} indicating an ill-behaved astrometric solution with respect to the expectations for a single source.

The detailed results of the variability analysis are given in Table \ref{tab_variability_analysis}. All the cases listed above are documented according to a flagging scheme \citep{van_der_swaelmen18,gilmore22}. Although the flags specific to the WG13 Li\`ege node (see Table \ref{tab_flags}) were provided on a star-to-star basis as part of the final public data release, we caution that they are superseded by those given here that rely on a more in-depth analysis. We also note that, because of the limited number of observations and inadequate time sampling over relatively short timescales, the flag reporting LPVs is often solely raised on the basis of profiles that are deemed to be asymmetric. Therefore, the identification of these variables is often not fully secure, especially in the cooler objects with fewer lines and a poorer S/N. Furthermore, the distinction between those and SB2's is generally ambiguous. For these reasons, the status of the intrinsically variable and SB2 candidates requires confirmation.

\begin{table*}[h!]
\centering
\caption{Flags specific to the WG13 Li\`ege node.}
\begin{tabular}{lcl} \hline\hline
  Flag  &        Results reported? & Description \\\hline
  {\bf Technical} &&  \\
10050-13-16-00  & No & Insufficient S/N ($\lesssim$ 30)\\	
10106-13-16-01	& Yes/No\tablefootmark{a} & Picket-fence pattern (only UVES)\\
10302-13-16-01	& No & Too poor fit of spectrum\\
10303-13-16-01	& No & $T_\mathrm{eff}$ below lowest bound of grid (10 kK)\\
  {\bf Stellar peculiarity} &&  \\
20010-13-16-00	& Yes & RV variations: SB1 binary motion\\	
20020-13-16-00	& Yes/No\tablefootmark{b} & SB$n$, $n$ $\geq$ 2\\
21100-13-16-00  & Yes/No\tablefootmark{b} & LPVs: intrinsic variability\tablefootmark{c}\\
25000-13-16-01	& Yes/No\tablefootmark{b} & Intrinsic emission in Balmer lines\tablefootmark{d}\\
\hline
\end{tabular}
\label{tab_flags}
\tablefoot{A suffix indicating the confidence level is added to each flag: A $\equiv$ probable, B $\equiv$ possible, C $\equiv$ tentative \citep{van_der_swaelmen18}.
  \tablefoottext{a}{Depends whether all the exposures are affected by this problem.}
  \tablefoottext{b}{See Sect.~\ref{sect_selection_spectral_analysis} for the criteria for further analysis.}
  \tablefoottext{c}{Either due to pulsations or rotational modulation.}
  \tablefoottext{d}{Double peaked or shell-like.}
}
\end{table*}

\begin{figure}[h!]
\centering
\includegraphics[trim=90 395 65 160,clip,width=0.75\hsize]{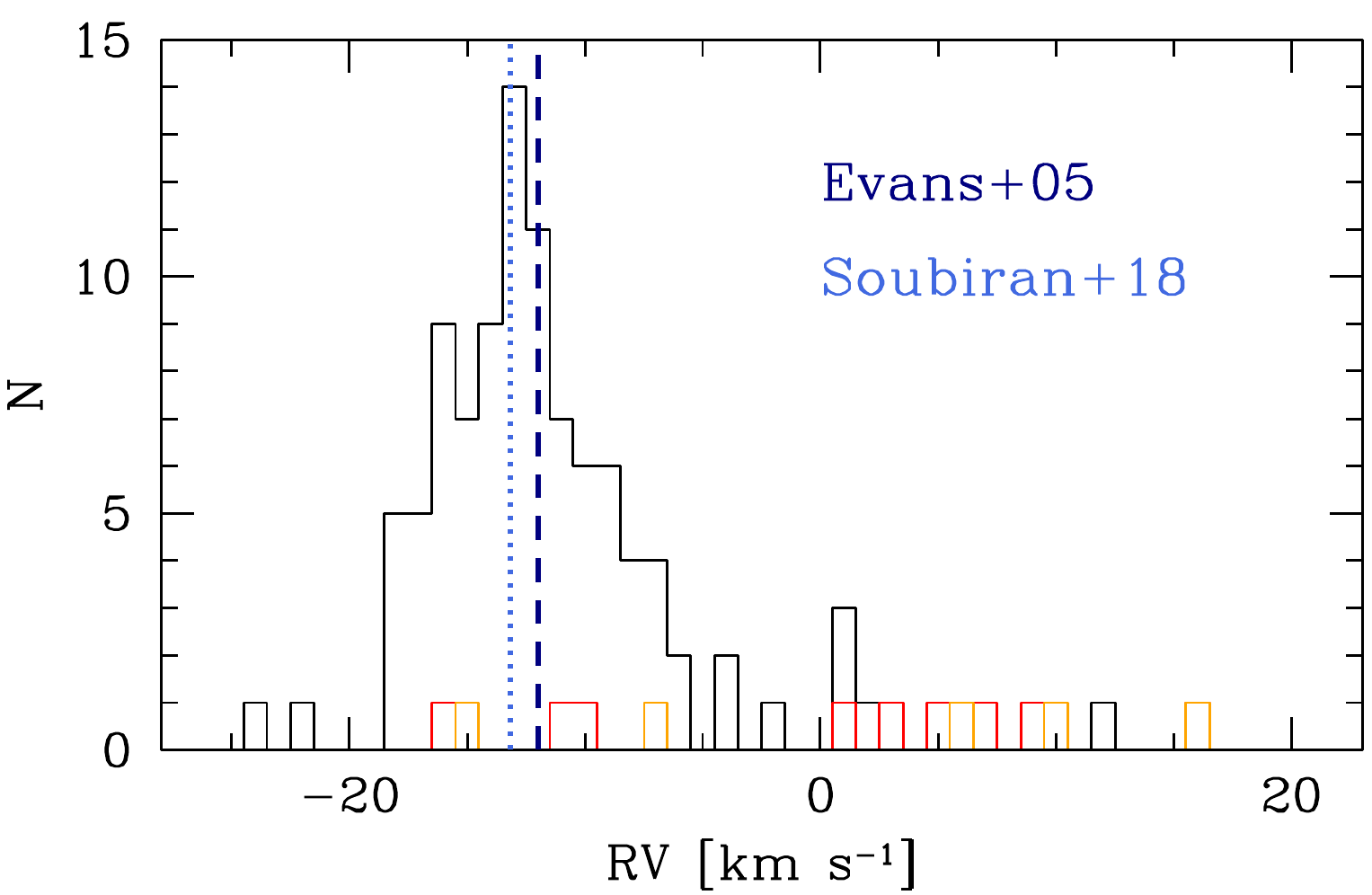}
\caption{Distribution of the mean RVs for the stars considered as constant. The red histogram is for stars assumed not to be cluster members, while the orange one is for the presumed members with the right parallax, but a discrepant proper motion (Sect.~\ref{sect_membership_gaia}). The other stars are shown in black. The mean values found by \citet{evans05} and \citet{soubiran18} are shown as dashed and dotted lines, respectively.}
\label{fig_vrad}
\end{figure}

 We count 113 objects (including 104 members) not listed in Table \ref{tab_variability_analysis} for which a lack of variations has been noticed with the data at hand.
 The breakdown of the mean RVs is shown in Fig.~\ref{fig_vrad}. The systemic velocity of the cluster is in agreement with previous estimates. For instance, from the FS \citep{evans05} or {\em Gaia} DR2 \citep{soubiran18}.

\subsection{Determination of chemical abundances}\label{sect_abundances}

The following chemical species were considered for the abundance analysis: He, C, N, Ne, Mg, and Si (both \ion{Si}{ii} and \ion{Si}{iii}). The non-LTE abundances were derived from a spectral synthesis of \ion{He}{i} $\lambda$4471, \ion{C}{ii} $\lambda$4267, \ion{N}{ii} $\lambda$4630, \ion{Ne}{i} $\lambda$6402, \ion{Mg}{ii} $\lambda$4481, \ion{Si}{ii} $\lambda$6371, and \ion{Si}{iii} $\lambda$4568-4575. Some illustrative fits are shown in Appendix A of \citet{blomme22}. These features were selected because they are relatively unblended and can be measured in the largest number of stars, even in case of a high rotation rate. Despite the much wider wavelength coverage of the UVES spectra, the same diagnostic lines were used to ensure consistency. 

The non-LTE code DETAIL-SURFACE \citep{giddings81,butler85} coupled to Kurucz LTE model atmospheres was used for the line modelling. See \citet{przybilla11} for a justification of such a hybrid method for stars for which wind effects can be neglected. The model atoms are described in \citet{morel06} and \citet{morel_butler08}. Synthetic \ion{C}{ii} $\lambda$4267 profiles were computed with the carbon model atom developed by \citet[][]{nieva08}. The line is not affected in NGC 3293 by nebular emission. Our carbon abundances are expected to be more reliable than those of the FS that were based on a more simplistic model ion and eventually corrected for a $T_{\mathrm{eff}}$ trend \citep{hunter09}. Metal lines blended with the diagnostic features (e.g. \ion{Al}{iii} $\lambda$4480) were modelled assuming abundances typical of B-type stars determined with the same code \citep[see table 6 of][]{morel08}. Oxygen abundances for all stars and carbon abundances for those with $T_\mathrm{eff}$ $<$ 17 kK are not reported because of suspiciously large values or unexpected trends with some stellar parameters.

Given our inability to constrain the microturbulence, either from global fitting (Sect.~\ref{sect_parameters}) or from the analysis of individual lines, it was fixed to sensible values. Because late B dwarfs largely dominate our sample, $\xi$ = 2 km\,s$^{-1}$ was adopted for the synthetic DETAIL-SURFACE grids. However, except for the carbon grid that was built for another purpose, this quantity for the relatively evolved, early B stars ($T_\mathrm{eff}$ $>$ 22 kK and $\log g$ $<$ 3.7 dex) was set to 5 km\,s$^{-1}$. The dependence as a function of the stellar parameters is based on previous determinations in the literature \citep[e.g.][]{hunter09,lefever10,lyubimkov13,nieva_przybilla12}. 

\begin{figure}[h!]
\centering
\begin{minipage}[l]{0.30\textwidth}
\includegraphics[trim=65 175 65 195,clip,width=\hsize]{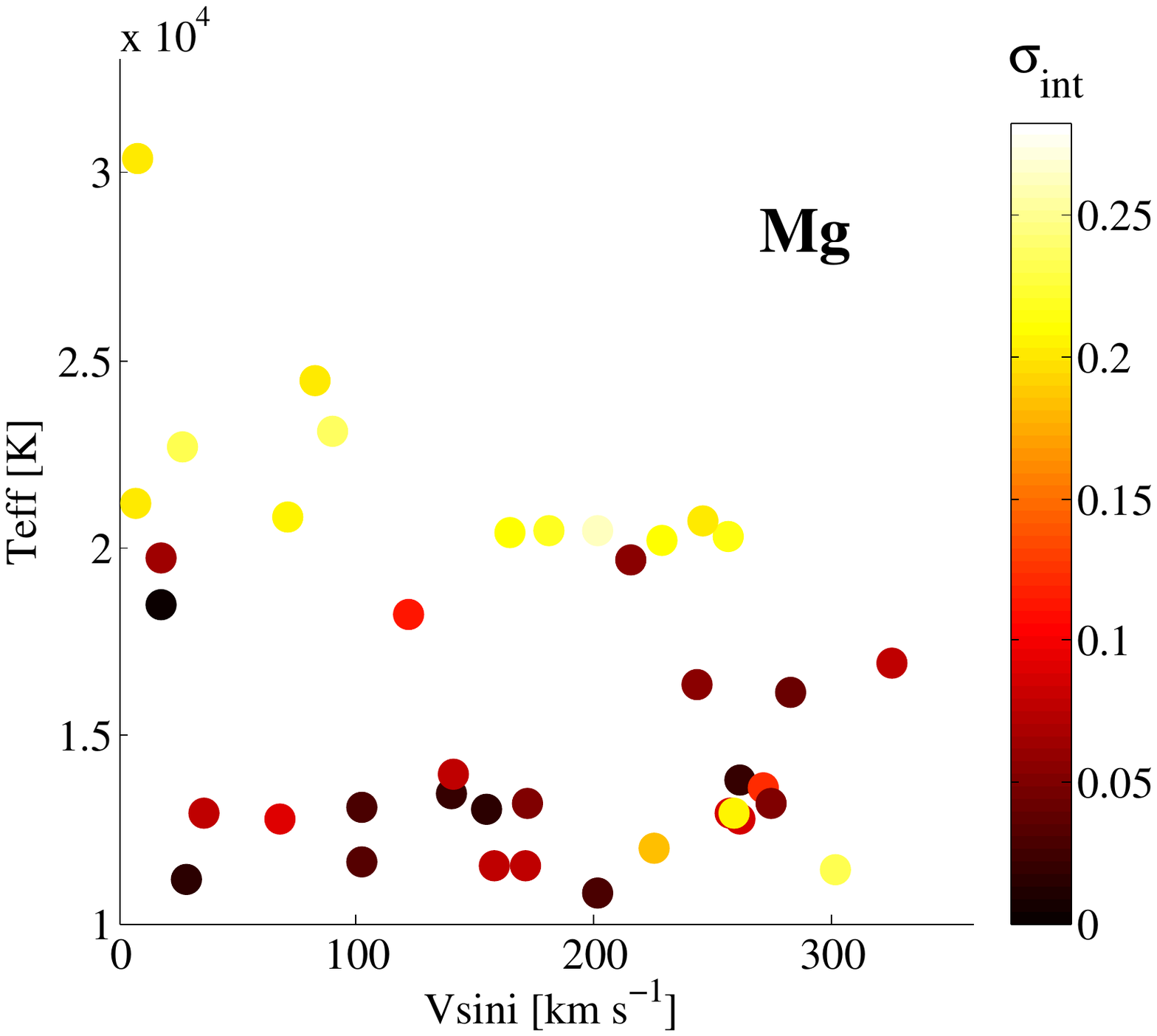}
\end{minipage}
\begin{minipage}[r]{0.30\textwidth}
\includegraphics[trim=65 175 65 195,clip,width=\hsize]{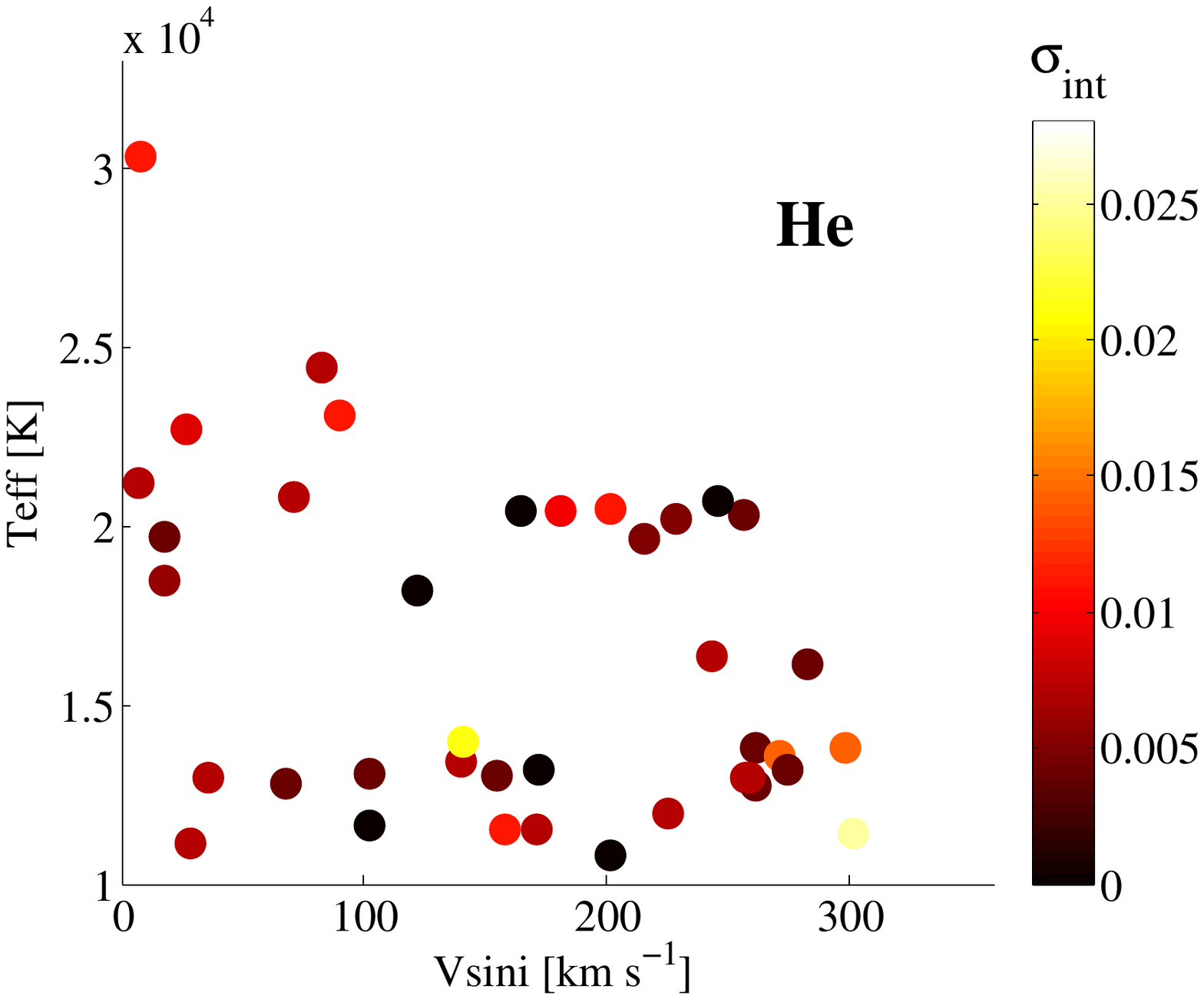}
\end{minipage}
\caption{Variations of the dispersion in the Mg and He abundances (colour coded) as a function of $T_{\mathrm{eff}}$ and $V\sin i$.}
\label{fig_errors}
\end{figure}

The abundance uncertainties were empirically estimated by comparing the results for stars having multiple determinations from GES and archival data. For the stars with $T_{\mathrm{eff}}$ $>$ 20 kK, we also compared the abundances obtained for $\xi$ = 2 and 5 km s$^{-1}$ to take the impact of the choice of the microturbulence into account. The various sources of error were added in quadrature. Figure \ref{fig_errors} shows the internal dispersion in the Mg abundances as a function of $T_{\mathrm{eff}}$ and $V\sin i$. In this particular case, irrespective of the $V\sin i$, an uncertainty of 0.15 and 0.20 dex was assigned to stars cooler and hotter than 20 kK, respectively. For the other elements, no clear dependence with the parameters was found (as illustrated in Fig.~\ref{fig_errors} for He) and a single value was adopted. The 1-$\sigma$ uncertainties lie in the range 0.1--0.3 dex for the metals and are fixed to 0.015 by number for helium\footnote{The He abundance, $y$, is defined as $\cal N$(He)/[$\cal N$(H)+$\cal N$(He)], where $\cal N$ is the number density of atoms.} (see Table \ref{tab_uncertainties}), but were arbitrarily inflated by a factor 1.5 when the fit of the line was poor or the spectrum contaminated by that of a calibration lamp.  

\begin{table}[h!]
\centering
\caption{Nominal abundance uncertainties in dex (except $y$: by number).
}
\begin{tabular}{ll} \hline\hline
  Quantity                       & Uncertainty\\
  \hline
  $y$                            & 0.015 \\
  $\log \epsilon$(\ion{C}{ii})   & 0.15 \\
  $\log \epsilon$(\ion{N}{ii})   & 0.20\\
  $\log \epsilon$(\ion{Ne}{i})   & 0.10 \\
  $\log \epsilon$(\ion{Mg}{ii})  & 0.15 (below 20 kK)\\
                                 & 0.20 (above 20 kK)\\ 
  $\log \epsilon$(\ion{Si}{ii})  & 0.15 \\
  $\log \epsilon$(\ion{Si}{iii}) & 0.30\\
  $[$N/C$]$\tablefootmark{a}     & 0.10\\
  \hline
\end{tabular}
\label{tab_uncertainties}
\tablefoot{
  The final uncertainties may differ because the nominal values were inflated in some cases, while repeated measurements were averaged.
  \tablefoottext{a}{Estimated from repeated measurements and not from a quadratic sum of the N and C uncertainties.}
}
\end{table}

\subsection{Validation and final results}\label{sect_final_results}

Through the analysis of repeated observations, we conclude that there is an overall satisfactory level of agreement between our parameters and abundances irrespective of the instrumental set-up (see Sect.~\ref{sect_internal_validation}). We therefore assume that these are independent measurements and weight them by their random uncertainties to obtain the final, mean values provided in Table \ref{tab_results}. A detailed comparison with respect to external sources (reference values for a set of benchmarks, results from other WG13 nodes or the FS for the stars in common) is provided in Sect.~\ref{sect_external_validation}. 

\begin{table*}[h!]
\tiny
\caption{Final parameters and abundances.}
\centering
\begin{tabular}{lllcccc}
\hline\hline
GES ID                          & Main SIMBAD ID          & FS ID           &  Member? & $T_\mathrm{eff}$ [K]    & $\log g$              & $V\sin i$ [km s$^{-1}$]  \\
\hline
\multicolumn{7}{c}{NGC 3293}\\   
\hline
\object{GES\,10341195--5813066} &  ...                    & ...             & Y?       & 10\,520$\pm$605 (1000) & 4.18$\pm$0.18 (1000)  & 281$\pm$12 (1000)       \\
\object{GES\,10341702--5811419} &  ...                    &  ...            & N        & 10\,730$\pm$647 (1000) & 4.05$\pm$0.12 (1000)  & 175$\pm$14 (1000)       \\
\object{GES\,10341774--5809101} &  ...                    &  ...            & Y        & 10\,880$\pm$631 (1000) & 4.31$\pm$0.10 (1000)  & 150$\pm$8  (1000)       \\
\object{GES\,10342068--5814107} &  ...                    &  ...            & Y        & 13\,571$\pm$659 (1000) & 4.18$\pm$0.08 (1000)  & 180$\pm$7  (1000)       \\
\object{GES\,10342078--5813305} &  CPD --57$^{\degr}$3450  & 3293-049        & Y        & 18\,221$\pm$727 (1100) & 4.03$\pm$0.08 (1100)  & 122$\pm$5  (1100)       \\
\object{GES\,10342325--5808448} &  ...                    &  ...            & N        & 13\,690$\pm$783 (1000) & 4.00$\pm$0.10 (1000)  & 204$\pm$2  (1000)       \\
\object{GES\,10342859--5807396} &  ...                    &  ...            & Y        & ...                    & ...                   & ...                     \\
 ...                            &  ...                    &  ...            & ...      & ...                    & ...                   & ...                     \\
\hline
\multicolumn{7}{c}{Benchmarks}\\   
\hline                 
\object{GES\,00131415+1511008}  & $\gamma$ Peg            &  ...            & ...      & 21\,204$\pm$192 (1010) & 3.83$\pm$0.04 (1010)  &    7$\pm$2 (1010)       \\
\object{GES\,05280146+0117537}  & HD 35912                &  ...            & ...      & 19\,735$\pm$323 (1010) & 4.06$\pm$0.03 (1010)  &   18$\pm$2 (1010)       \\
\object{GES\,05493290+1239044}  & 134 Tau                 &  ...            & ...      & 11\,160$\pm$416 (1010) & 4.09$\pm$0.08 (1010)  &   28$\pm$3 (1010)       \\
\object{GES\,07173159--0549215} & HD 56613                &  ...            & ...      & 13\,073$\pm$407 (1010) & 4.16$\pm$0.08 (1010)  &  102$\pm$6 (1010)       \\
\object{GES\,10425736--6423398} & $\theta$ Car            &  ...            & ...      & 31\,590$\pm$400 (1000) & 4.12$\pm$0.08 (1000)  &   96$\pm$6 (1000)       \\
\object{GES\,16355294--2812579} & $\tau$ Sco              &  ...            & ...      & 30\,369$\pm$240 (1010) & 4.06$\pm$0.05 (1010)  &    8$\pm$2 (1010)       \\
\hline                 
\end{tabular}
\label{tab_results} 
\end{table*}

\addtocounter{table}{-1}
\begin{table*}[h!]
\tiny
\caption{continued.}
\centering
\begin{tabular}{lcccccc}
\hline\hline
GES ID                                           & $\langle RV \rangle$ [km s$^{-1}$] & $y$                       & $\log \epsilon$(\ion{C}{ii}) & $\log \epsilon$(\ion{N}{ii}) & $\log \epsilon$(\ion{Ne}{i}) & $\log \epsilon$(\ion{Mg}{ii})\\
\hline                                                                                                                                                                         
\multicolumn{7}{c}{NGC 3293} \\                                                                                                                                                
\hline                                                                                                                                                                         
\object{GES\,10341195--5813066}                  &   +5.80$\pm$2.10  (1.48)           &  0.120$\pm$0.015 (1000)   &  ...                         &  ...                         & ...                          &  7.85$\pm$0.15 (1000)        \\
\object{GES\,10341702--5811419}                  &   +9.33$\pm$3.10  (1.79)           &  0.125$\pm$0.015 (1000)   &  ...                         &  ...                         & ...                          &  7.78$\pm$0.15 (1000)        \\
\object{GES\,10341774--5809101}                  & --16.43$\pm$7.05 (4.07)            &  0.080$\pm$0.015 (1000)   &  ...                         &  ...                         & ...                          &  7.69$\pm$0.15 (1000)        \\
\object{GES\,10342068--5814107}                  &   +0.83$\pm$3.29  (1.65)           &  0.085$\pm$0.015 (1000)   &  ...                         &  ...                         & ...                          &  7.56$\pm$0.15 (1000)        \\
\object{GES\,10342078--5813305}                  & --18.21$\pm$1.15 (0.43)            &  0.085$\pm$0.011 (1100)   &  8.34$\pm$0.11 (1100)        &  7.73$\pm$0.14 (1100)        & 7.89$\pm$0.10 (1000)         &  7.48$\pm$0.11 (1100)        \\
\object{GES\,10342325--5808448}                  &   +4.93$\pm$14.22 (8.21)           &  0.110$\pm$0.015 (1000)   &  ...                         &  ...                         & ...                          &  7.22$\pm$0.15 (1000)        \\
\object{GES\,10342859--5807396}                  & ...                                &  ...                      &  ...                         &  ...                         & ...                          &  ...                         \\
 ...                                             & ...                                &  ...                      &  ...                         &  ...                         & ...                          &  ...                         \\
\hline                                                                                                                                                                         
\multicolumn{7}{c}{Benchmarks} \\                                                                                                                                              
\hline                                                                                                                                                                         
\object{GES\,00131415+1511008}                   & ...                                &  0.085$\pm$0.011 (1010)   &  8.03$\pm$0.15 (0010)        &  7.63$\pm$0.14 (1010)        &  7.87$\pm$0.10 (1000)        &  7.16$\pm$0.14 (1010)        \\
\object{GES\,05280146+0117537}                   & ...                                &  0.082$\pm$0.011 (1010)   &  8.17$\pm$0.15 (0010)        &  7.70$\pm$0.14 (1010)        &  7.93$\pm$0.10 (1000)        &  7.49$\pm$0.11 (1010)        \\
\object{GES\,05493290+1239044}                   & ...                                &  0.085$\pm$0.011 (1010)   &  ...                         &  ...                         &  ...                         &  7.60$\pm$0.11 (1010)        \\
\object{GES\,07173159--0549215}                  & ...                                &  0.098$\pm$0.011 (1010)   &  ...                         &  ...                         &  ...                         &  7.47$\pm$0.11 (1010)        \\
\object{GES\,10425736--6423398}                  & ...                                &  0.160$\pm$0.015 (1000)   &  ...                         &  8.85$\pm$0.20 (1000)        &  ...                         &  7.76$\pm$0.20 (1000)        \\
\object{GES\,16355294--2812579}                  & ...                                &  0.122$\pm$0.011 (1010)   &  7.93$\pm$0.15 (0010)        &  8.05$\pm$0.14 (1010)        &  7.85$\pm$0.10 (1000)        &  7.42$\pm$0.14 (1010)        \\
\hline                 
\end{tabular}
\end{table*}

\addtocounter{table}{-1}
\begin{table*}[h!]
\tiny
\caption{continued.}
\centering
\begin{tabular}{lcccll}
\hline\hline
GES ID                                           & $\log \epsilon$(\ion{Si}{ii}) & $\log \epsilon$(\ion{Si}{iii}) & [N/C]                  & Technical flag   & Stellar peculiarity flag\\
\hline                                           
\multicolumn{6}{c}{NGC 3293} \\                  
\hline                                           
\object{GES\,10341195--5813066}                  &  ...                          &  ...                           & ...                    & ...              &  ...                    \\
\object{GES\,10341702--5811419}                  &  ...                          &  ...                           & ...                    & ...              &  ...                    \\
\object{GES\,10341774--5809101}                  &  ...                          &  ...                           & ...                    & ...              &  ...                    \\
\object{GES\,10342068--5814107}                  &  ...                          &  ...                           & ...                    & ...              &  ...                    \\
\object{GES\,10342078--5813305}                  &  7.76$\pm$0.15 (1000)         &  7.36$\pm$0.21 (1100)          & --0.61$\pm$0.07 (1100) & ...              &  ...                    \\
\object{GES\,10342325--5808448}                  &  ...                          &  ...                           & ...                    & ...              &  ...                    \\
\object{GES\,10342859--5807396}                  &  ...                          &  ...                           & ...                    & 10050-13-16-00-A &  ...                    \\
 ...                                             &  ...                          &  ...                           & ...                    & ...              &  ...                    \\
\hline                                           
\multicolumn{6}{c}{Benchmarks} \\                
\hline                                           
\object{GES\,00131415+1511008}                   &  6.91$\pm$0.15 (1000)         &  7.33$\pm$0.21 (1010)          & --0.46$\pm$0.10 (0010) &  ...             &  ...                    \\
\object{GES\,05280146+0117537}                   &  7.34$\pm$0.15 (1000)         &  7.48$\pm$0.21 (1010)          & --0.58$\pm$0.10 (0010) &  ...             &  ...                    \\
\object{GES\,05493290+1239044}                   &  ...                          &  ...                           & ...                    &  ...             &  ...                    \\
\object{GES\,07173159--0549215}                  &  ...                          &  ...                           & ...                    &  ...             &  ...                    \\
\object{GES\,10425736--6423398}                  &  ...                          &  8.03$\pm$0.30 (1000)          & ...                    &  ...             &  ...                    \\
\object{GES\,16355294--2812579}                  &  ...                          &  7.45$\pm$0.21 (1010)          &  +0.20$\pm$0.10 (0010) &  ...             &  ...                    \\
\hline                 
\end{tabular}
\tablefoot{The uncertain cluster members are the nine stars with discrepant proper motions discussed in Sect.~\ref{sect_membership_gaia}. Except for $\langle RV \rangle$, the number in brackets is a flag indicating the origin of the determination: the first, second, third, and fourth digit indicates whether the value is based on GES GIRAFFE, FS GIRAFFE, GES UVES, or FS UVES, respectively. For instance, `1001' means that the determination is based on GES GIRAFFE and FS UVES spectra. For $\langle RV \rangle$, it is the uncertainty in the average: 1-$\sigma$ dispersion divided by $\sqrt{N}$, where $N$ is the number of spectra used. The flags reported in the last two columns are described in Table \ref{tab_flags}. The table is available in its entirety through the CDS. A portion is shown here for guidance regarding its form and content.}
\end{table*}

\section{Discussion of apparent fundamental parameters and their parent non-rotating counterparts}\label{sect_discussion_pnrc}

Given that most targets are fast rotators, it is relevant to consider the influence of rotation on the observed stellar properties. For instance, stellar rotation can lead to a blurring of the main-sequence turn-off and mimic an age spread in young and intermediate age open clusters \citep[e.g.][]{marino18,bastian18}. There is therefore a need to distinguish the physical quantities that are based on stellar model atmospheres and evolutionary tracks neglecting or not rotation.

\subsection{Methodology}\label{sect_discussion_methodology}

The first stellar quantities are called `apparent' parameters and have been the subject of the previous sections. Those resulting from the use of models where the effects induced by the stellar rotation are taken into account are called `parent non-rotating counterparts' \citep[hereafter $pnrc$;][]{fremat05,Zorec2016,Cochetti2020} and represent the objects as if they were at rest. These parameters for the sub-sample of 137 cluster members with spectroscopic parameters are discussed below. We consider as apparent quantities $T_\mathrm{eff}$, $\log g_\mathrm{eff}$, $V\!\sin i$, as well as the bolometric luminosity, $L$. The last quantity is obtained from the apparent bolometric magnitude, $M_\mathrm{bol}$, which is in turn obtained from the absolute magnitude in the $V$ band, $M_\mathrm{V}$, following
 
\begin{equation} 
\displaystyle M_\mathrm{bol} = M_\mathrm{V}+BC(T_\mathrm{eff}),
\label{eqMv}
\end{equation}  

\noindent where $M_\mathrm{V}$ is calculated thanks to the {\em Gaia} EDR3 parallaxes and $(G,G_\mathrm{BP},G_\mathrm{RP})$ photometric data \citep{gaia21,gaia_c}. The {\em Gaia} parallaxes were corrected following Sect.~\ref{sect_membership_gaia}, while the magnitudes were transformed into the Johnson-Cousins $UBV$ system using the relations of \citet{riello21}. We call also `apparent' the stellar mass, $M$, and age, $t$, when they are obtained from the apparent $T_\mathrm{eff}$, $\log g_\mathrm{eff}$, and $L$ through evolutionary models without rotation. \par
The intrinsic $UBV$ colours are needed to estimate $M_\mathrm{V}$ and the interstellar colour excess, $E(B-V)$. They were interpolated as a function of the apparent parameters ($T_\mathrm{eff},\log g_\mathrm{eff}$) in the tables of \citet{caskur2003}, which were updated in 2011. We obtain $\langle E(B-V) \rangle$ = 0.29$\pm$0.11 mag. The individual values (given in Table \ref{tab_rotation}) show quite a large spread. The patchy nature of the extinction has long been known \citep[e.g.][]{turner80} and arises from dust clouds obscuring part of this young cluster \citep[][]{preibisch17}. The bolometric correction, $BC(T_\mathrm{eff})$, is taken from LTE model atmospheres \citep{caskur2003} for $T_\mathrm{eff}\lesssim 15$~kK and from non-LTE ones \citep{hula1995,Lanz+Hubeny07} when $T_\mathrm{eff}\gtrsim 15$~kK according to the recommendations of \citet{pedersen20}. The (unavoidable) slight inconsistencies at the $T_\mathrm{eff}$ boundary do not lead to appreciable errors. In both cases, the warnings put forward by \citet{torr2010} were taken into account.\par

 The set of fundamental parameters corrected for effects carried by rotation is made of the {\em pnrc} $T_{\rm eff, pnrc}(M,t)$, $g_{\rm eff, pnrc}(M,t)$, and $L_{\rm pnrc}(M,t)$. The quantity $V(M,t) \sin i$ is corrected for the overestimation discussed by \citet{sto1968} induced by gravity darkening \citep[GD;][]{zeipel1924,espinosa_lara11}. The actual stellar mass $M$, the age $t$ of the star as a rotating object, and the inclination angle $i$ of the rotation axis are considered $pnrc$, as opposed to those derived from the set of apparent $(T_\mathrm{eff}, \log g_\mathrm{eff}, L$) parameters and stellar models without rotation. All mentioned {\em pnrc} parameters and the $V\!\sin i$ corrected from GD effect are derived solving the following system of four equations:
  
 \begin{equation}
   \begin{cases} \displaystyle P_\mathrm{app}  =  \displaystyle P_\mathrm{pnrc}(M,t)\ C_\mathrm{P}(M,t,\eta,i) \\
\displaystyle \frac{(V\!\sin i)_\mathrm{app}}{V_\mathrm{c}(M,t)} = \displaystyle \left[\frac{\eta}{R_\mathrm{e}(M,t,\eta)/R_\mathrm{c}(M,t)}\right]^{1/2}\!\!\!\!\sin i-\frac{\Sigma(M,t,\eta,i)}{V_\mathrm{c}(M,t)}, 
   \end{cases}
\label{eqsy}
 \end{equation}

 \noindent where $P$ $\equiv$ $T_\mathrm{eff}$, $\log g$, or $L$. The variables $P_\mathrm{app}$ and $P_\mathrm{pnrc}$ stand for the apparent and $pnrc$ stellar parameters, while $V_\mathrm{c}$ is the critical velocity. The present-day, actual and critical stellar equatorial radii, $R_\mathrm{e}(M/M_{\odot},t/t_\mathrm{MS},\eta)$ and $R_\mathrm{c}(M/M_{\odot},t/t_\mathrm{MS})$,  are determined using 2D models of rigidly rotating stars \citep{Zorec2011,Zorec2012}. The quantity, $\eta=(\Omega/\Omega_\mathrm{c})^2[R_\mathrm{e}/R_\mathrm{c}]^3$, is the ratio of the centrifugal to the gravitational acceleration at the equator. On the right-hand side of Eq.~\ref{eqsy}, the functions $C_\mathrm{P}(M,t,\eta,i)$ carry all the information relative to the change of parameters due to the oblateness of the rotating star and the concomitant GD effect over the observed hemisphere \citep{fremat05,Zorec2016}. The term $\Sigma(M/M_{\odot},t/t_\mathrm{MS},\eta,i)$ is Stoeckley's correction that takes GD into account \citep{espinosa_lara11,Zorec2017b}. The current angular velocity is the result of the loss and redistribution of angular momentum undergone in the star since the pre-main-sequence phase. As no observational information exists about the internal angular velocity profile, the behaviour at the stellar surface, which is probably differential, is nevertheless taken here uniform over the whole area (rigid surface rotation). However, we adopt the time dependence, $\Omega \equiv \Omega(t)$, as predicted by the Geneva evolutionary models with rotation \citep{Maeder2000,Ekstrom2008,ekstroem12,georgy13}. A moderate core convective overshoot of 0.1$H_\mathrm{p}$, where $H_\mathrm{p}$ is the local pressure scale height, is adopted for the relevant mass range \citep{ekstroem12}. 

\begin{figure}[h!] 
\centerline{\includegraphics[trim=-60 285 150 150,clip,scale=0.9]{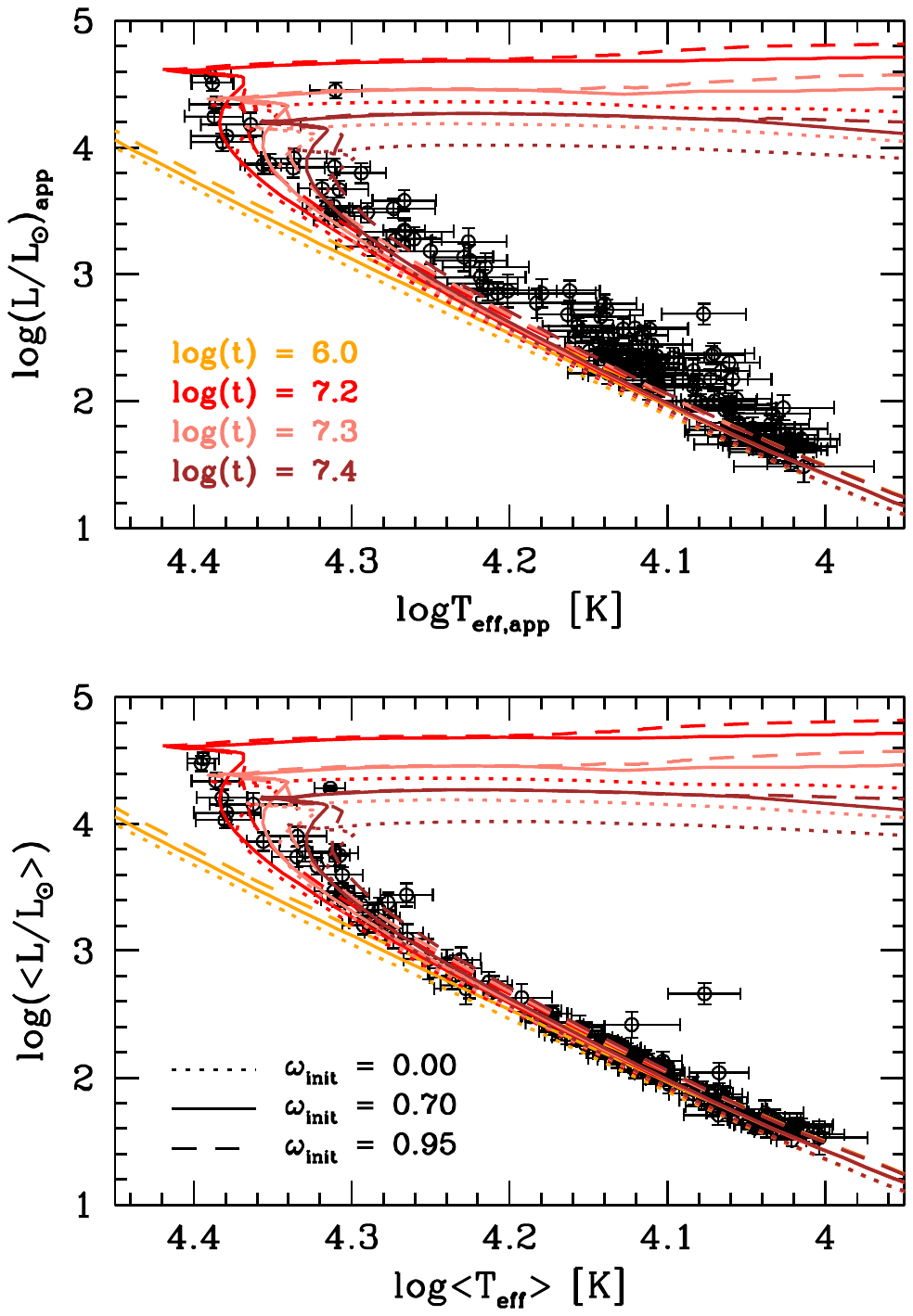}} 
\caption{\label{fig_HR} HR diagrams using apparent ({\it top panel}) and averaged over the stellar surface ({\it bottom panel}) stellar parameters. Isochrones for various ages and initial rotation rates are overlaid \citep{georgy13}.}
\end{figure}

 For each star, Eq.~\ref{eqsy} is solved for $10^4$ Monte Carlo trials. Each time, a new uncertainty, $\epsilon_\mathrm{P}$, corresponding to a given apparent parameter, $P_\mathrm{app}^\mathrm{eff}$, is drawn at random according to a Gaussian distribution with a standard deviation given in Tables \ref{tab_results} and \ref{tab_rotation}. Moreover, the system of equations must satisfy the condition that the $V$ predicted by the model reproduces the observed $V\!\sin i$ corrected for GD. At each iteration step, the $pnrc$ $T_\mathrm{eff}$ and $L$ that are input parameters to the evolutionary models with rotation are transformed into values averaged over the stellar surface deformed by rotation, so as to be similar in nature to the tabulated model quantities. The apparent and derived $pnrc$ quantities are given in Table \ref{tab_rotation}. We also provide the quantities averaged over the rotationally deformed stellar surface, $\log \langle T_\mathrm{eff} \rangle$, $\log \langle g \rangle$, and $\log$($\langle L/L_{\odot} \rangle$), that are best suited for a direct comparison with the model predictions. The Hertzsprung-Russell (HR) diagrams shown in Fig.~\ref{fig_HR} allow one to appreciate the impact of correcting for rotation-related effects. We note that using apparent stellar parameters leads to stars of the lower-main sequence lying significantly above the zero-age main sequence (ZAMS), which was a concern raised by \citet{dufton06} when analysing the FS data.\par

\begin{table*}
\centering
\caption{Colour excesses and stellar parameters ($pnrc$, averaged over the stellar surface, and apparent).}
\tiny
\begin{tabular}{l|c|ccccc} \hline\hline
  GES ID
& $E$($B-V$)
& $T_\mathrm{eff,pnrc}$   
& $\log g_\mathrm{pnrc}$       
& $\log$($L/L_{\odot}$)$_\mathrm{pnrc}$
& ($V\sin i$)$_\mathrm{pnrc}$   
& ($M/M_{\odot}$)$_\mathrm{pnrc}$
  \\
& [mag]
& [K]       
&       
&  
& [km s$^{-1}$]   
&  
\\
\hline
\object{GES\,10341195--5813066} &  0.369$\pm$0.054 & 10\,890$\pm$636  & 4.248$\pm$0.150 & 1.521$\pm$0.090  & 283$\pm$12  & 1.9$\pm$0.8\\
\object{GES\,10341774--5809101} &  0.415$\pm$0.053 & 10\,927$\pm$657  & 4.311$\pm$0.120 & 1.634$\pm$0.094  & 150$\pm$7   & 2.7$\pm$1.1\\
\object{GES\,10342068--5814107} &  0.407$\pm$0.046 & 13\,690$\pm$653  & 4.211$\pm$0.090 & 2.179$\pm$0.085  & 181$\pm$6   & 3.0$\pm$1.0\\
\object{GES\,10342078--5813305} &  0.236$\pm$0.043 & 18\,324$\pm$712  & 4.043$\pm$0.087 & 3.251$\pm$0.110  & 122$\pm$5   & 7.5$\pm$2.6\\
\object{GES\,10343505--5813506} &  0.201$\pm$0.045 & 13\,163$\pm$614  & 4.035$\pm$0.170 & 2.231$\pm$0.084  & 233$\pm$5   & 2.7$\pm$1.1\\
\object{GES\,10343562--5815459} &  0.243$\pm$0.057 & 10\,694$\pm$779  & 4.069$\pm$0.156 & 1.922$\pm$0.116  & 131$\pm$7   & 3.4$\pm$1.8\\
\object{GES\,10344202--5815419} &  0.290$\pm$0.046 & 15\,232$\pm$1196 & 3.899$\pm$0.153 & 2.818$\pm$0.143  &  95$\pm$7   & 4.5$\pm$2.5\\
...                             & ...              & ...              & ...             & ...              & ...         & ...        \\
\hline
\end{tabular}
\label{tab_rotation}
\end{table*}

\addtocounter{table}{-1}
\begin{table*}
\centering
\caption{continued.}
\tiny
\begin{tabular}{lcccccc} \hline\hline
  GES ID
&  $V_\mathrm{c,pnrc}$         
&  $i$             
&  ($V/V_\mathrm{c}$)$_\mathrm{pnrc}$      
&  $\omega_\mathrm{pnrc}$
&  ($t/t_\mathrm{MS}$)$_\mathrm{pnrc}$        
&  $\log$($t_\mathrm{pnrc}$)
  \\
&  [km s$^{-1}$]       
&  [$\degr$]               
&     
&  
&     
&  [yr]         
\\
\hline
\object{GES\,10341195--5813066} & 416$\pm$29 & 46$\pm$13 &  0.925$\pm$0.067 &  0.999$\pm$0.013 & 0.016$\pm$0.004 & 7.3927$\pm$0.0920  \\
\object{GES\,10341774--5809101} & 429$\pm$25 & 69$\pm$17 &  0.372$\pm$0.017 &  0.540$\pm$0.022 & 0.024$\pm$0.015 & 7.1817$\pm$0.2058  \\
\object{GES\,10342068--5814107} & 443$\pm$21 & 55$\pm$14 &  0.489$\pm$0.025 &  0.686$\pm$0.027 & 0.072$\pm$0.029 & 7.3028$\pm$0.1389  \\
\object{GES\,10342078--5813305} & 458$\pm$21 & 60$\pm$16 &  0.305$\pm$0.015 &  0.451$\pm$0.020 & 0.409$\pm$0.139 & 7.3914$\pm$0.1068  \\
\object{GES\,10343505--5813506} & 403$\pm$30 & 75$\pm$17 &  0.593$\pm$0.025 &  0.792$\pm$0.024 & 0.079$\pm$0.026 & 7.3436$\pm$0.1170  \\
\object{GES\,10343562--5815459} & 379$\pm$28 & 71$\pm$18 &  0.360$\pm$0.021 &  0.529$\pm$0.026 & 0.039$\pm$0.021 & 7.2471$\pm$0.1797  \\
\object{GES\,10344202--5815419} & 397$\pm$30 & 86$\pm$21 &  0.244$\pm$0.015 &  0.369$\pm$0.021 & 0.168$\pm$0.095 & 7.2517$\pm$0.1840  \\
...                             & ...        & ...       & ...              & ...              & ...             & ...                \\
\hline
\end{tabular}
\end{table*}

\addtocounter{table}{-1}
\begin{table*}
\centering
\caption{continued.}
\tiny
\begin{tabular}{l|ccc|cccc} \hline\hline
  GES ID 
&  $\log \langle T_\mathrm{eff} \rangle$
&  $\log \langle g \rangle$
&  $\log$($\langle L/L_{\odot} \rangle$)    
&  $T_\mathrm{eff,app}$
&  $\log$($L/L_{\odot}$)$_\mathrm{app}$     
&  ($t/t_\mathrm{MS}$)$_\mathrm{app}$    
&  $\log$($t_\mathrm{app}$)        \\
&  [K]          
&     
&
&  [K]
&
&
&  [yr]           \\
\hline
\object{GES\,10341195--5813066} &  4.0150$\pm$0.0195 & 4.1370$\pm$0.1000 & 1.6122$\pm$0.0686 & 10\,520$\pm$605  &  1.704$\pm$0.081 &  0.0234$\pm$0.0232 &  7.0545$\pm$0.2952 \\
\object{GES\,10341774--5809101} &  4.0426$\pm$0.0143 & 4.2906$\pm$0.0642 & 1.6322$\pm$0.0746 & 10\,880$\pm$631  &  1.660$\pm$0.081 &  0.0183$\pm$0.0216 &  6.9580$\pm$0.3355 \\
\object{GES\,10342068--5814107} &  4.1321$\pm$0.0119 & 4.1738$\pm$0.0518 & 2.1783$\pm$0.0655 & 13\,571$\pm$659  &  2.220$\pm$0.078 &  0.0679$\pm$0.0511 &  7.1397$\pm$0.2385 \\
\object{GES\,10342078--5813305} &  4.2788$\pm$0.0111 & 4.0308$\pm$0.0542 & 3.1812$\pm$0.1006 & 18\,221$\pm$727  &  3.279$\pm$0.092 &  0.5685$\pm$0.1224 &  7.4181$\pm$0.0736 \\
\object{GES\,10343505--5813506} &  4.1293$\pm$0.0109 & 3.9784$\pm$0.0867 & 2.1893$\pm$0.0630 & 13\,005$\pm$632  &  2.304$\pm$0.077 &  0.0936$\pm$0.0521 &  7.2502$\pm$0.1850 \\
\object{GES\,10343562--5815459} &  4.0769$\pm$0.0157 & 4.0502$\pm$0.0742 & 1.8263$\pm$0.0818 & 10\,633$\pm$794  &  1.949$\pm$0.100 &  0.0516$\pm$0.0273 &  7.2494$\pm$0.1788 \\
\object{GES\,10344202--5815419} &  4.2252$\pm$0.0183 & 3.8910$\pm$0.0786 & 2.7298$\pm$0.1072 & 15\,120$\pm$1190 &  2.852$\pm$0.115 &  0.2079$\pm$0.1303 &  7.2332$\pm$0.2044 \\
...                             & ...                & ...               & ...               &                & ...              & ...                & ...                \\
\hline
\end{tabular}
\tablefoot{The table is available in its entirety through the CDS. A portion is shown here for guidance regarding its form and content.}
\end{table*}

 Given the unknown incidence of magnetic stars in NGC 3293, specific models \citep[e.g.][]{keszthelyi19,keszthelyi20} are not discussed. Furthermore, only about 10\% of all O and early B stars \citep{grunhut17,morel15} or late B stars \citep{donati09} have a detected, large-scale magnetic field. Unlike the majority of our targets, they are usually (very) slow rotators because of magnetic braking or show some sort of chemical peculiarities, as illustrated by the case of \object{CPD --57$^{\degr}$3509} \citep{przybilla16}.

\subsection{Rotational velocity distribution}\label{sect_discussion_rotational_velocity}

\begin{figure*}[h!] 
  \centering
  \includegraphics[trim=0 10 0 20,clip,width=0.85\hsize]{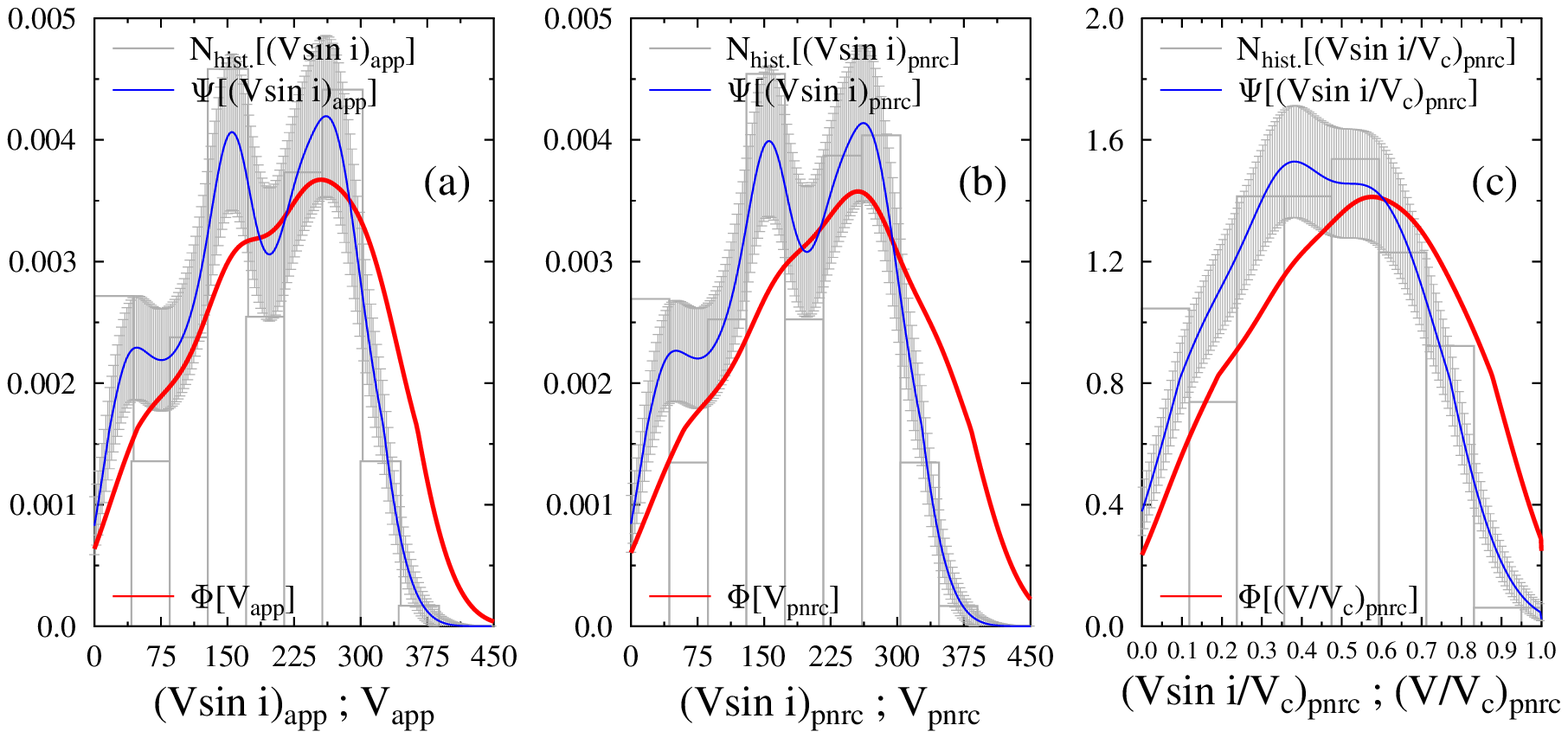}
\caption{\label{fig_lucy} Rotational velocity distributions. (a) Histogram ({\em grey}) and smoothed distribution ({\em blue}) of  $(V\!\sin i)_\mathrm{app}$. The $V_\mathrm{app}$ distribution is shown in red; (b) same as left panel, but for rotational velocities corrected for GD; (c) histogram ({\em grey}) and smoothed distribution ({\em blue}) of velocity ratios corrected for GD, $(V\!\sin i/V_\mathrm{c})_\mathrm{pnrc}$. The distribution of GD-corrected velocity ratios, $(V/V_\mathrm{c})_\mathrm{pnrc}$, is shown in red.}
\end{figure*}

The differences between the distributions of apparent and corrected surface rotational velocities, as obtained from the solutions of Eq.~\ref{eqsy}, are illustrated in Fig.~\ref{fig_lucy}. The distributions of the observed (apparent) $V\!\sin i$ are shown in the left panel, where the histogram corresponds to the raw values (Table \ref{tab_results}). The class-steps of the histogram are established according to the bin-width optimisation method of \citet{2shi07}. The smoothed version of the $V\!\sin i$ frequency density distribution corrected for measurement uncertainties, $\Psi(V\!\sin i)$, was calculated using kernel estimators \citep{boazz97}. Each observed $V\!\sin i$ is represented by a Gaussian distribution, whose dispersion is given by the standard deviation of individual $V\!\sin i$ estimates. Also shown is the $\Phi(V)$ distribution of the apparent true velocities $V$, which was obtained using the Richardson-Lucy deconvolution method \citep{rich72,lucy74} under the assumption of a random distribution of viewing angles. The middle panel of Fig.~\ref{fig_lucy} depicts the $V\!\sin i$'s corrected for Stoeckley's overestimation and the true velocities, $V_\mathrm{pnrc}$, as they result from the solution of Eq.~\ref{eqsy}. Finally, the distributions of the ratios, $(V\!\sin i/V_\mathrm{c})_\mathrm{pnrc}$, and true velocities, $(V/V_\mathrm{c})_\mathrm{pnrc}$, are shown in the right panel of Fig.~\ref{fig_lucy}. The error bars for the $\Psi$ and $\Phi$ distributions are of similar magnitude. According to the distribution of the $(V/V_\mathrm{c})_\mathrm{pnrc}$ ratios, most of the studied stars in the cluster have values in the range 0.2--0.8. The maximum of the histogram is at $(V/V_\mathrm{c})_\mathrm{pnrc}$ $\sim$ 0.55, which translates into an equatorial acceleration ratio, $\eta=0.23$. It is far from the critical ratio, $\eta=1.0$. The transformation of $(V/V_\mathrm{c})_\mathrm{pnrc}=0.55$ into a ratio of angular velocities leads to $\omega_\mathrm{pnrc}=0.73$, where $\omega$ = $\Omega/\Omega_\mathrm{c}$. It is also below the values closer to critical rotation commonly observed in Be stars.

\begin{figure*}[h!] 
  \centering
  \includegraphics[trim=0 5 0 25,clip,width=0.7\hsize]{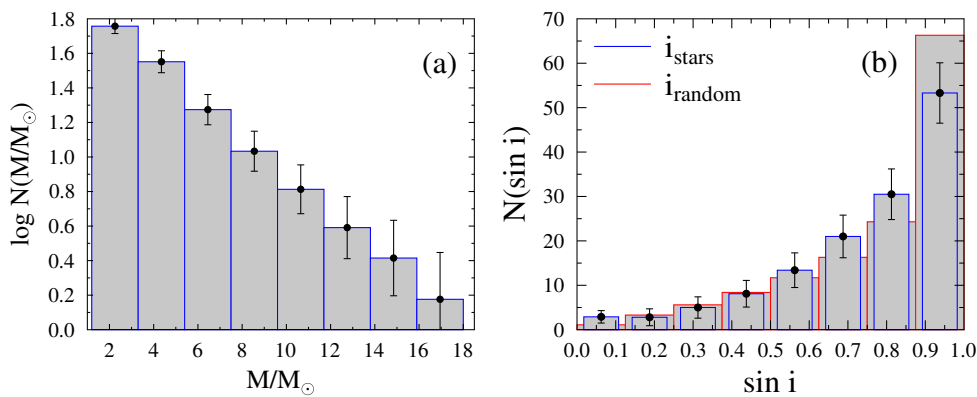}
\caption{\label{fig_mass_i} Stellar mass and inclination angle distributions. (a) $pnrc$ masses; (b) inclination angles compared to the theoretical distribution for values at random.}
\end{figure*}

The B stars lying on the high-velocity tail of the $(V/V_\mathrm{c})_\mathrm{pnrc}$ distribution (i.e. with values above 0.7) are almost all of rather low mass, that is $M$ $\lesssim$ 4--5 M$_{\odot}$. They certainly need a very long time to display the Be phenomenon through the redistribution towards the surface of their internal angular momentum \citep{zoetal2005}. However, these stars could be considered  members of the Bn class, which is an old terminology to designate rapidly rotating B stars and potential candidates to display the Be phenomenon \citep{vanbe1997,baade2000,zfmct2007}. As shown in Fig.~\ref{fig_mass_i}, most of the stars in the cluster have masses $M$ $\lesssim$ 6 M$_{\odot}$. While the maximum frequency of Be stars occurs for spectral type B2 \citep{be1997}, there is only 7\% of stars with masses $M/M_{\odot}\sim 8.0\pm2.1$ and $(V/V_\mathrm{c})_\mathrm{pnrc}\sim 0.74\pm0.15$. This may partially explain that the number of classical Be stars is very low despite the fact that they are common in open clusters with an age in the range 13--25 Myrs \citep{fabregat00}. We only found five stars showing clear evidence for emission in Balmer lines (one shell-like), which is very similar to the tally already reported by the FS \citep{evans05} or \citet{mc_swain09} despite our larger sample. A transition from an absorption to a strong double-peaked H$\alpha$ emission profile is observed between the FS and GES observations of \object{CPD --57$^{\degr}$3531} (\object{GES\,10360595--5814270}), confirming its strongly transient nature \citep{mc_swain09}. Although stars with stable discs are relatively straightforward to detect, even with snapshot observations, other transients might have evidently escaped detection.
 The seven probable Be stars lying in the outskirts of the cluster proposed by \citet{baume03} from narrow-band photometric indices were not observed by the GES. We do not expect any Herbig Ae/Be stars in our sample. Although star formation is still believed to be operating in NGC 3293, according to \citet{baume03} or \citet{delgado16} only lower-mass stars with spectral types later than about A5 may still be contracting on their way to the ZAMS.

 The solution of Eq.~\ref{eqsy} also provides the inclination angle of the stellar rotation axis whose sine distribution is compared with that for angles drawn at random in Fig.~\ref{fig_mass_i}. On account of the uncertainties that plague the angle estimates, there is no indication in this cluster for significant deviations from a global isotropic distribution. The case of other clusters is discussed, for instance, by \citet{jackson10}.

\subsection{Age of the cluster}\label{sect_discussion_age}

\begin{figure}[h!]
  \centering
  \includegraphics[trim=0 0 0 5,clip,width=0.75\hsize]{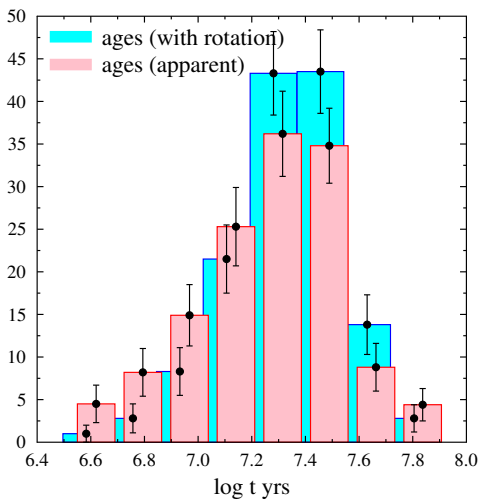}
\caption{\label{fig_ages} Distribution of stellar ages derived from `apparent' ({\em red}) and corrected for GD ({\em blue}) fundamental parameters.}
\end{figure}

From models of stellar evolution, it is apparent that the relationships for stars with masses $M$ $\lesssim$ 6 M$_{\odot}$ between evolutionary time and either $L$ or $T_\mathrm{eff}$ are near `vertical' loci. It implies that the interpolation of ages for stars in this range of masses is highly sensitive to uncertainties in the above input parameters (i.e. to small deviations in $\epsilon_{L}$ or $\epsilon_{T_\mathrm{eff}}$). In some extreme cases, it can lead to assign the objects to either the ZAMS or the terminal age main-sequence (TAMS). To avoid spurious estimates of ages, we impose two limitations to the values obtained from solving Eq.~\ref{eqsy}. Thus, the lowest accepted ages should represent an epoch after the nominal beginning of the ZAMS, which corresponds to the stabilisation phase of the initial angular momentum redistribution in the star \citep{Maeder2000}. To the opposite side of the age distribution, ensuring the consistency of all stellar fundamental parameters for the given rotational rate also acts to rule out extreme values. These limitations produce the truncated shape of the age histograms shown in Fig.~\ref{fig_ages}. Four objects without trustworthy solutions were rejected. The lack of dependence between stellar mass and age ensures that the low-mass stars do not bias our estimate (see above). The fractional ages, $t/t_\mathrm{MS}$, derived from evolutionary models with rotation are smaller that those obtained from models neglecting it. Nevertheless, the time $t_\mathrm{MS}$ that a rotating star spends on the main sequence is substantially longer compared to its non-rotating counterpart. For this reason, the mean cluster age, $\log$($t$) [yr] = 7.31$\pm$0.26, is $\sim$10\% larger than the value when rotation is not taken into account. The age uncertainty is not the standard deviation of the mean, but is computed based on the outcome of the Monte Carlo simulations (Sect.~\ref{sect_discussion_methodology}). As it takes all the sources of error thoroughly into account, it must be regarded as a rather conservative estimate.

\section{Discussion of abundance results}\label{sect_discussion_abundances}

Before discussing the surface chemical properties of the cluster stars, we note that a self-consistent analysis would have required a (tedious) correction of our abundances for GD effects, as described in the previous section for the atmospheric parameters. According to \citet{fremat05}, however, we anticipate that such changes would be small for the relevant rotation rates \citep[see also][]{cazorla17}.

\subsection{General abundance properties}\label{sect_discussion_abundance properties}

\begin{figure*}[h!]
\centering
\includegraphics[trim=50 190 15 135,clip,width=\hsize]{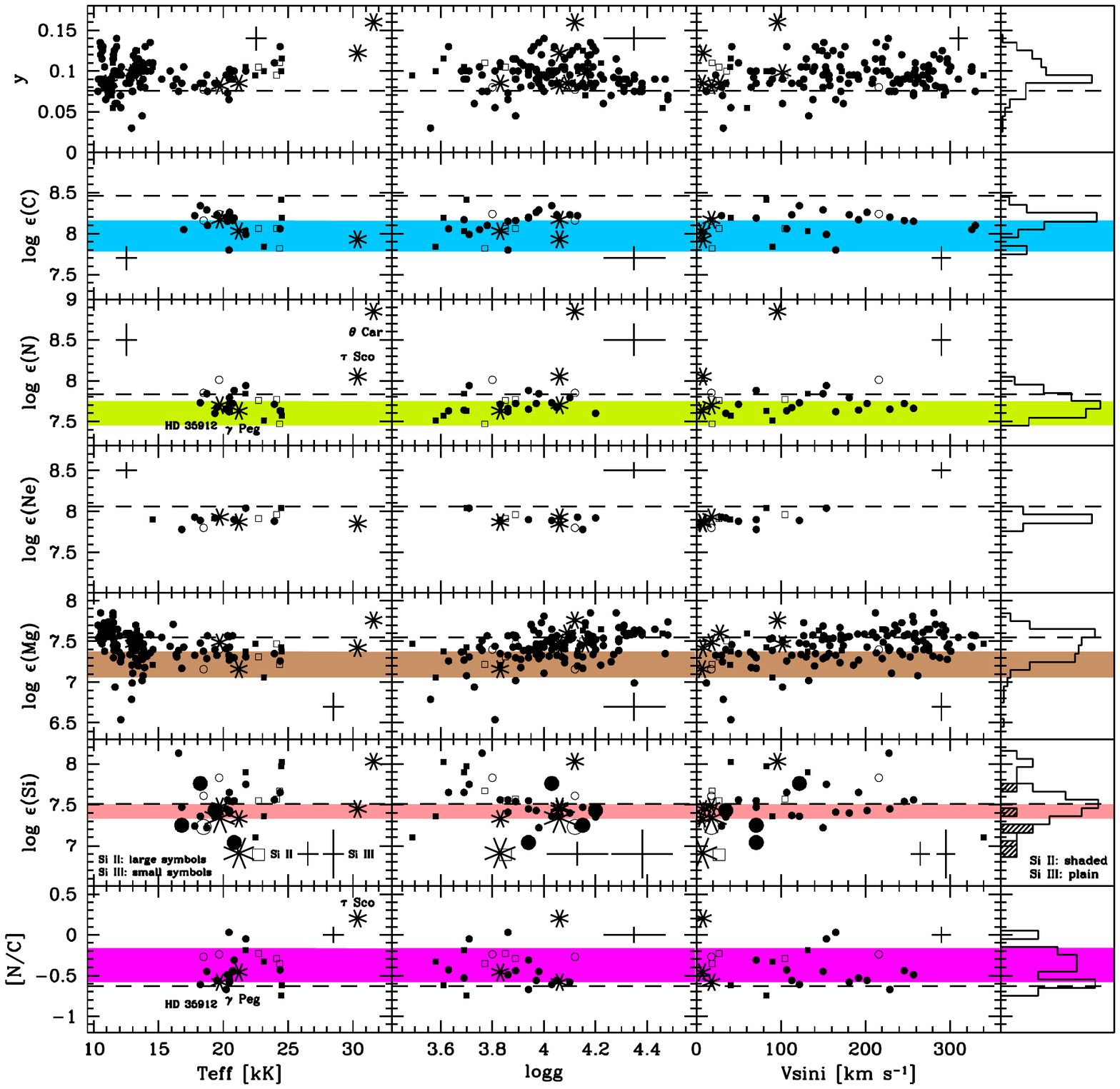}
\caption{Variations of the abundances as a function of the apparent stellar parameters. Non cluster-members (Sect.~\ref{sect_membership_gaia}) were excluded. Stars with or without a binary flag are shown with open and filled symbols, respectively. Stars flagged or not as line-profile variables are plotted as squares and circles, respectively. For both variability types, a confidence level `A' or `B' is required. The benchmarks are shown with star symbols (for convenience, the position of those discussed in Sect.~\ref{sect_discussion_internal_mixing} is indicated in the panels showing the behaviour of the N and [N/C] abundances as a function of $T_\mathrm{eff}$). Crosses show illustrative error bars. The \ion{Si}{ii} and \ion{Si}{iii} data are plotted together in the same panels. The horizontal, dashed line indicates the solar abundance \citep{asplund21}, while the horizontal stripe shows the mean values ($\pm$1$\sigma$) for stars in NGC 3293 determined by \citet{hunter09}. The values are provided in Table \ref{tab_mean_abundances}. The rightmost panels show the breakdown of our abundance data for the NGC 3293 sample.}
\label{fig_abundances_vs_parameters}
\end{figure*}

Figure~\ref{fig_abundances_vs_parameters} shows the variations of the abundances as a function of the stellar parameters for both the cluster stars and the benchmarks. The data are compared to the mean values for NGC 3293 reported by the FS \citep{hunter09} and the most recent set of solar abundances \citep{asplund21}. We find systematically larger average abundances than the FS (see also Fig.~\ref{fig_comparison_parameters_FS}). However, except for Si, our estimates for the metals still lie significantly below the solar values, as also found by  \citet{mathys02} or \citet{niemczura09_ngc3293}. Yet there is no reason to believe that this young cluster is metal poor \citep{strobel91,niemczura05}. A mean LTE iron abundance fully consistent with solar was also reported by \citet{trundle07}. The observed discrepancies might be remedied by the use of better model atoms \citep{nieva_przybilla12}, although we note that a state-of-the-art modelling was employed for \ion{C}{ii} $\lambda$4267 \citep[][]{nieva06,nieva08}.

As a preamble to discussing the dependence between the chemical abundances and the stellar parameters, we recall that our sample spans an exceptionally wide $T_\mathrm{eff}$ and $V\sin i$ range. As a consequence, our analysis is prone to systematic errors, for instance, because of $T_\mathrm{eff}$-dependent deficiencies in the modelling or biases in the treatment of the narrow-lined vs fast-rotating stars. It is therefore important to interpret the results cautiously. This danger is illustrated by the behaviour of helium. As can be seen in Fig.~\ref{fig_abundances_vs_parameters}, we obtain supersolar He abundances and an increase as a function of $T_\mathrm{eff}$ for the early B stars. We chose \ion{He}{i} $\lambda$4471 as our unique diagnostic because this strong, diffuse line can be measured in virtually all stars. However, as part of the GES iDR3 processing cycle, we experimented with \ion{He}{i} $\lambda$4713, which is intrinsically much weaker and cannot be measured in many objects. We found that the difference between the \ion{He}{i} $\lambda$4471- and \ion{He}{i} $\lambda$4713-based abundances increases as a function of $T_\mathrm{eff}$, and reaches up to $\Delta y$ $\sim$ 0.04 at $\sim$25 kK. A similar tendency was noticed during our past abundance studies of nearby B-type stars also based on DETAIL-SURFACE \citep{morel06,morel08}. We conclude that relatively large line-to-line abundance differences may be expected and that the $T_\mathrm{eff}$ trend is likely an artefact. A helium excess in the B0--B0.2 V stars $\tau$ Sco and $\theta$ Car is not supported by previous studies \citep[e.g.][]{hubrig08}, and might instead be found in dwarfs more massive and rotating dramatically faster \citep[e.g.][]{howarth01,cazorla17}. We note that \object{GES\,10354901--5814541} (\object{CPD --57$^{\degr}$3509}) that is known to be a strongly magnetic, He-rich star \citep[][]{przybilla16} was flagged as a line-profile variable and later discarded given the unsuitability of standard models.

 There are only six stars with both \ion{Si}{ii} and \ion{Si}{iii} abundances. The mean difference shows a large scatter (--0.21$\pm$0.40 dex; \ion{Si}{ii} minus \ion{Si}{iii}), as expected because of the large uncertainties affecting the abundances. On an individual basis, the largest (at the 2- to 3-$\sigma$ level) discrepancies are found for the two early B stars where \ion{Si}{ii} $\lambda$6371 begins to vanish and is barely measurable. Given the limited information at hand, it is unclear whether any meaningful conclusions about the $T_\mathrm{eff}$ scale can be drawn from Si ionisation balance. 

\subsection{Chemically peculiar objects}\label{sect_discussion_chemically_peculiar objects}

\begin{figure}[h!]
\centering
\includegraphics[trim=155 360 115 150,clip,width=1.0\hsize]{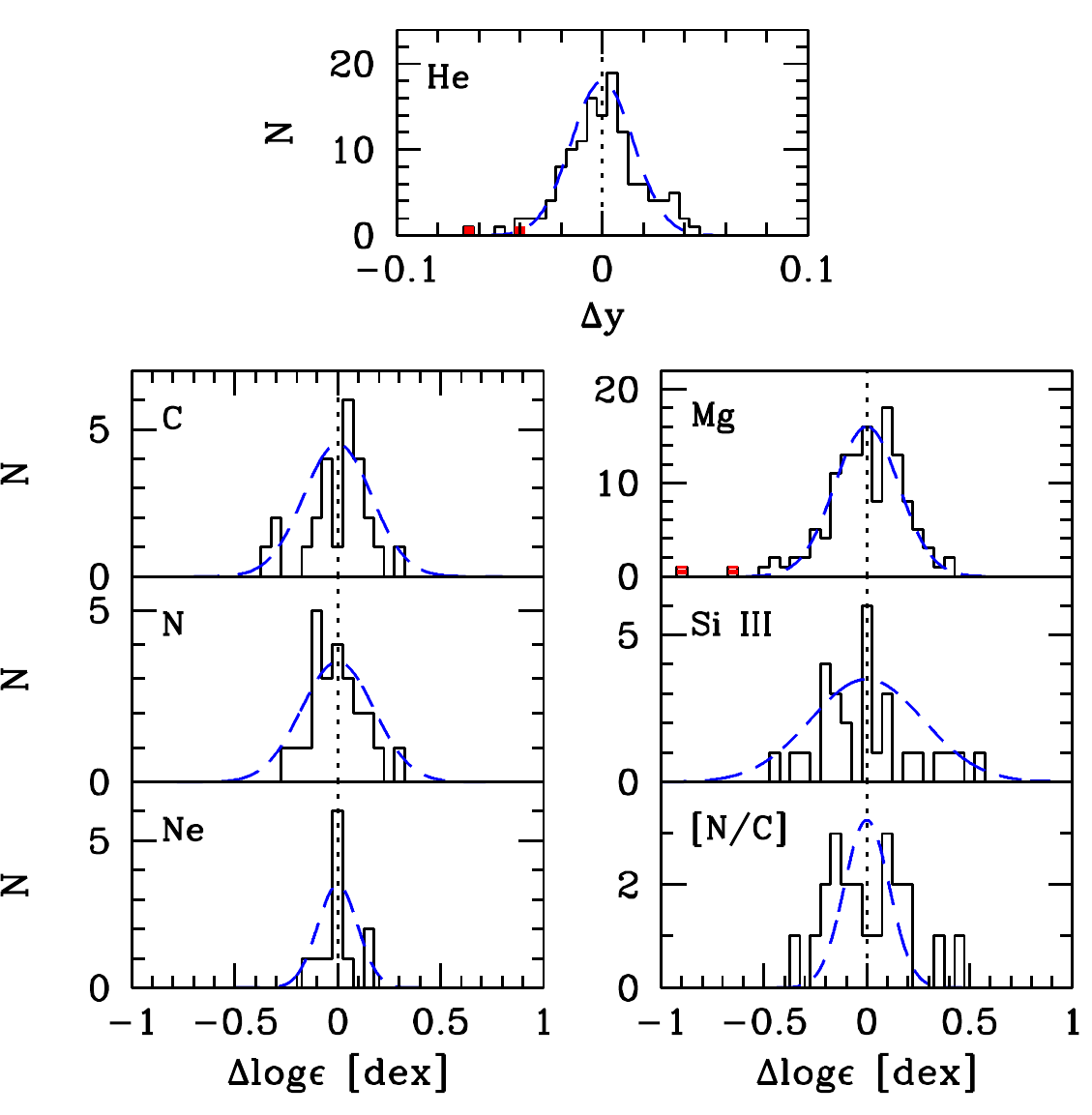}
\caption{Distributions for each abundance ratio of the residuals with respect to the mean values (Table \ref{tab_mean_abundances}). Non cluster-members (Sect.~\ref{sect_membership_gaia}) were excluded. A normal distribution with a standard deviation corresponding to the mean of our final random uncertainties is overplotted with a dashed line. The two stars with peculiar abundances are indicated in red.}
\label{fig_distribution_abundances}
\end{figure}

Figure~\ref{fig_distribution_abundances} shows the distributions for each abundance ratio of the residuals with respect to the mean value, which was estimated by iteratively removing any stars deviating by more than 3$\sigma$. The \ion{Si}{ii}-based abundances are ignored because they are only estimated for a handful of stars. The mean values for the whole sample are given in Table \ref{tab_mean_abundances} and compared to previous estimates in the literature. The distributions of the residuals in Fig.~\ref{fig_distribution_abundances} appear compatible with the measurement errors. It shows that the cluster is globally chemically homogeneous to within our level of precision.

\begin{table*}[h!]
\centering
\caption{Mean abundances for NGC 3293 compared to the solar photospheric values \citep{asplund21} and previous non-LTE spectroscopic studies in the literature.}
\begin{tabular}{l|c|ccc} \hline\hline
\multicolumn{1}{c}{} & \multicolumn{1}{c}{Sun}          & \multicolumn{3}{c}{NGC 3293} \\  
                     &                                  & This study                          & \citet{hunter09}                    &   \citet{mathys02}     \\       
\hline
$y$                  & 0.076$\pm$0.003\tablefootmark{a} & 0.096$\pm$0.019 (129)               & ...                                 &   0.124$\pm$0.018 (6)\tablefootmark{b}  \\
$\log \epsilon$(C)   & 8.46$\pm$0.04                    & 8.13$\pm$0.16 (25)                  & 7.97$\pm$0.19 (27)                  &   8.20$\pm$0.10 (6)   \\
$\log \epsilon$(N)   & 7.83$\pm$0.07                    & 7.72$\pm$0.14 (24)                  & 7.60$\pm$0.15 (27)                  &   7.77$\pm$0.10 (6)   \\
$\log \epsilon$(O)   & 8.69$\pm$0.04                    & ...                                 & 8.65$\pm$0.17 (26)                  &   8.50$\pm$0.13 (6)   \\               
$\log \epsilon$(Ne)  & 8.06$\pm$0.05\tablefootmark{c}   & 7.91$\pm$0.08 (12)                  & ...                                 &   ...                  \\
$\log \epsilon$(Mg)  & 7.55$\pm$0.03                    & 7.45$\pm$0.18 (128)                 & 7.22$\pm$0.16 (26)                  &   ...                 \\
$\log \epsilon$(Si)  & 7.51$\pm$0.03                    & 7.56$\pm$0.25 (28)\tablefootmark{d} & 7.42$\pm$0.09 (27)\tablefootmark{e} &   ...                 \\
  \hline                                                                                                   
$[$N/C$]$            & --0.63$\pm$0.09                  & --0.40$\pm$0.21 (22)                & --0.37$\pm$0.21 (27)                & --0.43$\pm$0.15 (6)   \\
$[$N/O$]$            & --0.86$\pm$0.09                  & ...                                 & --1.05$\pm$0.26 (26)                & --0.73$\pm$0.17 (6)   \\
\hline
\end{tabular}
\tablefoot{Non cluster-members (Sect.~\ref{sect_membership_gaia}) were excluded. The number of stars the estimate is based on is given in brackets. 
\tablefoottext{a}{Based on helioseismology.}
\tablefoottext{b}{Under LTE.}
\tablefoottext{c}{Based on solar wind data.}
\tablefoottext{d}{Only based on \ion{Si}{iii}.}
\tablefoottext{e}{The microturbulence was adjusted to derive the same Si abundance for each star in the cluster \citep[see][]{hunter07,trundle07}.}
} 
\label{tab_mean_abundances}
\end{table*}

There are only two slowly rotating, quite evolved late B stars (\object{GES\,10355513--5811053} and \object{GES\,10362103--5810530}) clearly at odds with the statement made above that the cluster is chemically homogeneous. This conclusion is primarily based on account of their very low Mg abundances, but they are He-poor as well (Fig.~\ref{fig_distribution_abundances}). Late B stars with a severe Mg depletion are discussed, for instance, by \citet[][]{hempel03}. There are no other abundances available and they are not identified as binaries. The weak decline of the Mg abundances for lower $V\sin i$ (Fig.~\ref{fig_abundances_vs_parameters}) is remarkably similar to that observed in other late B-type samples \citep{niemczura09} and may be ascribed to mild diffusion effects in the slow rotators. It is unclear whether the lower abundances for hotter or more evolved objects is also of  physical origin \citep[for a possible correlation as a function of $T_\mathrm{eff}$, see][]{fossati11}, but empirically correcting for these trends does not erase the dependence with $V\sin i$. The two peculiar stars fall about 0.5--1.0 dex below the values for stars with similar $T_\mathrm{eff}$ or $\log g$, which supports their classification as chemically peculiar.

The \ion{Hg}{ii} $\lambda$3984 line, which is one of the prime diagnostics for a HgMn classification is not covered by our observations. However, a cursory inspection of the spectra allowed us to confidently detect several \ion{Mn}{ii} lines (at $\lambda$4137, $\lambda$4363, $\lambda$4365, and $\lambda$4479 \AA) in \object{GES\,10355440--5812563} (\object{FS\,3293-064}). This star was too cool to be processed with our code and was classified as A0 II by \citet{evans05}. Identifying chemically peculiar stars in clusters with a precise age estimate is particularly valuable. However, its {\em Gaia} EDR3 data clearly identify it as a foreground object with a discrepant proper motion.

It is well known that radiative levitation and gravitational settling can lead in late B stars to surface abundances dramatically departing from solar \citep[e.g.][]{hempel03,niemczura09}. The high spin rates of our targets (Fig.~\ref{fig_lucy}) may inhibit the development of diffusion processes, although some chemically peculiar objects are unexpectedly quite fast rotators \citep[e.g.][]{gonzalez21}. As mentioned in Sect.~\ref{sect_selection_spectral_analysis}, we only processed stars selected on the basis that \ion{Ti}{ii} $\lambda$4468.5 is weaker than \ion{He}{i} $\lambda$4471.5. The pitfall of such a simple approach is that it may be biased against the selection of He-weak objects. Stars with peculiar spectra could also have been rejected  because they were poorly fit with our synthetic spectra computed for a scaled-solar chemical composition. It would also contribute to the apparent lack of stars with strongly unusual surface abundances.

\subsection{C and N abundances as proxies of internal mixing}\label{sect_discussion_internal_mixing}

 Rotation triggers the transport of angular momentum and chemicals in stellar interiors. It notably leads to changes in the chemical abundances seen at the surface of massive stars and, in particular, a nitrogen excess accompanied by a lower-amplitude carbon depletion \citep[e.g.][]{daflon01}. Observations by the FS have unveiled two unevolved stellar populations in the Magellanic Clouds that exhibit surface nitrogen abundances not predicted by single-star evolutionary models incorporating rotational mixing \citep{hunter08b,hunter09,brott11}: namely, slow rotators with an unexpected excess of nitrogen and, conversely, fast rotators with no or little nitrogen enrichment at their surface \citep[see also, e.g.][]{rivero_gonzalez12,grin17,dufton20}. The inability of evolutionary models to reproduce these two populations has been questioned \citep{maeder09,maeder14}, but there are clear examples where they fail to reproduce the observations \cite[e.g.][]{keszthelyi21}. The origin of these two populations is a matter of speculation, but might result from the action of magnetic fields \citep[e.g.][]{keszthelyi19} or mass-transfer processes in binaries \citep[e.g.][]{de_mink13,song18,mahy20}. However, although slowly rotating, N-rich B dwarfs have long been known in the field \cite[][]{gies92}, they were not clearly detected in the Galactic clusters observed by the FS, including NGC 3293 \citep[see fig.~6 of][]{hunter09}. 

 As seen in Fig.~\ref{fig_abundances_vs_parameters}, the dramatic nitrogen overabundance in the primary of the post-mass transfer binary \object{$\theta$ Car} is confirmed \citep[][]{hubrig08}. If similar spun-up objects following an accretion event were present in our sample, it is very likely that they would have been detected, but none is found. We only discuss below the [N/C] abundance ratio because it is a more robust indicator of the dredge up of core-processed material at the surface of OB stars. First, because the evolutionary changes affecting C and N are inversely correlated. Second, because the \ion{C}{ii}- and \ion{N}{ii}-based abundances have the same qualitative sensitivity to errors in $T _\mathrm{eff}$, for instance. We adopt as baseline for this ratio our mean value, [N/C] $\sim$ --0.5, found for the two benchmarks $\gamma$ Peg and HD 35912. They have both been shown from high-precision studies to have a [N/C] fully compatible with solar \citep[][]{nieva_simon_diaz11,nieva_przybilla12}. A few stars show some observational evidence for a modest N enhancement exceeding the 3-$\sigma$ level. Two of them (\object{GES\,10355539--5812197} and \object{GES\,10360491--5810433}) with [N/C] $\sim$ --0.1 are about boron normal and therefore very unlikely to have experienced deep mixing \citep{proffitt16}. The N-rich status of some of these candidates is therefore questionable. Nonetheless, we confirm earlier claims \citep[e.g.][]{hunter09} that the cluster lacks a population of strongly N-enriched stars. This conclusion is supported by the fact that an analogue of $\tau$ Sco that is known to show a relatively mild enhancement would be spotted quite easily. As shown in Fig.~\ref{fig_abundances_vs_parameters}, we recover the well-known nitrogen overabundance of this star \citep[e.g.][]{martins12}. 

\begin{figure}[h!]
\centering
\includegraphics[trim=30 265 50 240,clip,width=\hsize]{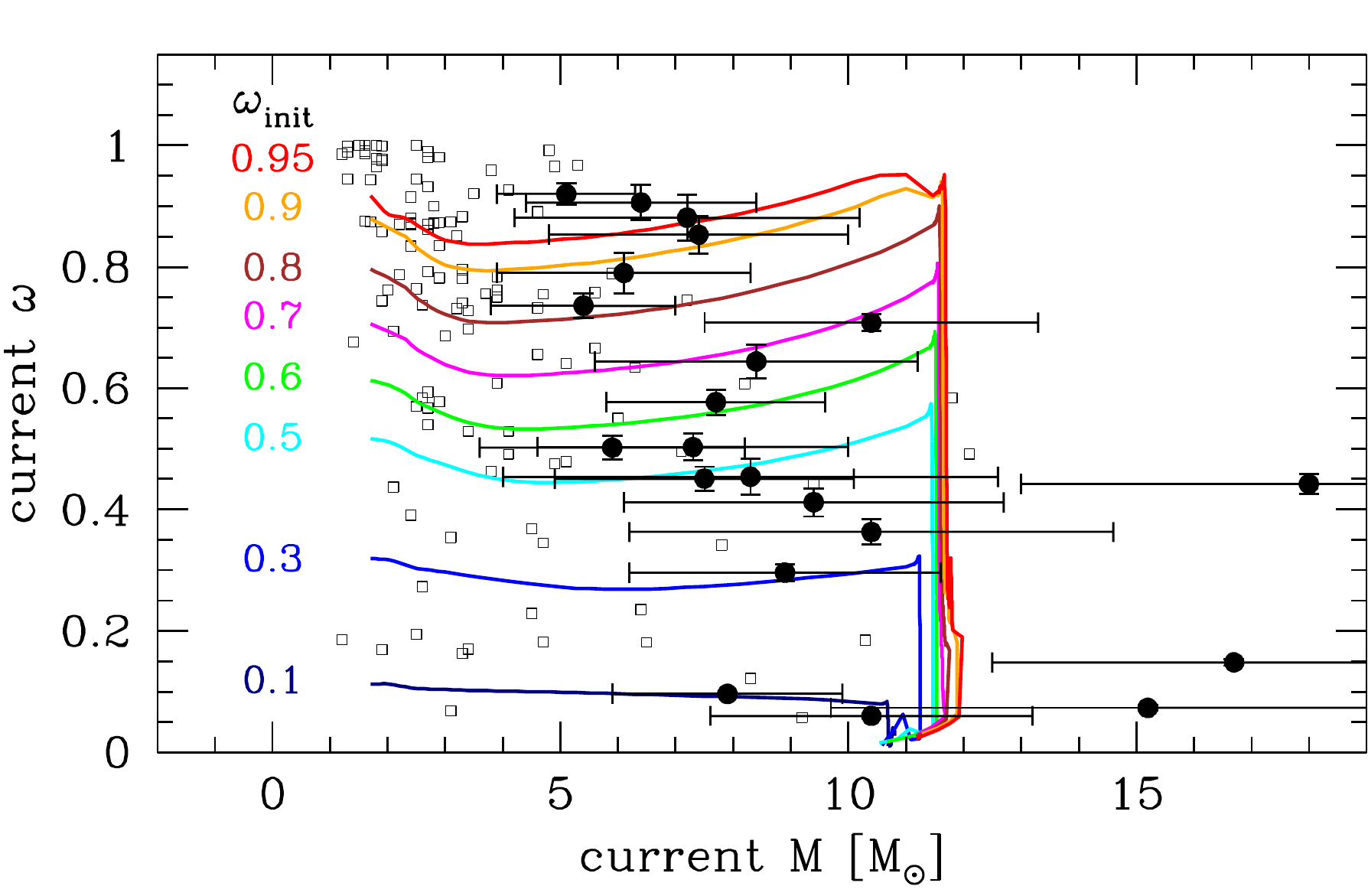}
\caption{Model dependence between the current $\omega$ and mass for $\log$($t$) = 7.3 \citep{georgy13}. The behaviour is shown for rotation rates at birth, $\omega_\mathrm{init}$, ranging from 0.1 to 0.95. The observed $\omega$ and $M$ values for the stars with and without [N/C] data are overplotted as filled circles and open squares, respectively. For clarity, only the former are shown with error bars. The star \object{GES\,10355661--5812407} is off scale because of a spuriously large mass.}
\label{fig_omega_vs_M}
\end{figure}

 We now compare our [N/C] measurements to the expectations from solar-metallicity evolutionary models that incorporate the effects of rotation on the internal stellar structure \citep{georgy13}. The predicted abundance ratios were scaled such that the baseline value on the ZAMS is [N/C]$_\mathrm{ZAMS}$ = --0.5 (see above). As discussed by \citet{ekstroem12}, the default value of the models at the onset of main-sequence evolution is [N/C]$_\mathrm{ZAMS}$ = --0.61 \citep{asplund05}. The stars with [N/C] data constitute a heterogeneous sample in terms of $pnrc$ mass and $\omega$ values. A basic property of models of massive stars is that the amount of core-processed material dredged up to the surface strongly depends on the rotational velocity. To ensure a meaningful comparison, the wide range of $\omega$ values determined in Sect.~\ref{sect_discussion_pnrc} thus requires the use of a set of models matching the current stellar rotation rates. As a result, it is first necessary to associate for each star the current, observed $\omega$ to the appropriate value at birth (see, for instance, \citealt{Ekstrom2008} for the evolution of $\omega$ along the evolution for the Geneva models). As shown in Fig.~\ref{fig_omega_vs_M}, the two quantities are roughly equivalent for the fiducial age of the cluster irrespective of the mass. For simplicity, we therefore assume in the following that the present-day $\omega$ is representative of the initial value on the ZAMS, $\omega_\mathrm{init}$.

\begin{figure*}[h!]
\centering
\includegraphics[trim=20 200 40 220,clip,width=\hsize]{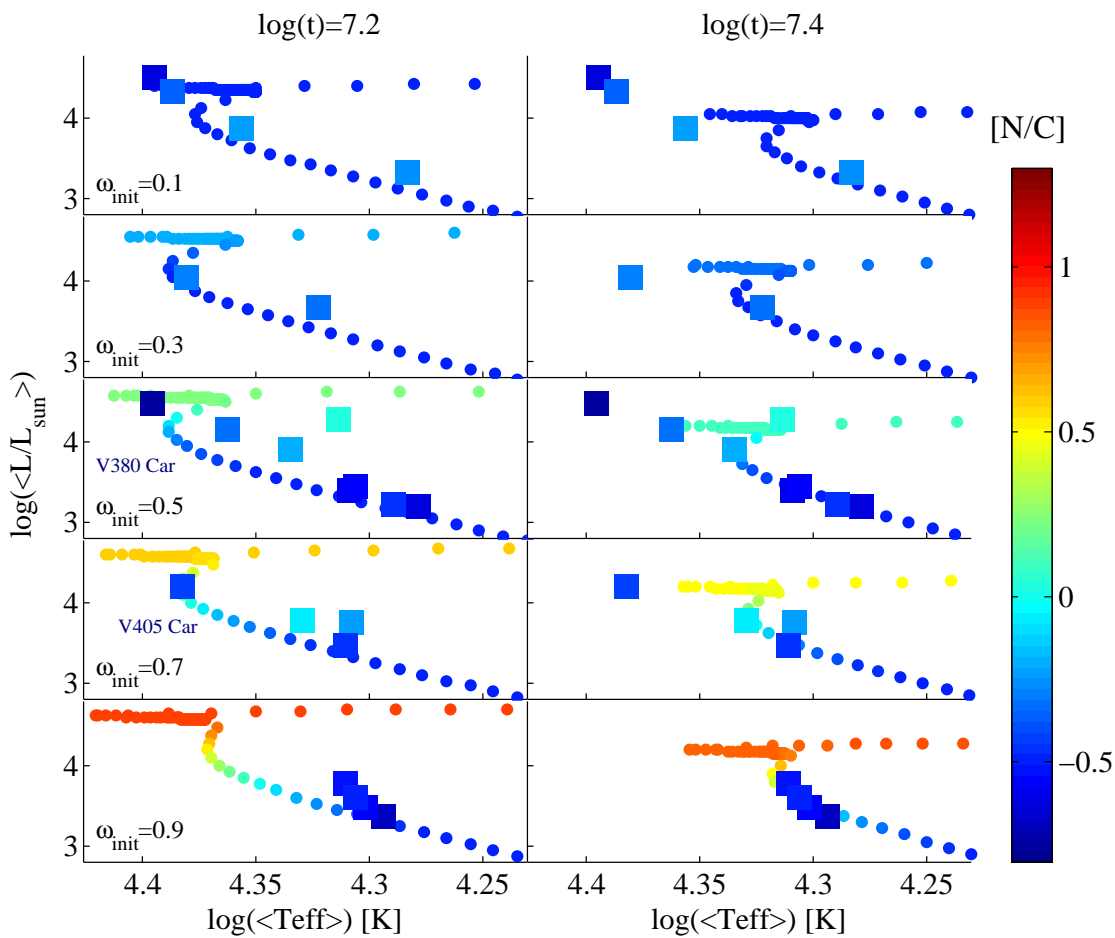}
\caption{Comparison in a HR diagram between the [N/C] data ({\em squares}) and the predictions ({\em circles}) of solar-metallicity evolutionary models for various initial rotation rates and two cluster ages: $\log$($t$) = 7.2 ({\it left panels}) and 7.4 ({\it right panels}). The model predictions are taken from \citet{georgy13}. The stars selected in each panel fulfil the condition $|\omega-\omega_\mathrm{init}|$ $<$ 0.1. The observational error bars are smaller than the square symbols. The [N/C] abundance ratio is colour coded. The stars \object{V380 Car} -- and to a lesser extent \object{V405 Car} -- whose observations are at odds with the model predictions are indicated.}
\label{fig_NC}
\end{figure*}

 The [N/C] data are compared in Fig.~\ref{fig_NC} to the theoretical values for various initial rotation rates and two cluster ages, $\log$($t$) = 7.2 and 7.4, that bracket our mean estimate. Accounting for the uncertainty in the cluster age, overall we do not detect outstanding, systematic discrepancies with respect to the model predictions. However, one exception is \object{GES\,10360160--5815096} (\object{FS\,3293-012} or \object{V380 Car}) for which [N/C] is much lower than expected. Our C abundance is large and quite uncertain, but other studies also reported a low [N/C] ratio \citep{hunter09,niemczura09_ngc3293}. Furthermore, it is not at all boron depleted \citep{proffitt16}. A somewhat less convincing case for a similar behaviour is provided by \object{GES\,10354822--5812329} (\object{FS\,3293-019} or \object{V405 Car}), which is another confirmed $\beta$ Cep star \citep[][]{stankov05}. A low [N/C] is once again confirmed \citep{hunter09}, but it is slightly boron depleted \citep{proffitt16} contrary to \object{V380 Car}. These two stars are indicated in Fig.~\ref{fig_NC}. It thus appears that a consistent picture is not entirely achieved under the assumption of a wide \citep[possibly mass-dependent;][]{huang10} distribution of spin rates at birth. It is in particular telling that the only noteworthy disagreement is found for two fast-rotating, massive stars ($\omega$ $\sim$ 0.44 and $M/M_{\odot}$ $\sim$ 18 for V380 Car, while $\omega$ $\sim$ 0.64 and $M/M_{\odot}$ $\sim$ 8 for V405 Car) that are among the most sensitive probes of rotational mixing in our sample. The others are either slow rotators or not massive enough.

\section{Summary and conclusions}\label{sect_conclusions}

We present a homogeneous analysis of the Galactic open cluster NGC 3293 based on GES and FS VLT-FLAMES observations of about 160 B-type member candidates spanning a wide range of physical properties. To our knowledge, it is the most comprehensive spectroscopic study of this cluster to date.

We infer the present-day intrinsic distribution of the rotational velocities for stars spanning the whole B1--B9.5 domain through a deconvolution algorithm. Our analysis supports the results from the FS \citep[][]{dufton06}\footnote{We note that their estimate of the underlying rotational velocity distribution is also based on data for \object{NGC 4755}, which presumably has similar characteristics.}, and suggests for the full sample a Gaussian-like velocity distribution that peaks around 200--250 km s$^{-1}$. We do not find evidence for a bimodal distribution as claimed, for instance, for the relatively unevolved, early B-type stars in \object{30 Doradus} \citep[][]{dufton13}. However, our sample does not contain enough B0--B3 stars to investigate the shape of the distribution for this particular population. Most stars in NGC 3293 appear to rotate at $\sim$50--60\% of their critical velocity, as is also the case for other samples dominated by largely unevolved, late B-type stars \citep{huang10}. In contrast, however, significantly lower spin rates are observed for more massive members (Fig.~\ref{fig_omega_vs_M}). Similar mass-dependent distributions are observed in \object{h \& $\chi$ Per} \citep{strom05}, which is a solar-metallicity open cluster of about the same age. The distribution for the O stars in the Carina nebula peaks at lower velocities (Berlanas et al., in preparation). Our results are not directly comparable, however, because it is a younger population in which spin-down effects due to stellar winds are much more important. A caveat resulting from the incompleteness of the binary census in NGC 3293 is the unknown importance of physical effects potentially experienced by close binaries, such as tides or mass-transfer processes (see, e.g. \citealt{ramirez_agudelo15,mahy20}). Notwithstanding, a more fundamental difficulty is that post-interaction objects may be truly single (i.e. mergers) or go unnoticed through a standard RV monitoring \citep{de_mink14}.

We obtain a cluster nuclear age of $\sim$20 Myrs based on a realistic distribution of the spin rates and a detailed correction on a star-to-star basis for the effect of stellar rotation. It is larger than the typical value obtained from photometric studies that made use of non-rotating isochrones \citep[$\sim$10--15 Myrs; e.g.][]{baume03,bisht21}: for instance, the Padova-Trieste suite of models \citep{girardi00,marigo17} employed by the two papers cited above. However, there is a case in young clusters from a comparison with independent estimates from the lithium depletion boundary method for an underestimation of ages from isochrone fitting when the models do not take the full range of effects induced by rotation into account \citep[][]{cummings18}: namely, the combination of changes in the evolutionary paths and in the photometric properties because of GD. This raises the issue of a possible upward revision of the age of young open clusters that host a large proportion of fast rotators, as has been invoked in older ones \citep[e.g.][]{brandt15a}. Whether physical processes other than rotation significantly contribute to the spread observed near the turn-off of very young clusters is still debated \citep[e.g.][]{li19}. But, in any case, rotating models (let alone because of their extended core hydrogen-burning lifetime) are an essential ingredient of any attempt to interpret the observational properties of an ensemble of fast-rotating massive stars \citep[e.g.][]{Maeder2000,georgy14}. The extent of the apparent age spread in clusters spanning a very wide range of ages is well reproduced by rotating Geneva models \citep[][]{georgy19}.

By quite a significant margin, NGC 3293 appears to be  the oldest stellar aggregate in the Carina Nebula complex \citep[see][]{preibisch17}. Although not observed by the GES, we note that the position of the red supergiant \object{V361 Car} (M1.5 Iab--Ic) in the HR diagram is also compatible with our inferred age when adopting the spectroscopic parameters of \citet{arentsen19}. Its membership is confirmed by {\em Gaia} EDR3 data and the RV of \citet{feast58}. To the opposite side of the mass spectrum, the status of the O7 V((f))z star \object{HD 91824} has for long been debated \citep[e.g.][]{feinstein80}. Although its {\em Gaia} EDR3 proper motion is indistinguishable from that of the bona fide members, it is likely a foreground object based on its parallax that is discrepant at a level exceeding 5$\sigma$. It is a known SB1 \citep{sota14}, but its {\tt RUWE} does not indicate problems with the astrometric solution. Unless it is a rejuvenated binary product, the existence of an unevolved O Vz star in such a moderate aged cluster is not expected \citep[e.g.][]{arias16}.

Finally, this cluster appears to be to a large extent devoid of objects exposing core-processed material at their surface despite the fact that most of them are fast rotators. We argue it is primarily the consequence of most members being low-mass B dwarfs. Overall, the lack of widespread deviations from the baseline CN abundances (e.g. Fig.~\ref{fig_distribution_abundances}) is in agreement with the theoretical expectations for the cluster age once the rotation rate of the evolutionary model is matched to the observations. However, noteworthy exceptions are two quite rapidly rotating, apparently single sub-giants with little, if any, evidence for internal mixing (even shallow) based on their boron and nitrogen abundances. It suggests that the efficiency of rotational mixing is overestimated in these two objects. We note that our conclusion about the lack of strongly N-enriched stars in NGC 3293 extends to the more massive and evolved stars not observed by the GES. Only the two brightest B-type members classified as supergiants by \citet{evans05}, whose membership is confirmed from {\em Gaia} EDR3, show some evidence for a mild N enrichment at similar levels as found in our sample \citep{hunter09}\footnote{Two caveats can be noted: the FS $T _\mathrm{eff}$ scale has been claimed to be too cool for early B-type stars \citep{proffitt16} and the N abundance might be slightly revised following improvements in the data reduction \citep{dufton18}.}. Concerning the abundance properties in more general terms, macroscopic transport in (non magnetic) stars is known to inhibit the development of diffusion processes (e.g. \citealt{michaud15} and \citealt{niemczura09} for theoretical and observational arguments, respectively). Meridional circulation arising from fast rotation may thus largely account for the low occurrence of chemical peculiarities for helium and the metals not affected by evolutionary effects. From a completely different perspective, it might also explain the dearth of faint, high-order $g$-mode B pulsators, as speculated by \citet{balona94}.

 \begin{acknowledgements}

 We would like to thank the anonymous referee for useful comments that led to a clearer presentation of our results. T.M. acknowledges financial support from Belspo for contracts PRODEX {\em Gaia}-DPAC and PLATO mission development. He is grateful to Keith Butler for making the DETAIL-SURFACE code available to him. E.G. is grateful to Belgian F.R.S.-FNRS for multiple supports. A.B. and T.S. are thankful for grants of Concerted Research Actions (ARC) financed by the Federation Wallonie-Brussels. A.L. acknowledges funding received in part from the European Union Framework Programme for Research and Innovation Horizon 2020 (2014--2020) under the Marie Sklodowska-Curie grant Agreement No. 823734.
M.F.N. acknowledges the support of the Austrian Science Fund (FWF) via a Lise-Meitner Fellowship under project number N-1868-NBL. L.M. thanks the European Space Agency (ESA) and the Belgian Federal Science Policy Office (Belspo) for their support in the framework of the PRODEX Programme. W.S. acknowledges FAPERJ for a Ph.D. fellowship. M.B. is supported through the Lise Meitner grant from the Max Planck Society. She acknowledges support by the Collaborative Research centre SFB 881 (projects A5, A10), Heidelberg University, of the Deutsche Forschungsgemeinschaft (DFG, German Research Foundation).  This project has received funding from the European Research Council (ERC) under the European Union’s Horizon 2020 research and innovation programme (Grant agreement No. 949173).

This paper is based on data products from observations made with ESO Telescopes at the La Silla Paranal Observatory under programme ID 188.B-3002. These data products have been processed by the Cambridge Astronomy Survey Unit (CASU) at the Institute of Astronomy, University of Cambridge, and by the FLAMES/UVES reduction team at INAF/Osservatorio Astrofisico di Arcetri. This work was partly supported by the European Union FP7 programme through ERC grant number 320360 and by the Leverhulme Trust through grant RPG-2012-541. We acknowledge the support from INAF and Ministero dell'Istruzione, dell'Universit\`a e della Ricerca (MIUR) in the form of the grant `Premiale VLT 2012'. The results presented here benefit from discussions held during the {\em Gaia}-ESO workshops and conferences supported by the ESF (European Science Foundation) through the GREAT Research Network Programme.

This work has made use of data from the European Space Agency (ESA) mission {\em Gaia} (\url{https://www.cosmos.esa.int/gaia}), processed by the {\em Gaia} Data Processing and Analysis Consortium (DPAC, \url{https://www.cosmos.esa.int/web/gaia/dpac/consortium}). Funding for the DPAC has been provided by national institutions, in particular the institutions participating in the {\em Gaia} Multilateral Agreement.

This research has made use of NASA's Astrophysics Data System Bibliographic Services, the SIMBAD database operated at CDS, Strasbourg (France), and the WEBDA database, originally developed by Jean-Claude Mermilliod, and now operated at the Department of Theoretical Physics and Astrophysics of the Masaryk University.
  
\end{acknowledgements}

\bibliographystyle{aa} 
\bibliography{aa_2022_44112} 

\begin{appendix}

\section{Comparison with GES homogenised results}\label{appendix_comparison_homogeneised_results}

A total of 584 stars in the field of \object{NGC 3293} have homogenised stellar parameters. They are only based on WG13 data products and are released to the community as part of the last GES public data release.

\begin{figure}[h!]
\centering
\includegraphics[trim=45 530 190 175,clip,width=\hsize]{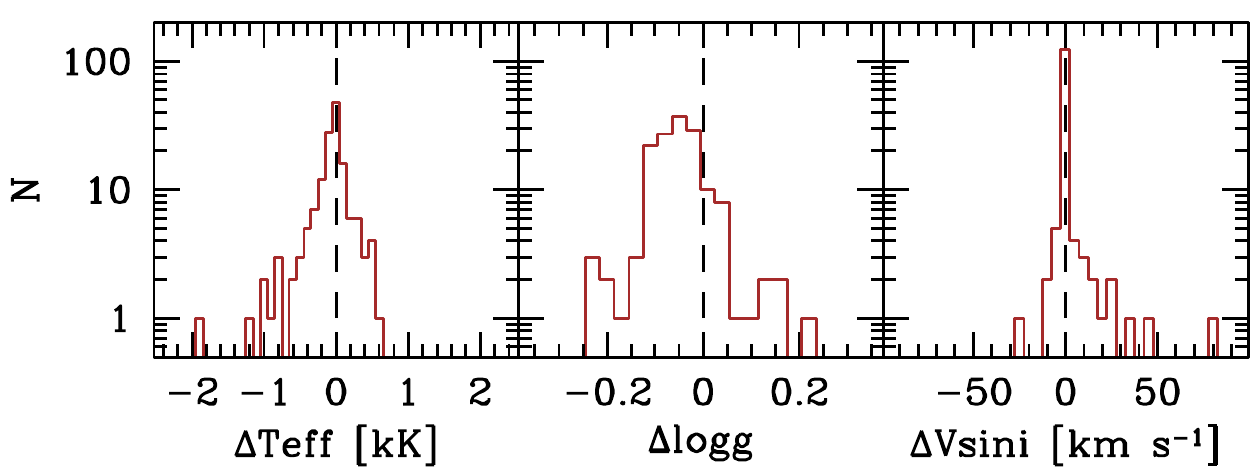}
\caption{Comparison between our stellar parameters and those recommended by the GES. The differences are expressed as this study minus recommended. All the stars are shown irrespective of their membership status (Sect.~\ref{sect_membership_gaia}).}
\label{fig_WG15}
\end{figure}

The Li\`{e}ge node provided all the WG13 abundances for this cluster and the GES abundance scale for massive stars is not anchored to that determined by other WGs for the FGK stars. The recommended abundances for NGC 3293 are therefore fully based on our results. Slight differences between the two datasets only arise because of different choices for rounding off or averaging multiple measurements of the same star (the GES procedures for homogenisation are described in Hourihane et al., in preparation). The comparison shown in Fig.~\ref{fig_WG15} is therefore restricted to the stellar parameters. The discrepancies are due to differences with respect to the other WG13 nodes, especially ROBGrid because it was the other main provider of data for this cluster (Sect.~\ref{sect_external_validation_other_nodes}).

\section{Impact of LPVs}\label{appendix_impact_pulsations}

Several observational studies have demonstrated that B stars commonly exhibit LPVs that arise from non-radial pulsations \citep[e.g.][]{telting06}. Although some stars show a hybrid character, two broad classes can be defined: slowly pulsating B stars (SPBs) and $\beta$ Cephei-like variables with mid-late and early spectral types, respectively. As indicated in Table \ref{tab_variability_analysis}, some bright stars in our sample intensively monitored with UVES are confirmed $\beta$ Cep stars \citep[][]{stankov05} and indeed show clear evidence for LPVs. Although such variations are also likely to take place in others, they may escape detection because of a poor S/N or a lack of repeated observations with an adequate time sampling. The cluster is rich in $\beta$ Cep stars, although SPBs are apparently rarer \citep{balona94}. However, a handful of candidates discovered by \citet{handler08} through a ground-based photometric survey are listed in Table \ref{tab_variability_analysis}.

\begin{figure}[h!]
\centering
\includegraphics[trim=20 220 220 200,clip,width=0.9\hsize]{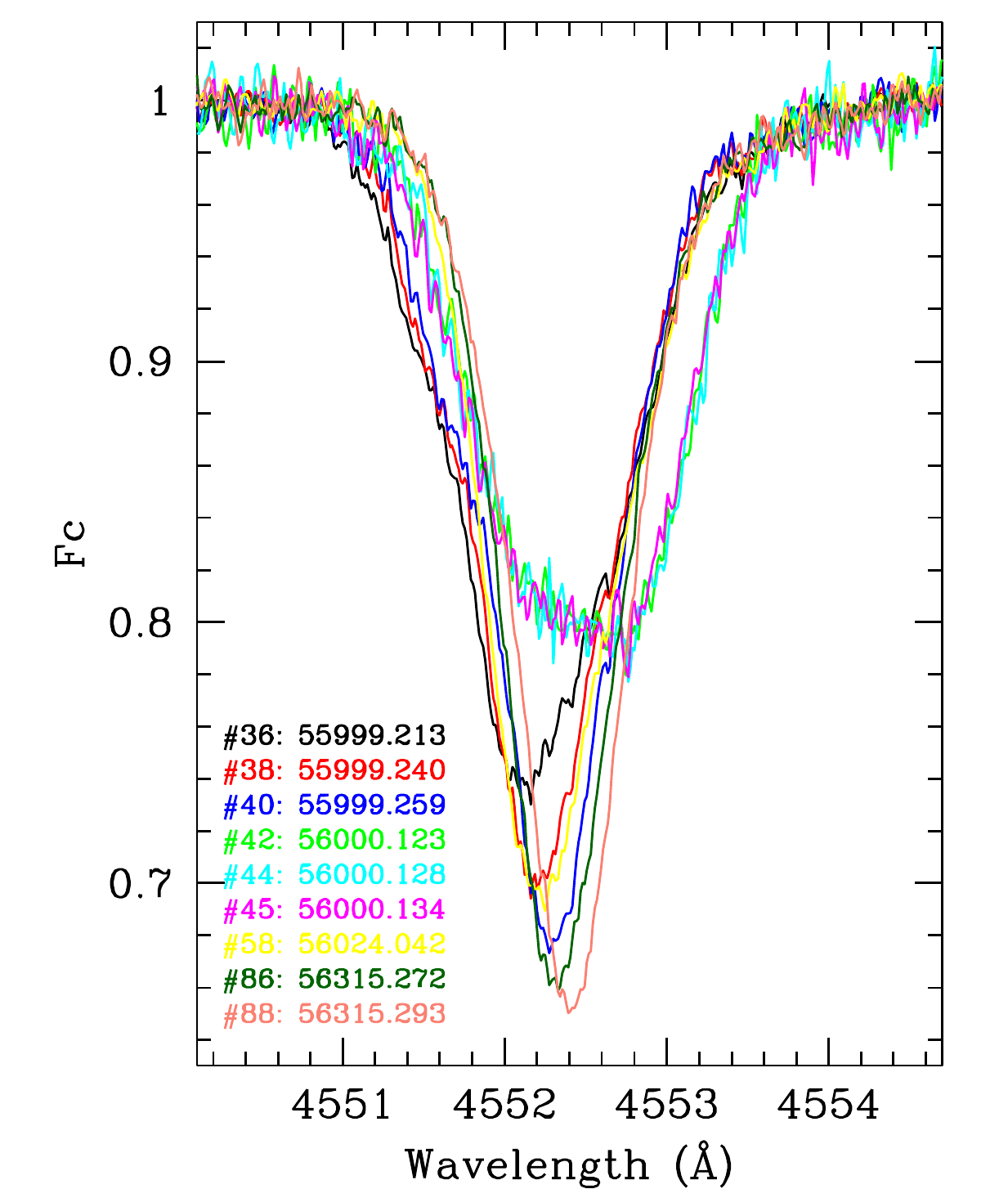}
\caption{Superposition of the GES UVES exposures of GES\,10354072--5812440 for the spectral range encompassing the \ion{Si}{iii} $\lambda$4553 line profile. The exposure ID and MJD are indicated.}
\label{fig_LPVs_V403_Car}
\end{figure}

To roughly quantify the impact of LPVs on the derived atmospheric parameters and abundances for the B-type pulsators, we analysed three representative exposures of the well-known $\beta$ Cep star \object{GES\,10354072--5812440} (\object{V403 Car}; see, e.g. \citealt{engelbrech86}). As shown in Fig.~\ref{fig_LPVs_V403_Car} where all the exposures are overlaid, it is quite an extreme example in the sense that it displays conspicuous LPVs, even though they are accompanied by modest EW changes ($\lesssim$ 10\%). The results for the various exposures are given in Table \ref{tab_results_V403_Car}.

Our analysis shows that some derived parameters ($T_\mathrm{eff}$, $\log g$) and all the abundances remain within the (admittedly quite large) uncertainties irrespective of the phase of observation along the pulsation cycle. This conclusion is in line with previous studies of this kind \citep[e.g.][and references therein]{morel06}. However, as anticipated because of the large changes in the line width and skewness, significant differences are obtained for the radial and rotational velocities. These results are based on iDR3 data and analysis procedures \citep{semaan15}, but can be regarded as being representative.

\begin{table*}[h!]
\centering
\caption{Results of the spectroscopic analysis based on iDR3 data for three representative exposures of \object{GES\,10354072--5812440}.}
\vspace*{0.3cm}
\begin{tabular}{cccccc} \hline\hline
Exposure & $T_\mathrm{eff}$ & $\log g$ & $\xi$        & $V\sin i$    & RV \\ 
         & [K]         & [cgs]    & [km s$^{-1}$] & [km s$^{-1}$] & [km s$^{-1}$] \\ 
\hline 
\#36 & 23\,500$\pm$1000 & 3.43$\pm$0.05 & 10 & 50$\pm$3 & --23$\pm$2\\  
\#40 & 23\,500$\pm$1000 & 3.46$\pm$0.06 & 10 & 40$\pm$2 & --20$\pm$2\\  
\#45 & 23\,500$\pm$1000 & 3.43$\pm$0.05 & 10 & 64$\pm$5 & --11$\pm$2\\  
\hline
\end{tabular}
\label{tab_results_V403_Car}
\end{table*}

\begin{table*}[h!]
\centering
\vspace*{0.3cm}
\begin{tabular}{ccccccc} \hline\hline
Exposure & $y$ & $\log \epsilon$(\ion{C}{ii}) & $\log \epsilon$(\ion{N}{ii}) & $\log \epsilon$(\ion{Mg}{ii}) & $\log \epsilon$(\ion{Si}{iii}) & [N/C]\\ 
\hline 
\#36 & 0.130$\pm$0.015 & 8.16$\pm$0.23 & 7.44$\pm$0.20 & 7.25$\pm$0.20 & 7.52$\pm$0.30 & --0.72$\pm$0.29\\  
\#40 & 0.125$\pm$0.015 & 8.25$\pm$0.23 & 7.47$\pm$0.20 & 7.25$\pm$0.20 & 7.52$\pm$0.30 & --0.78$\pm$0.29\\  
\#45 & 0.120$\pm$0.015 & 8.17$\pm$0.23 & 7.42$\pm$0.20 & 7.23$\pm$0.20 & 7.53$\pm$0.30 & --0.75$\pm$0.29\\  
\hline
\end{tabular}
\end{table*}

\FloatBarrier

\section{Example of spectral fits}\label{appendix_spectral_fits}

\begin{figure*}[t!]
\begin{minipage}[t]{0.5\textwidth}
\centering
\includegraphics[trim=0 0 0 0,clip,width=1.1\textwidth]{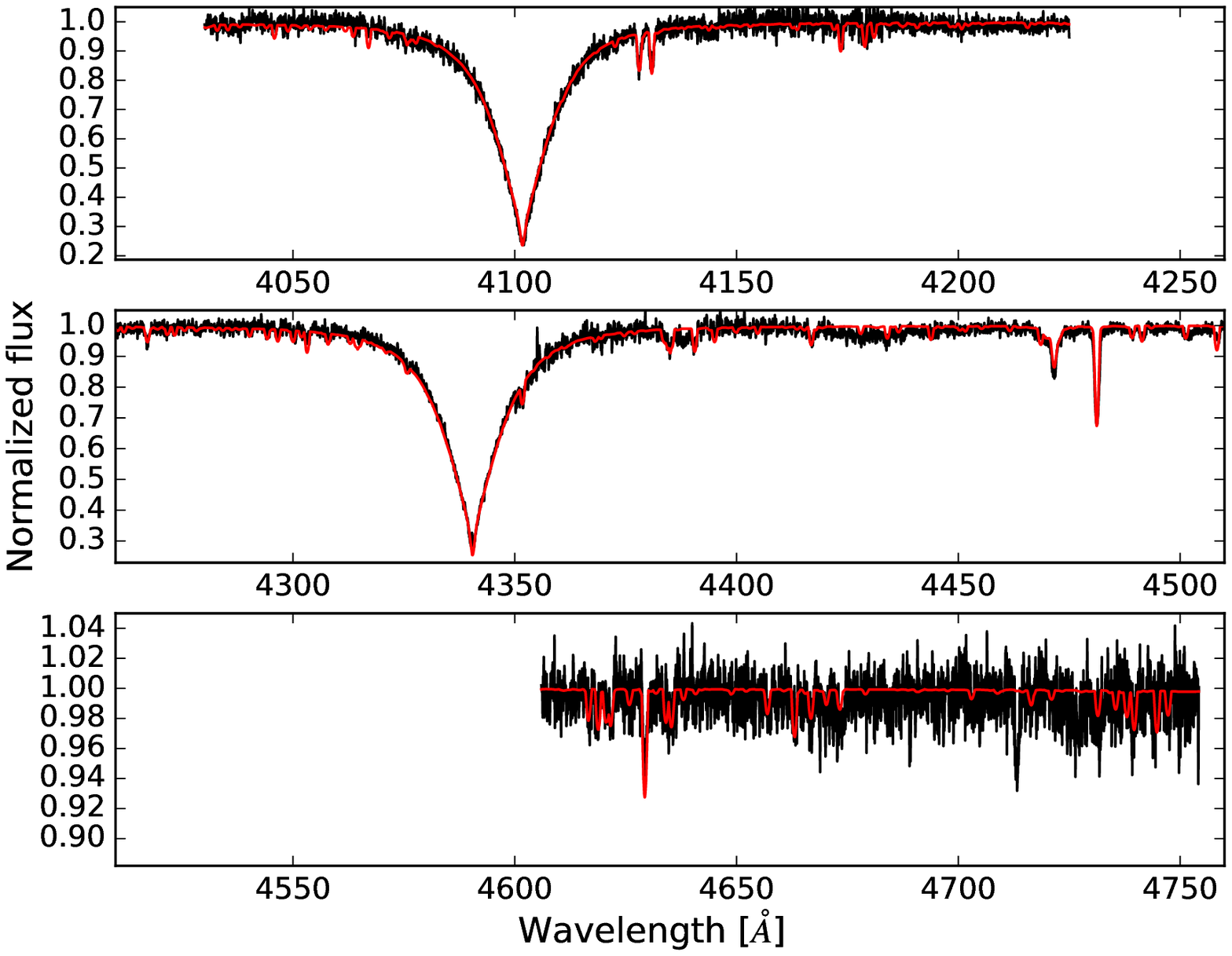}
\end{minipage}
\begin{minipage}[t]{0.5\textwidth}
\centering
\includegraphics[trim=0 0 0 0,clip,width=1.1\textwidth]{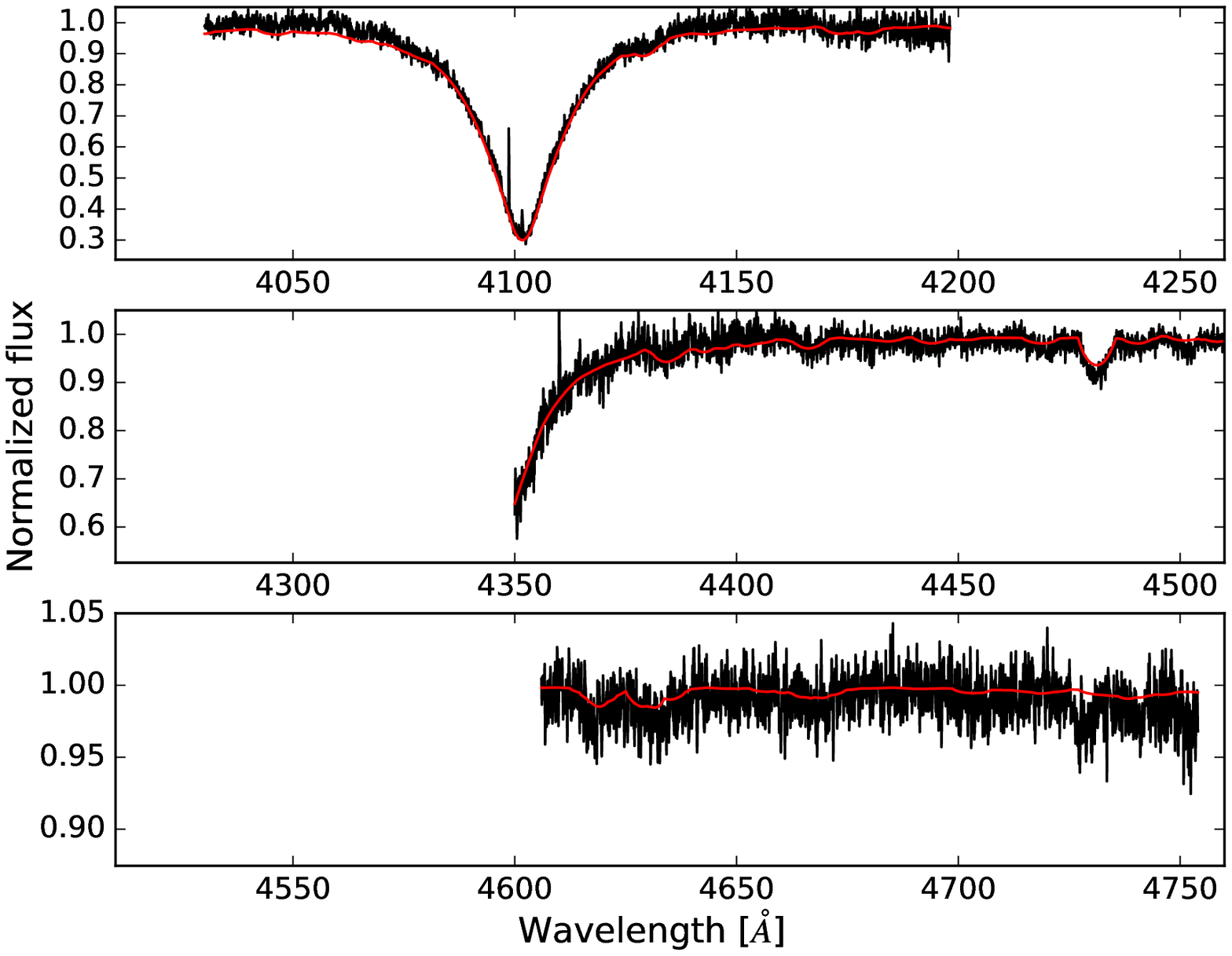}
\end{minipage}
\begin{minipage}[t]{0.5\textwidth}
\centering
\includegraphics[trim=0 0 0 0,clip,width=1.1\textwidth]{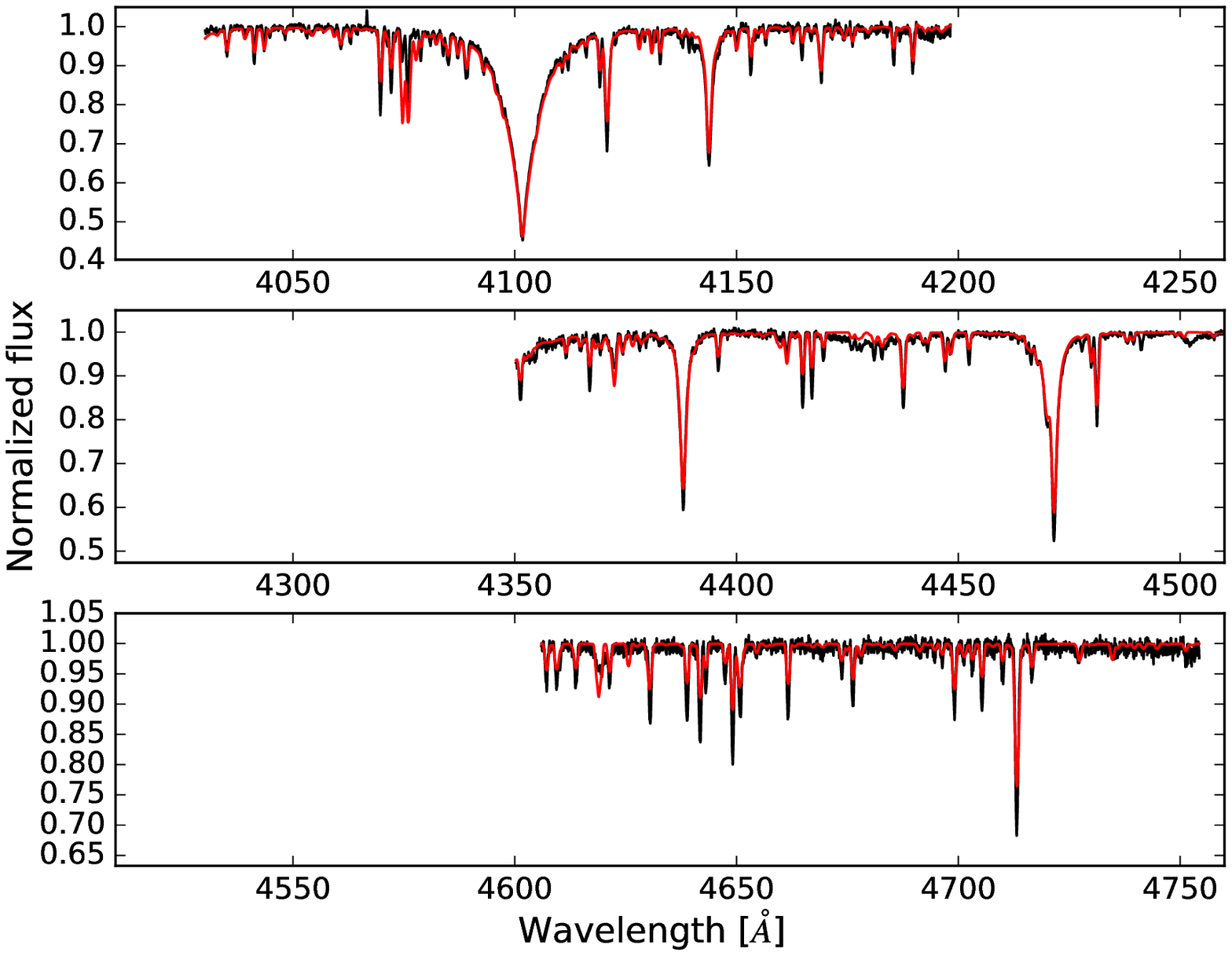}
\end{minipage}
\begin{minipage}[t]{0.5\textwidth}
\centering
\includegraphics[trim=0 0 0 0,clip,width=1.1\textwidth]{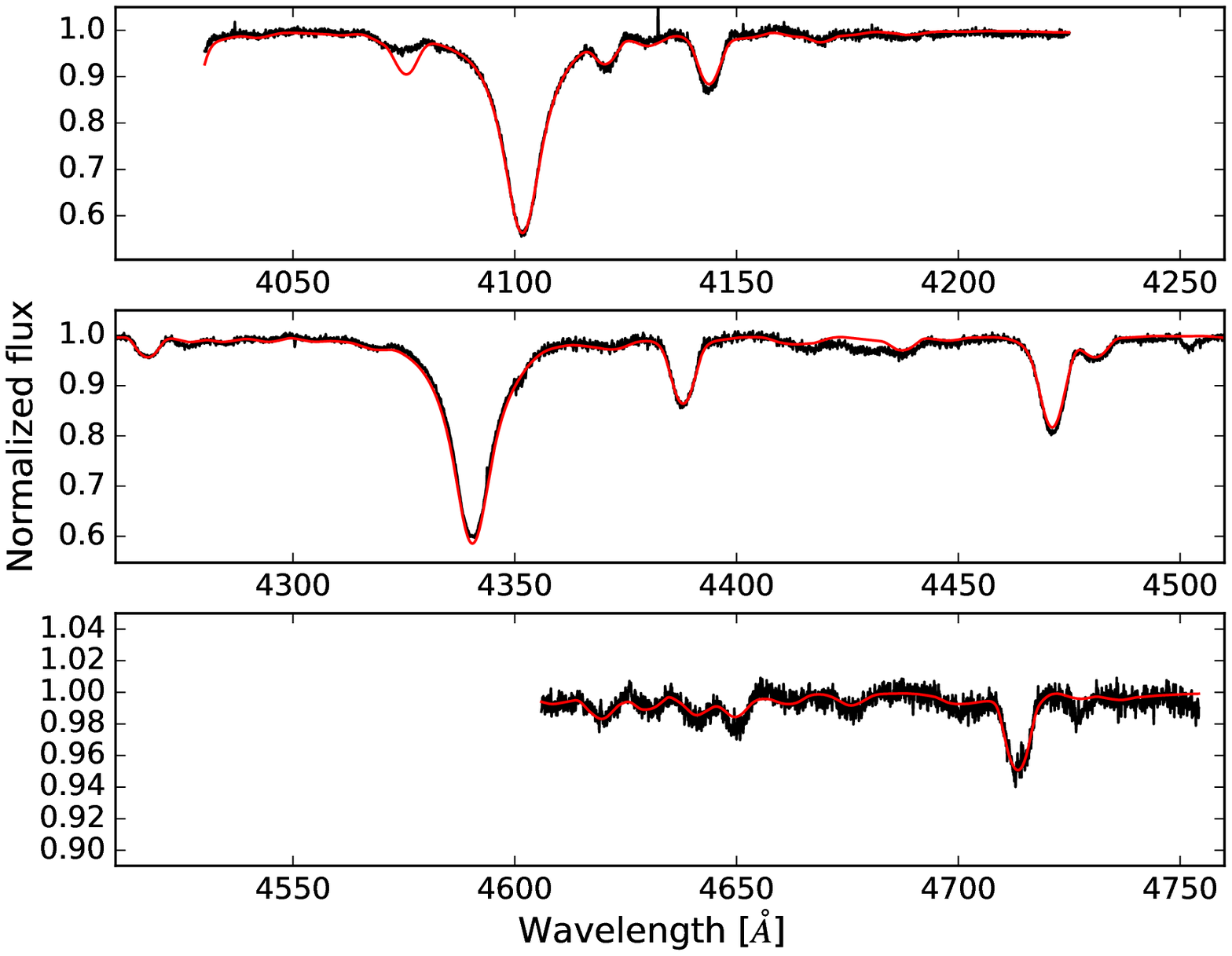}
\end{minipage}
\caption{Example of spectral fits for the cool, slow rotator \object{GES\,10353450--5813461} with $T_\mathrm{eff}$ = 11\,760 K and $V\sin i$ = 42 km s$^{-1}$ ({\em top left panel}), the cool, fast rotator \object{GES\,10353820--5811092} with $T_\mathrm{eff}$ = 10\,800 K and $V\sin i$ = 276 km s$^{-1}$ ({\em top right panel}), the hot, slow rotator \object{GES\,10355660--5811314} with $T_\mathrm{eff}$ = 22\,701 K and $V\sin i$ = 27 km s$^{-1}$ ({\em bottom left panel}), and the hot, fast rotator \object{GES\,10360349--5814401} with $T_\mathrm{eff}$ = 20\,322 K and $V\sin i$ = 257 km s$^{-1}$ ({\em bottom right panel}). The observations are shown in black and the fits in red. Two stars do not have HR04 data available.}
\label{fig_spectral_fits}
\end{figure*}

Figure \ref{fig_spectral_fits} presents illustrative fits to the GES GIRAFFE data obtained as part of the determination of the atmospheric parameters (Sect.~\ref{sect_parameters}).

\FloatBarrier

\section{Results of variability analysis}\label{appendix_variability_analysis}

Table \ref{tab_variability_analysis} provides full details for the stars discussed in Sect.~\ref{sect_binarity}.

\begin{sidewaystable*}
  \centering
  \tiny{
    \caption{Details of the variability analysis.}
\begin{tabular}{l|c|ccccccc|c|l}
\hline\hline 
                            &                & \multicolumn{5}{c}{GIRAFFE only}          & UVES                  & UVES +                   &  & \\
GES ID                                  & Eye inspection & HR03-HR04 & HR03-HR05 & HR03 & HR04& HR05 & only\tablefootmark{a} & GIRAFFE\tablefootmark{a} &  Remarks\tablefootmark{b}                        & Flags\\
\hline 
\object{GES\,10342859--5807396} & Noisy & ... & ...  & ... & ... & ... & ... & ... & ... & 10050-13-16-00-A\\                                    
\object{GES\,10343386--5812282} & ...   & ... & ...  & ... & ... & ... & ... & ... & ... & 10302-13-16-01-A \\                                  
\object{GES\,10344856--5804574} & SB2?   & ... & ...  & ... & ... & ... & ... & ... & ... & 10302-13-16-01-A, 20020-13-16-00-B \\                       
\object{GES\,10344868--5809012} & ...   & ... & 1    & ... & ... & 1   & ... & ... & SH05 & 21100-13-16-00-C \\
\object{GES\,10344903--5810069} & ...   & 1   & 1    & ... & ... & 1   & ... & ... & ... & 20010-13-16-00-A, 21100-13-16-00-A\\
\object{GES\,10344954--5825058} & ...   & ... & 1    & ... & ... & 1   & ... & ... & ... & 21100-13-16-00-C \\
\object{GES\,10345274--5823545} & Noisy & ... & ...  & ... & ... & ... & ... & ... & ... & 10050-13-16-00-A\\
\object{GES\,10350957--5807477} & ...  & 1 & ...  & ...  & ...  & ...  & ...  & ...  & ...  & ...\\
\object{GES\,10351611--5812597} & SB2?? & ...  & ...  & ...  & ...  & ...  & ...  & ...  & ...  & 10303-13-16-01-A\\
\object{GES\,10352356--5821069} & SB2?? & ...  & ...  & ...  & ...  & ...  & ...  & ...  & ...  & 20020-13-16-00-C, 21100-13-16-00-C \\
\object{GES\,10352795--5805382} & ...  & ...  & 1 & ...  & ...  & ...  & ...  & ...  & ...  & 21100-13-16-00-C \\
\object{GES\,10352847--5825069} & Noisy & ...  & ...  & ...  & ...  & ...  & ...  & ...  & ...  & 10050-13-16-00-A\\
\object{GES\,10352851--5812496} & ...  & 2 & 3 & 1 & 1 & 2 & ...  & VAR A, SB1A & ...  & 20010-13-16-00-A \\
\object{GES\,10353007--5812080} & ...  & 2 & 1 & 1 & ...  & 2 & ...  & VAR A, LPV A & SH05  & 21100-13-16-00-A \\
\object{GES\,10353230--5815220} & SB2?? & ...  & ...  & ...  & ...   & ...   & ...   & ...   & Be   & 10302-13-16-01-A, 25000-13-16-01-A \\   
\object{GES\,10353549--5807090} & ...  & ...  & ...  & ...  & ...  & ...  & ...  & ...  & ...  & 21100-13-16-00-C\\
\object{GES\,10353568--5813564} & ...  & ...  & ...  & ...  & ...   & ...   & ...   & ...  & FS only  &  ...\\
\object{GES\,10353645--5808103} & Noisy, SB2?? & ...  & ...  & ...  & ...  & ...  & ...  & ...   & ...  & 10050-13-16-00-A, 10302-13-16-01-A \\
\object{GES\,10353883--5804260} & ...  & ...  & ...  & ...  & ...  & ...  & ...  & ...  & ...  & 10302-13-16-01-A\\
\object{GES\,10353997--5813569} & SB2 & ...  & ...  & ...  & ...  & ...  & ...  & VAR A  & H08  & 20020-13-16-00-A \\
\object{GES\,10354072--5812440} & ...  & ...  & ...  & ...  & ...  & ...  & VAR A, LPV A & ...  & SH05  & 21100-13-16-00-A \\
\object{GES\,10354132--5815392} & Noisy  & ...  & ...  & ...  & ...  & ...  & ...  & ...   & ...  & 10302-13-16-01-A \\
\object{GES\,10354331--5813334} & SB2? & ...  & ...  & ...  & ...  & ...  & ...  & VAR A, LPV A & SH05  & 20020-13-16-00-B, 21100-13-16-00-A \\
\object{GES\,10354393--5821573} & SB1!, SB2? & ...  & ...  & ...  & ...  & ...  & ...  & ...  & ...  & 20010-13-16-00-A\tablefootmark{c}, 20020-13-16-00-B\\
\object{GES\,10354724--5810167} & ...  & ...  & 1 & ...  & ...  & ...  & ...  & ...  & ...  & ... \\
\object{GES\,10354752--5812471} & Noisy & ...  & ...  & ...  & ...  & ...  & ...  & ...  & ...  & ... \\
\object{GES\,10354901--5814541} & SB2?? & ...  & ...  & ...  & ...  & ...  & ...  & VAR B & P16, H17 & 21100-13-16-00-B\\
\object{GES\,10354954--5815131} & Noisy & ...  & ...  & ...  & ...  & ...  & ...  & ... & ...  & 10050-13-16-00-A\\
\object{GES\,10355282--5813117} & SB2! & ...  & ...  & ...  & ...  & ...  & ...  & ...  & ...  & 20020-13-16-00-A \\
\object{GES\,10355301--5812168} & SB1? & 1 & ...  & ...  & 1 & 1 & ...  & VAR C long-term & ...  & 20010-13-16-00-B, 21100-13-16-00-C \\
\object{GES\,10355349--5810428} & ...  & ...  & 1 & ...  & ...   & ...   & ...   & ...   & ...   & ...\\
\object{GES\,10355363--5814478} & ...  & ...  & ...  & ...  & ...   & ...   & VAR A, LPV A & ... & SH05  & 21100-13-16-00-A \\
\object{GES\,10355376--5813033} & ...  & 1 & 1 & ...  & ...   & ...   & ...   & ...   & ...  & 21100-13-16-00-A \\
\object{GES\,10355422--5815267} & SB2 (UVES)? & ...  & ...  & ...  & ...  & ... & SB2 B  & ...  & Single UVES spectrum & 20020-13-16-00-B \\
\object{GES\,10355440--5812563} & ...  & ...  & ...  & ...  & ...  & ...  & ...  & ...  & ...  & 10303-13-16-01-A \\
\hline
\end{tabular}}
\end{sidewaystable*}

\addtocounter{table}{-1}
\begin{sidewaystable*}
  \centering
  \tiny{
    \caption{continued.}
\begin{tabular}{l|c|ccccccc|c|l}
\hline\hline 
                            &                & \multicolumn{5}{c}{GIRAFFE only}          & UVES                  & UVES +                   &  & \\
GES ID                                  & Eye inspection & HR03-HR04 & HR03-HR05 & HR03 & HR04& HR05 & only\tablefootmark{a} & GIRAFFE\tablefootmark{a} &  Remarks\tablefootmark{b}                        & Flags\\
\hline 
\object{GES\,10355469--5812371} & ...  & ...  & ...  & ...  & ...  & ...  & ...  & ...  & ...  & 10302-13-16-01-A \\
\object{GES\,10355491--5812591} & ...  & ...  & ...  & ...  & ...  & ...  & VAR A, LPV A  & ...  & SH05  & 21100-13-16-00-A \\
\object{GES\,10355660--5811314} & ...  & 1 & 2 & 1 & ...  & 2 & ...  & VAR A, weak & ...  & 20010-13-16-00-A, 21100-13-16-00-A\\
\object{GES\,10355698--5817447} & Noisy  & ...  & ...  & ...  & ...  & ...  & ...  & ...  & ...  & ... \\
\object{GES\,10355711--5815218} & SB1!, SB2?? & ...  & ...  & ...  & ...  & ...  & ...  & ...  & ...  & 10303-13-16-01-A, 20010-13-16-00-A\tablefootmark{c},\\
                                & ...         & ...  & ...  & ...  & ...  & ...  & ...  & ...  & ...  & 20020-13-16-00-C   \\ 
\object{GES\,10355781--5812213} & ...  & ...  & ...  & ...  & ...  & ...  & VAR A, SB1 A & ...  & SH05  & 20010-13-16-00-A, 21100-13-16-00-A\\
\object{GES\,10355849--5814148} & ...   & ...  & ...  & ...  & ...  & ...  & ... & ...  & Be, single UVES spectrum       & 25000-13-16-01-A \\
\object{GES\,10360019--5813303} & ...  & 1 & ...  & ...  & ...  & ...  & ...  & ...  & ...  & 21100-13-16-00-C \\
\object{GES\,10360116--5812128} & ...  & 1 & 1 & ...  & ...  & ...  & ...  & ...  & ...  & 21100-13-16-00-A \\
\object{GES\,10360160--5815096} & ...  & ...  & ...  & ...  & ...  & ...  & ...  & VAR A, LPV A & SH05  & 21100-13-16-00-A \\
\object{GES\,10360171--5807557} & SB2! & ...  & ...  & ...  & ...  & ...  & ...  & ...  & ...  & 20020-13-16-00-A \\
\object{GES\,10360262--5813199} & Noisy & ...  & ...  & ...  & ...  & ...  & ...  & ...  & ...  & 10050-13-16-00-A\\
\object{GES\,10360382--5811196} & Noisy & ...  & ...  & ...  & ...  & ...  & ...  & ...  & ...  & 10050-13-16-00-A\\
\object{GES\,10360491--5810433} & ...  & ...  & ...  & ...  & ...  & ...  & ...  & VAR A, LPV B & ...  & 21100-13-16-00-B \\
\object{GES\,10360523--5813221} & Noisy & ...  & ...  & ...  & ...  & ...  & ...  & ...  & ...  & 10050-13-16-00-A\\
\object{GES\,10360525--5816455} & SB2?? & 1 & 1 & ...  & ...  & ...  & ...  & ...  & ...  & 20020-13-16-00-C, 21100-13-16-00-C\\
\object{GES\,10360528--5820598} & Noisy        & ...  & ...  & ...  & ...  & ...  & ...  & ...  & ...  & ... \\
\object{GES\,10360595--5814270} & LPV? & ...  & ...  & ...  & ...  & ...  & ...  & VAR A, LPV B & Be, H08  & 21100-13-16-00-B, 25000-13-16-01-A \\
\object{GES\,10360657--5817538} & ...        & ...  & ...  & ...  & ...  & ...  & ...  & ...  & ...  & 10302-13-16-01-A \\
\object{GES\,10360764--5815204} & ...        & 1 & ...  & ...  & ...  & ...  & ...  & ...  & ...  & 21100-13-16-00-C \\
\object{GES\,10360834--5813041} & ...        & ...  & ...  & ...  & ...  & ...  & ...  & ...  & SH05, EB  & 10106-13-16-01-A \\
                                & ...        & ...  & ...  & ...  & ...  & ...  & ...  & ...  & all UVES exposures & ... \\
                                & ...        & ...  & ...  & ...  & ...  & ...  & ...  & ...  & with picket fence  & ... \\
\object{GES\,10360976--5810579} & ...        & 1 & 2 & 1 & ...  & 2 & ...  & ...  & ...  & 20010-13-16-00-A, 21100-13-16-00-A\\
\object{GES\,10360986--5805441} & ...        & 2 & ...  & 1 & ...  & ...  & ...  & ...  & ...  & 21100-13-16-00-B \\
\object{GES\,10361290--5813250} & Noisy        & ...  & ...  & ...  & ...  & ...  & ...  & ...  & ...  & ... \\
\object{GES\,10361339--5816514} & Noisy, SB2?? & ...  & ...  & ...  & ...  & ...  & ...  & ...  & ...  & 10303-13-16-01-A\\ 
\object{GES\,10361348--5811207} & SB2?? & ...  & ...  & ...  & ...  & ...  & ...  & ...  & ...  & ...\\
\object{GES\,10361370--5817327} & ...   & ...   & ...  & ...  & ...  & ...  & ...  & ...  & Be  & 25000-13-16-01-A\\
\object{GES\,10361385--5819052} & SB2!  & ...  & ...  & ...  & ...  & ...  & ...  & ...  & M17 & 20020-13-16-00-A \\
\object{GES\,10361503--5808043} & ...        & 2 & 1 & 1 & 1 & ...  & ...  & ...  & ...  & 21100-13-16-00-A \\
\object{GES\,10361562--5818519} & ...        & ...  & 1 & ...  & ...  & 1 & ...  & ...  & ...  & 21100-13-16-00-B \\
\object{GES\,10361733--5809031} & ...        & ...  & 2 & ...  & ...  & ...  & ...  & ...  & ...  & 21100-13-16-00-B \\
\object{GES\,10361791--5814296} & SB1! SB2? & 2 & 2 & 1 & 1 & 2 & ...  & ...  & M17 & 20010-13-16-00-A, 20020-13-16-00-C,\\
                                & ...          &...  &...  & ... & ... & ... & ...  & ...  & ... & 21100-13-16-00-C \\
\object{GES\,10362586--5814362} & SB2?? & ...  & ...  & ...  & ...  & ...  & ...  & ...  & ...  & 10303-13-16-01-A\\
\object{GES\,10363025--5822144} & Noisy & ...  & ...  & ...  & ...  & ...  & ...  & ...  & ...  & 10050-13-16-00-A \\
\object{GES\,10363044--5819516} & Noisy & ...  & ...  & ...  & ...  & ...  & ...  & ...  & ...  & 10050-13-16-00-A \\
\object{GES\,10363792--5824198} & ...  & ...  & 1 & ...  & ...  & ...  & ...  & ...  & ...  & 21100-13-16-00-C \\
\object{GES\,10364205--5819028} & ...  & ...  & 1 & ...  & ...  & 1 & ... & ...  & ...  & 21100-13-16-00-B \\
\object{GES\,10365274--5809555} & SB1??  & ...  & ...  & ...  & ...  & ...  & ...  & ...  & ...  & 10302-13-16-01-A \\
\hline
\end{tabular}
\label{tab_variability_analysis} 
\tablefoot{
      The spectral morphology as judged from eye inspection is given in the second column (a blank indicates that no clear peculiarities were detected). Columns 3 to 7 indicate the number of pairs of GIRAFFE spectra with discrepant RVs. Columns 8 and 9 provide the binarity and variability information for stars with UVES or UVES+GIRAFFE spectra. The status from selected studies (an exhaustive review of the literature was not attempted) and general comments are given in Column 10. Finally, our flags are given in the last column (see nomenclature in Table \ref{tab_flags}).
  \tablefoottext{a}{VAR: exhibits a significant RV variation, but without any conclusion about the origin. The star is flagged as SB1 when there is evidence for a global, coherent motion of the lines without marked changes in their profile, which is strongly reminiscent of an orbital motion; LPV: exhibits changes in the line shape, most probably due to pulsations, but other possibilities (e.g. rotational modulation of a spotted photosphere) cannot be ruled out. In each case, a confidence level is assigned (see Table \ref{tab_flags}).}
  \tablefoottext{b}{M17: identified as SB2 based on GES iDR4 data by \citet{merle17}; H08: SPB candidate according to \citet{handler08}; SH05: confirmed $\beta$ Cep star \citep{stankov05}; EB: $\beta$ Cep variable in an eclipsing binary \citep[][]{engelbrecht86}; P16: strongly magnetic, He-rich star \citep[][]{przybilla16}; H17: lack of SB2 signature, but rotational modulation because of a spotted surface \citep[][]{hubrig17}.}
\tablefoottext{c}{GIRAFFE RVs inadvertently not measured, but SB1 status obvious.}}}
\end{sidewaystable*}

\FloatBarrier

\section{Validation of results}\label{sect_validation}

\subsection{Internal validation}\label{sect_internal_validation}

\begin{figure}[h!]
\centering
\includegraphics[trim=60 285 135 130,clip,width=\hsize]{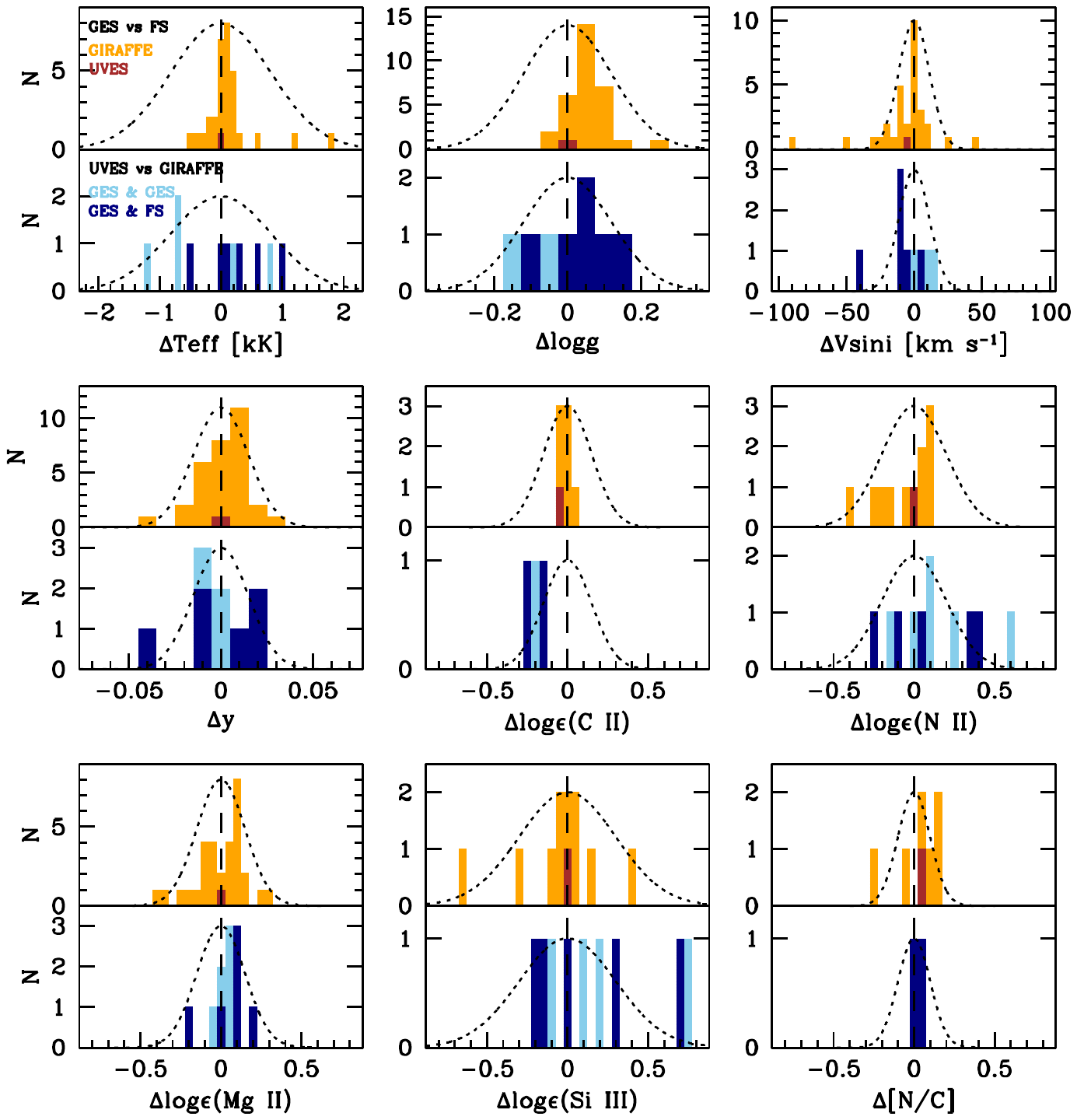}
\caption{Comparison of results from repeated observations. {\em Top panel of each sub-figure}: differences between the results obtained using GES or FS data based on either GIRAFFE or UVES. {\em Bottom panel of each sub-figure}: differences between the results obtained using either UVES or GIRAFFE data. Two cases are considered: only GES data or observations taken from a different source (FS or GES). All the differences are expressed as GES minus FS and UVES minus GIRAFFE. The dotted curve is a normal distribution with a variance set to the nominal random uncertainty.}
\label{fig_histogram_multiple_results}
\end{figure}

Figure \ref{fig_histogram_multiple_results} shows a comparison between our results obtained with GIRAFFE vs UVES or GES vs FS data. There are no relevant abundance data for \ion{Si}{ii}, while only three stars have a \ion{Ne}{i} value from FS and GES GIRAFFE that differ by at most $\sim$0.2 dex. There is evidence for a slight offset at the $\sim$0.05-dex level between the surface gravities estimated from FS and GES GIRAFFE data. It is present irrespective of the availability of GES HR04 spectra. Such an offset could have various causes that are difficult to pinpoint. For instance, the different quality of the GIRAFFE HR03 and HR04 data between the two surveys (Sect.~\ref{sect_data}) might lead to a systematic bias in the placement of the continuum level for H$\delta$ and H$\gamma$ that are our main gravity diagnostics. As argued below, however, because a $\log g$ error of this magnitude has a limited impact on our results, its exact origin was not explored in great detail.

As expected, the determination of $T_\mathrm{eff}$ and $\log g$ is degenerate. Based on our data, we find that $\Delta T_\mathrm{eff}$ (K) $\sim$ 8 $\times$ 10$^3$ $\Delta\log g$. Correlations between the deviations in surface gravity and abundances are not clearly seen and may be buried in the noise. Because of the small impact of the $\log g$ differences in the vast majority of cases and of our inability to assess the accuracy of the determinations at this level, we proceed by averaging the results regardless of the instrumental set-up (GIRAFFE or UVES) or origin of the data (FS or GES)\footnote{The only exception to the rule is \object{GES\,10361503--5808043} for which the $V\sin i$ based on GES data appears underestimated (210 vs 302 km s$^{-1}$; Fig.~\ref{fig_histogram_multiple_results}) from the inspection of the fit to the metal features used for the abundance analysis (Sect.~\ref{sect_abundances}). In addition, ROBGrid obtained 300 km s$^{-1}$. All the GES results for this star are therefore ignored.}.

\subsection{External validation}\label{sect_external_validation}

\subsubsection{Against high-precision studies in the literature}\label{sect_external_validation_benchmarks}

\begin{figure}[h!]
\centering
\includegraphics[trim=60 165 40 380,clip,width=\hsize]{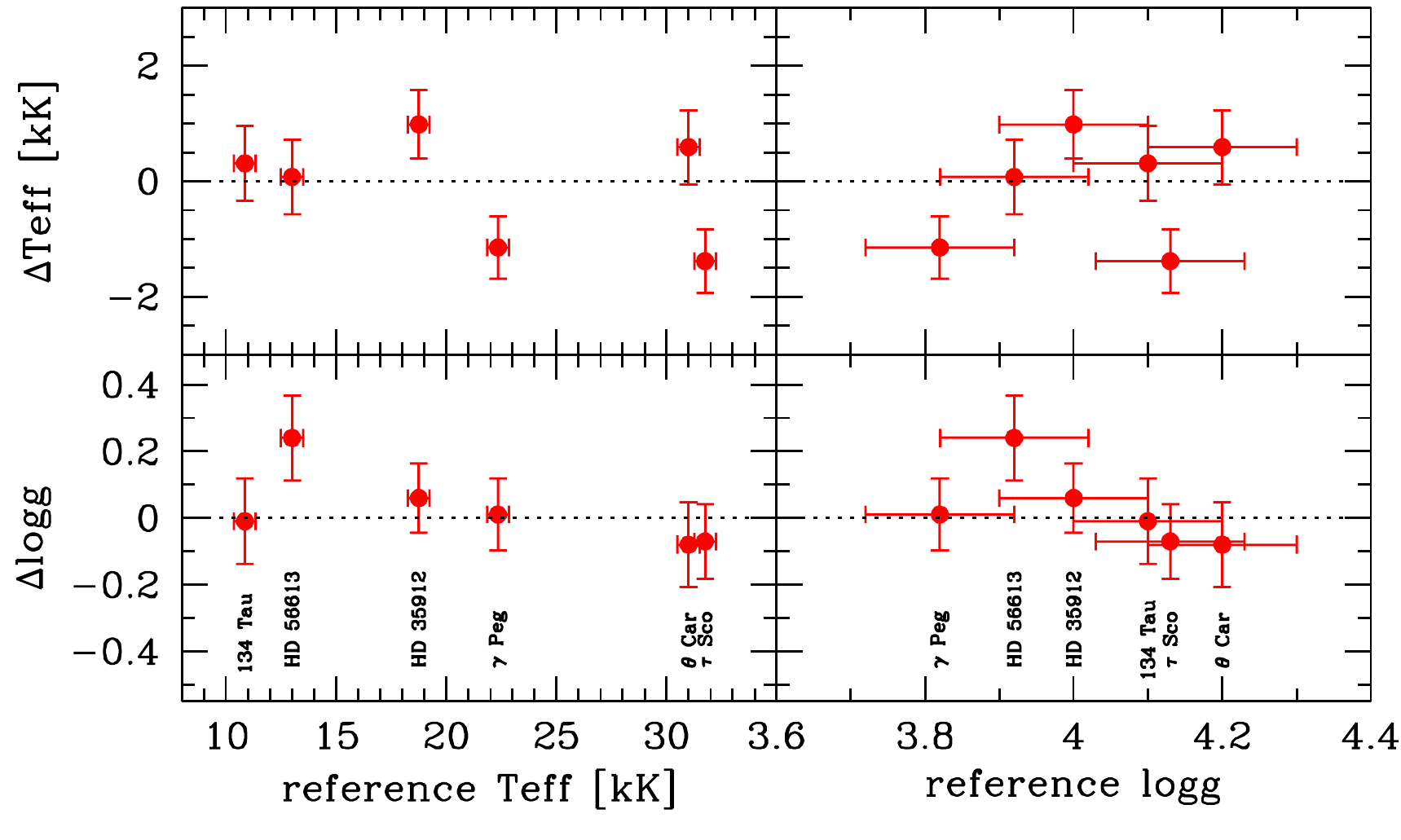}
\caption{Comparison for a set of GES benchmarks between our mean results and those taken from the literature \citep{pancino17,blomme22}. The differences are expressed as this study minus reference. The error bars along the vertical axis are the quadratic sum of our internal uncertainties and those in the reference values. The latter are assumed to be 500 K and 0.1 dex for $T_\mathrm{eff}$ and $\log g$, respectively.}
\label{fig_benchmarks}
\end{figure}

Our validation sample is made up of six well-studied, warm GES benchmarks \citep{pancino17,blomme22}. The reference, literature values are based on similar spectroscopic methods, but were obtained in the vast majority of cases completely independently \citep{smith_dworetsky93,mokiem05,simon_diaz06,morel_butler08,simon_diaz10,nieva_simon_diaz11,hubrig08,lefever10,martins12,nieva_przybilla12}. They are based on high-quality data analysed using state-of-the-art techniques and, as such, are believed to be of high precision. More details about the values adopted are given in Appendix~\ref{appendix_benchmarks}.  Owing to the lack of (nearly) model-independent measurements, it is worth emphasising that the reference values are not necessarily accurate. Evaluating the {\em accuracy} of our results would require the analysis of certain types of bona fide benchmarks \citep[e.g. individual components of detached eclipsing binaries;][]{pavlovski18}. Despite being less extensively studied, \object{HD 56613} and \object{$\theta$ Car} are valuable for validating our results for the NGC 3293 sample mainly made up of fast rotators because of their relatively high rotation rate \citep[$V\sin i$ $\sim$ 100 km s$^{-1}$;][]{lefever10}. Yet the general mismatch in terms of spin rates between our targets and the benchmark control sample must be kept in mind. A satisfactory agreement with the reference $T_\mathrm{eff}$ and $\log g$ data is found without any dependence as a function of these parameters (Fig.~\ref{fig_benchmarks}). Furthermore, our $V\sin i$ values are fully compatible with those determined in the papers above.

\subsubsection{Against results from other WG13 nodes}\label{sect_external_validation_other_nodes}

Stars in NGC 3293 were analysed by three other WG13 nodes: ON (Observat\'orio Nacional, Brazil) and two groups both based at the Royal Observatory of Belgium, ROB and ROBGrid, but that used different techniques. The reader is referred to \citet{blomme22} for full details about the analyses performed by these teams. In brief, ROBGrid followed a similar approach as in this study to estimate the parameters and performed a global spectral synthesis over wide spectral ranges, whereas ON and ROB relied on the detailed modelling of a set of selected lines. While ROBGrid provided results for most stars in the cluster, ON exclusively focused on slowly rotating, early B stars and ROB only considered A- and late B-type stars. Therefore, it should be kept in mind that each comparison sample occupies a distinct region of the parameter space.  

\begin{figure*}[h!]
\centering
\includegraphics[trim=50 175 0 415,clip,width=0.9\hsize]{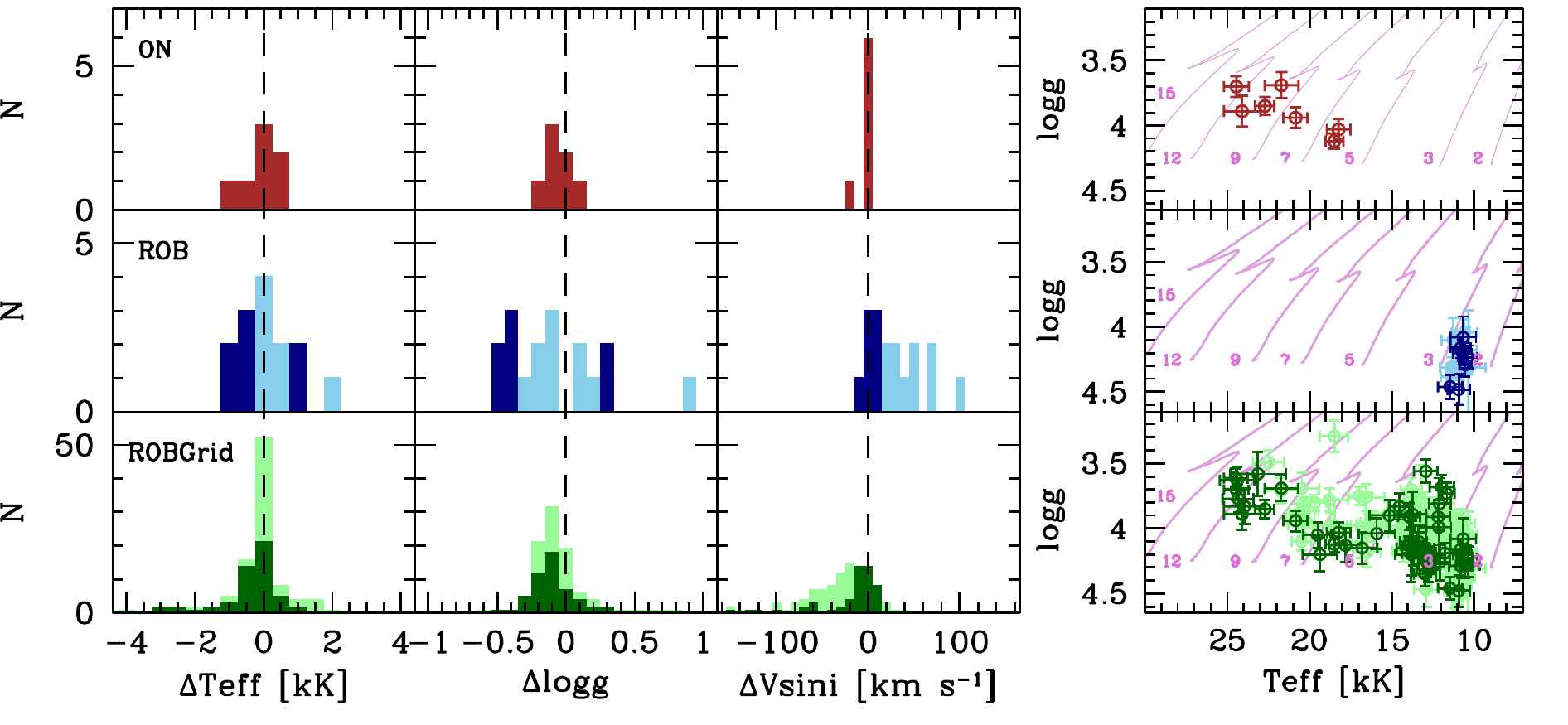}
\caption{Comparison with the results of other WG13 nodes. {\em Left panels}: differences between the stellar parameters expressed as this study minus other node. In each panel, fast rotators for which we estimate $V\sin i$ above 150 km s$^{-1}$ are indicated by lighter colours. Only one colour is used for the ON node because only slow rotators below that threshold were analysed. {\em Right panels}: position of each sub-sample in a Kiel diagram based on our parameters. Theoretical evolutionary tracks at solar metallicity and with $\omega_\mathrm{init}$ = 0.8 (see Sect.~\ref{sect_discussion_pnrc} for definition) are overlaid \citep{georgy13}. The initial stellar mass (in solar units) is indicated.}
\label{fig_other_nodes}
\end{figure*}

The comparison between our stellar parameters and those from the other nodes is shown in Fig.~\ref{fig_other_nodes}. While the $T_\mathrm{eff}$ scales are similar, there is some evidence -- especially with respect to the ROBGrid results -- that our surface gravities are systematically lower by $\sim$0.1 dex. The $V\sin i$ values are directly comparable because in all cases the contribution from macroturbulence is not separated out\footnote{The ON node convolved their synthetic spectra assuming a macroturbulence fixed to 5 km s$^{-1}$, but rotation largely dominates the line broadening for the seven stars in Fig.~\ref{fig_other_nodes}.}. As discussed by \citet{blomme22}, our estimates are dramatically lower and larger than those from ROBGrid and ROB, respectively. As seen in Fig.~\ref{fig_other_nodes}, it mostly concerns the fast rotators with $V\sin i$ $\ga$ 150 km s$^{-1}$. Whether it is related is unclear, but a limitation plaguing the ROBGrid results is that the sampling of their theoretical grid is sparse in this regime ($V\sin i$ step of 50 km s$^{-1}$). We find that adopting the ROB or ROBGrid value for the fast-rotating objects leads to considerably poorer fits to the metal features (e.g. \ion{Mg}{ii} $\lambda$4481). Furthermore, we do not find a systematic offset for this sub-sample with respect to the results obtained by the FS (see below for a more general comparison).

Flags indicating a possible single-lined binary were also set by ROBGrid. It supports our conclusions in about 60\% of cases, while for the eight remaining stars a flag was raised by one node only. It can be understood by the widely different methods and detection criteria adopted.

\subsubsection{Against results from the `VLT-FLAMES survey of massive stars'}\label{sect_external_validation_FS}

\begin{figure*}[h!]
\centering
\includegraphics[trim=20 585 0 150,width=\hsize]{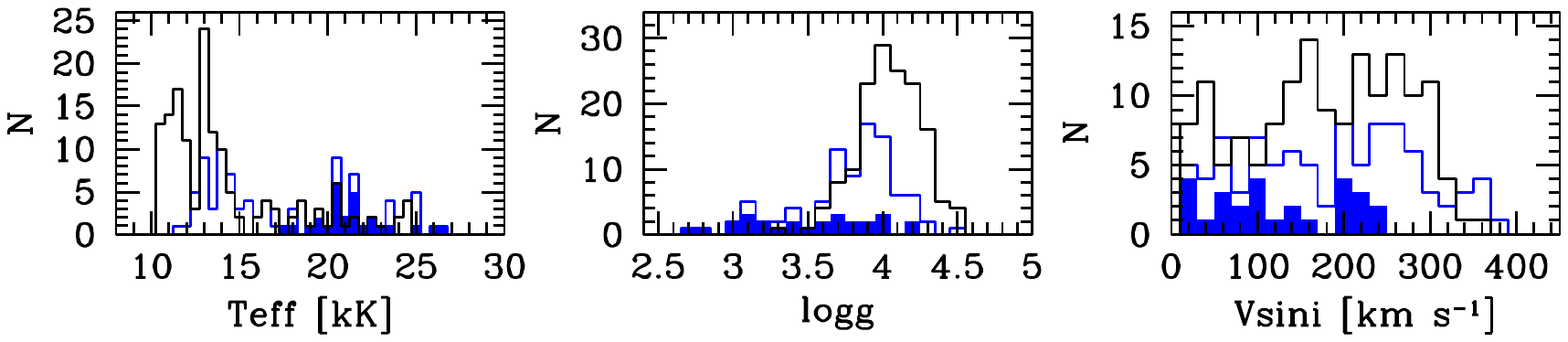}
\caption{Basic physical properties of the full GES  ({\em black}) and FS ({\em blue}) stellar samples. The distribution of the stars with C, N, Mg, or Si abundances from the FS is shown as a filled, blue histogram.}
\label{fig_parameters_FS_vs_GES}
\end{figure*}

Given that the FS carried out a comprehensive study of the brightest end of the cluster population and that this ambitious project attracted much attention in the community, it is of interest to confront our results to those of their final data release \citep{hunter09}. The basic properties of the full GES and FS samples are compared in Fig.~\ref{fig_parameters_FS_vs_GES}. Our targets are on average cooler, while the apparent difference in evolutionary status may be explained by a zero-point offset between both sets of $\log g$ estimates (see below). Noteworthy is that the GES went to much fainter magnitudes (down to $V$ $\sim$ 18 mag) and observed a large sample of late B dwarfs for which the FS did not provide any abundance data. A property shared by the two samples is fast rotation, with $\langle$$V\sin i$$\rangle$ $\sim$ 200 km s$^{-1}$.

\begin{figure*}[h!]
\centering
\includegraphics[trim=50 405 0 190,clip,width=0.85\hsize]{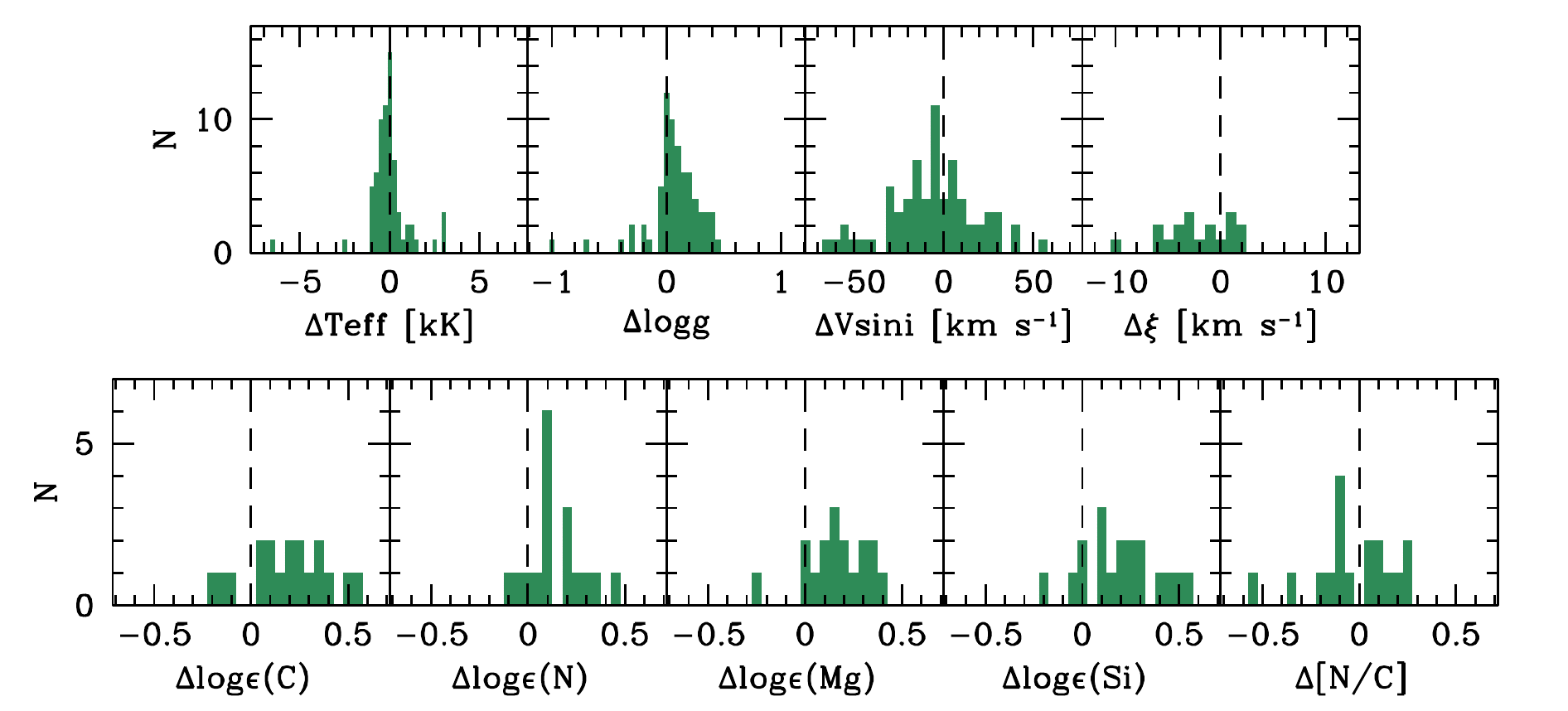}
\caption{Comparison between our results and those from the FS. The differences are expressed as this study minus FS. Our silicon abundances are based on \ion{Si}{iii}. The microturbulence was adjusted by \citet{hunter09} to derive the same Si abundance for each star in the cluster \citep[see][]{hunter07,trundle07}. Only the $\Delta \xi$ data for the stars with abundances from both this study and the FS are shown.}
\label{fig_comparison_parameters_FS}
\end{figure*}

As seen in Fig.~\ref{fig_comparison_parameters_FS}, a comparison of the parameters for the stars in common leads to the conclusion that our surface gravities and abundances with respect to hydrogen are systematically larger by $\sim$0.15--0.20 dex on average. The abundance offset could (partly) arise from the larger microturbulences often adopted by the FS. The strongest outlier by far in terms of $T_\mathrm{eff}$ and $\log g$ (\object{GES\,10355467--5813486} or \object{FS\,3293-032}) is one of the fastest rotators in both samples ($V\sin i$ $\sim$ 350 km s$^{-1}$) and a possible Be transient \citep{mc_swain09}. No abundances are determined in either case. Our much cooler $T_\mathrm{eff}$ and lower $\log g$ are clearly in better agreement with the ROBGrid results. As discussed by \citet{hunter_thesis}, the $T_\mathrm{eff}$ adopted by the FS for this star is calibrated on the spectral type that is ill-determined (in the range B0.5--B1.5 Vn).

\FloatBarrier

\section{Reference parameters for benchmarks}\label{appendix_benchmarks}

\begin{figure*}[h!]
\centering
\includegraphics[trim=50 455 30 180,clip,width=0.9\hsize]{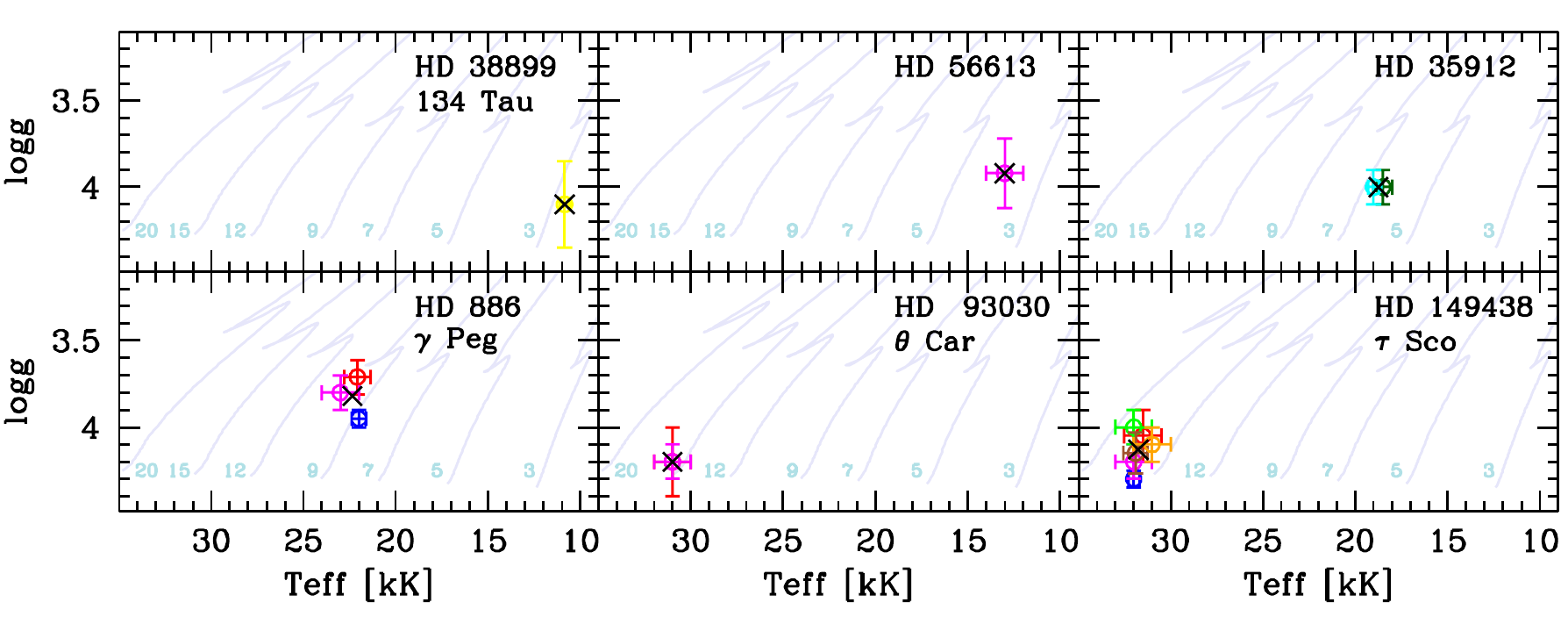}
\caption{Reference position of the benchmark stars used in this study in the $\log g$-$T_\mathrm{eff}$ plane. Colour coding for literature values: \citet[][{\em yellow}]{smith_dworetsky93}, \citet[][{\em magenta}]{lefever10}, \citet[][{\em green}]{simon_diaz06}, \citet[][{\em dark green}]{simon_diaz10}, \citet[][{\em cyan}]{nieva_simon_diaz11}, \citet[][{\em blue}]{nieva_przybilla12}, \citet[][{\em sienna}]{mokiem05}, \citet[][{\em red}]{morel_butler08}, \citet[][{\em red}]{hubrig08}, and \citet[][{\em orange}]{martins12}. The average, adopted parameters are shown as a black cross. Evolutionary tracks at solar metallicity and with $\omega_\mathrm{init}$ = 0.568 are overlaid \citep{ekstroem12}. The initial stellar mass (in solar units) is indicated.}
\label{fig_parameters_benchmarks}
\end{figure*}

Figure \ref{fig_parameters_benchmarks} shows a Kiel diagram for the stars used as benchmarks in this study.

\end{appendix}

\end{document}